\documentclass[conference]{IEEEtran}

\usepackage{00_preamble}
\usepackage{00_macros}

\begin{document}

\iffalse %enable this for NDSS
\IEEEoverridecommandlockouts
\makeatletter\def\@IEEEpubidpullup{6.5\baselineskip}\makeatother
\IEEEpubid{\parbox{\columnwidth}{
		Network and Distributed System Security (NDSS) Symposium 2027\\
		22--26 March 2027, Seoul, Republic of Korea\\
		ISBN 978-1-970672-09-1\\  
		https://dx.doi.org/10.14722/ndss.2027.[23$|$24]xxxx\\
		www.ndss-symposium.org
}
\hspace{\columnsep}\makebox[\columnwidth]{}}
% correct bad hyphenation here
\hyphenation{op-tical net-works semi-conduc-tor}
\fi

\title{SonicNudge: Controlled Displacement of Hovering UAVs via Estimator–Controller Coupling} % (UAVs)}

\author{%remove for submission
{\rm 
Shaocheng Luo,
Ashir Raza,
Haocheng Meng,
David Hunt,
Miroslav Pajic
}\\
Duke University
}

% make the title area
\maketitle

\begin{abstract}
UAV displacement attacks have traditionally relied on spoofing sensors that directly report position or translational motion, such as GNSS and optical flow.
In this work, we introduce \modelAbbr, a new attack primitive that instead targets the gyroscope and shows that low-level inertial errors can be transformed into controlled displacement of hovering or slow-moving UAVs.
The attack exploits estimator--controller coupling: a small gyroscope perturbation by ultrasonic resonance can persist as an attitude-estimation bias, and the flight controller can convert this biased estimate into a shifted hover point.
This behavior is especially relevant to UAV tasks that require hovering, station-keeping, slow approach, or precise final alignment, such as perimeter denial, inspection, docking, landing alignment, and close-proximity operation, where meter-scale position errors can be operationally meaningful.
We analyze this attack primitive in a PX4-style flight stack and validate it through 81 simulation runs and more than 10 indoor/outdoor physical experiments, showing that displacement is governed by estimator weighting, bias observability, and closed-loop position correction.
Our study suggests that UAV and vehicle-system security should look beyond direct navigation spoofing and pay closer attention to low-level inertial errors and estimator--controller coupling as a subtle but important attack surface.
\end{abstract}

\section{Introduction}
\label{sec:intro}

Unmanned aerial vehicles (UAVs) increasingly operate near buildings, private property, and restricted spaces. Many physical-layer UAV attacks aim to crash, destabilize, or trigger failsafes, but such outcomes are often difficult to reproduce in modern autopilots because of safety logic and recovery controllers. They are also poorly suited when the desired effect is predictable and directional rather than destructive. This paper studies a different operating regime: a controlled push-away that produces meter-scale displacement of a hovering or near-hovering UAV while preserving stable flight.

Existing UAV displacement attacks usually target sensors that directly encode position or translational motion. GNSS spoofing~\cite{tibaldo2025gnss, sathaye2022experimental} changes the vehicle's perceived global location, causing the navigation stack to follow false coordinates. Optical-flow spoofing~\cite{davidson2016controlling} manipulates apparent ground-plane motion, causing the UAV to compensate for fictitious lateral drift. In contrast, we ask whether displacement can arise from perturbing a gyroscope, a sensor that measures angular rate rather than position or translational velocity. This is not obvious: a gyro error may be rejected by sensor fusion, appear only as transient attitude noise, or cause instability rather than controlled motion. Demonstrating this pathway is significant because it expands UAV displacement attacks beyond sensors that directly report where the vehicle is or how it is translating. It shows that low-level inertial errors, if retained by the estimator and interpreted by the controller, can become position-level physical effects. Thus, \modelAbbr reveals a more subtle attack surface in the estimator--controller stack: UAV displacement can be induced not only by spoofing navigation inputs, but also by perturbing the attitude estimation that navigation control relies on.

Prior acoustic-injection work provides important foundations but does not answer the UAV-specific control question studied in this paper. Son et al.~\cite{son2015rocking} showed that acoustic interference with gyroscopes can destabilize and crash drones, establishing the security relevance of acoustic gyro attacks. However, their focus is disruption and loss of control rather than controlled displacement through estimator--controller coupling. Related studies such as WALNUT~\cite{trippel2017walnut} further showed that acoustic excitation can compromise MEMS sensor-output integrity, while Injected and Delivered~\cite{tu2018injected} and KITE~\cite{gao2022kite} demonstrated that out-of-band injection can manipulate embedded inertial sensors and create implicit control effects. These works primarily establish sensor-level vulnerability, system disruption, or general embedded-system controllability. They do not analyze how a small gyro perturbation propagates through a modern UAV flight stack consisting of EKF-based state estimation and cascaded position, attitude, and rate control. In particular, they do not explain when a gyro bias becomes a lateral displacement rather than merely sensor corruption, attitude error, or crash.

We present \modelAbbr, a UAV-focused acoustic displacement primitive for hovering UAVs. \modelAbbr exploits a well-known property of commodity MEMS gyroscopes: narrow ultrasonic resonances, often in the 20--30\,kHz band, that can be acoustically excited. By modulating an ultrasonic carrier near resonance, the injected component aliases into the estimator band as a bias-like term. The rate controller largely attenuates the ultrasonic carrier through low-pass behavior. Still, the EKF can integrate the aliased component as a residual roll/pitch estimation error, especially under common tuning practices such as down-weighted accelerometer tilt updates and a stiff gyro-bias random-walk model.

The key insight is that the UAV's estimator and controller convert this small inertial error into physical displacement. The EKF partially retains the injected gyro bias as a roll or pitch attitude-estimation error. In a cascaded control stack, the attitude controller regulates the vehicle's orientation using the estimated attitude, while the outer position controller uses that attitude response to correct horizontal motion. The inner attitude loop then tracks the biased attitude estimate, while the outer position loop interprets the resulting apparent tilt as horizontal acceleration and commands a compensating counter-tilt. Under normal position/velocity aiding, the closed loop converges to a displaced hover equilibrium: neither the physical tilt nor the position error can be driven to zero simultaneously under biased state feedback. 
For applications that depend on last-meter accuracy, even a tiny hover-point shift can push the UAV outside its authorized operating envelope.
This estimator--controller coupling yields a bounded, directionally programmable lateral offset---a controlled nudge rather than a crash.

The main contributions of this paper are as follows.

\begin{itemize}
    \item \textbf{Unveiling a gyro-to-position displacement pathway in UAV estimator--controller coupling.}
    We identify a critical but underexplored UAV security pathway: \textit{a small gyroscope perturbation can survive estimator  corrections and be converted by the cascaded flight controller into lateral displacement.} 
    This reveals that low-level inertial errors can become position-level physical effects even when navigation sensors, communication channels, and mission commands are not spoofed, shifting the focus of traditional security paradigms to the importance of estimator--controller coupling vulnerability in UAVs as well as more generic vehicle systems.

    \item \textbf{A control-level model and quantified simulation analysis of estimator--controller coupling.}
    We formalize how residual roll/pitch bias propagates through a PX4-style EKF and cascaded position--attitude--rate controller, explaining why the attack produces a displaced equilibrium and why the steady offset scales with residual tilt bias and outer-loop gains. Using a high-fidelity MATLAB/SIMULINK simulation, we conduct an 81-run EKF weighting sweep over a $9\times9$ grid of accelerometer trust and gyro-bias adaptation settings. The results show that the attack impact is strongest under weak accelerometer correction, a common estimator behavior in practical UAV operation, whereas slower gyro-bias adaptation can cause the nudged displacement to persist after excitation stops. This provides a systematic basis for reasoning about attack transferability, post-attack recovery, and estimator-level mitigation.

    \item \textbf{Alias-Gated Resonance for telemetry-driven signed gyro-bias injection.}
    Building on prior out-of-band inertial-sensor injection work, we design Alias-Gated Resonance (AGR), a phase-gated scheduling method for accumulating signed gyroscope bias and manipulating estimator--controller coupling. 
    AGR uses only standard low-rate UAV radio telemetry as feedback, which is broadly available in practical MAVLink-based operation, rather than high-speed Ethernet/USB telemetry or direct onboard access. It locks to the aliased gyro response and gates opposite-phase waveforms around predicted peak/trough intervals to reinforce the desired bias direction, avoiding real-time coherent twin-frequency switching while preserving direction-selective control. In physical flight tests, AGR operates on the observed $\sim$2--2.5\,Hz aliased oscillation and steers the UAV between displaced hover equilibria.
    Although realized here through ultrasound, AGR is a general phase-gated scheduling abstraction and can in principle apply to other mechanisms that induce phase-observable gyro perturbations, including laser-induced acoustic interference~\cite{shamsi2024wip} and transduction-based attacks~\cite{gao2022kite}.

    \item \textbf{Physical validation and cross-IMU boundary characterization.}
    We validate \modelAbbr on a real PX4-based UAV through indoor and outdoor physical experiments and characterize cross-IMU susceptibility across representative drone-grade IMUs/gyroscopes. In flight tests, \modelAbbr induces a peak roll-estimation bias of approximately $5.5^\circ$ and a lateral displacement of approximately $1.1$\,m while maintaining stable position control, echoing the predicted displaced-equilibrium behavior. AGR achieves a $46.7\%$ phase-aligned trigger rate in a representative run and steers the UAV between nearby displaced hover equilibria. We further evaluate 10 commonly used drone-grade IMUs/gyroscopes, including six not covered by prior datasets, and observe meter-scale effective standoff for several sensors in speaker-detached constrained tests. These results show that the gyro-to-position pathway is not limited to a single IMU, while clarifying practical boundaries such as bias persistence, axis asymmetry, yaw dependence, and attacker--victim geometry.

\end{itemize}

\section{Preliminaries}
\label{sec:preliminaries}

\begin{figure*}[t]
    \centering
    \begin{subfigure}{0.8\linewidth}
    \centering
        \includegraphics[width=0.99\linewidth]{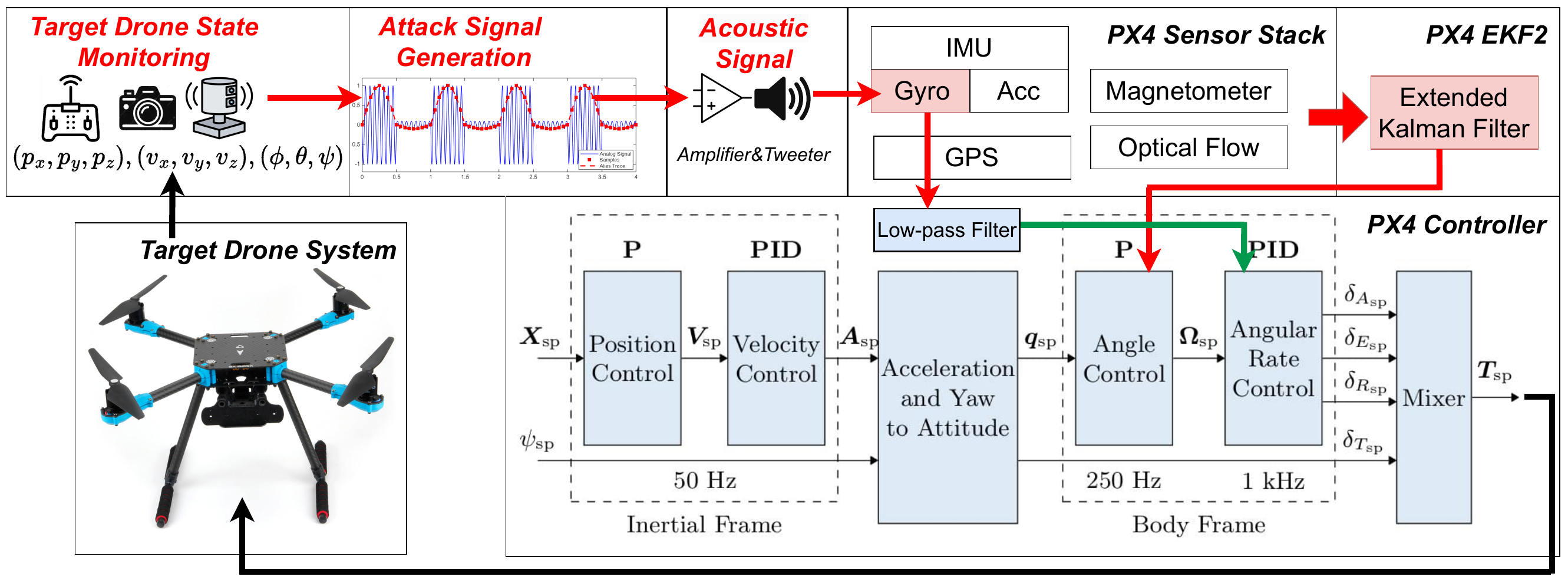}
        % \caption{}
    \end{subfigure}
    \caption{\modelAbbr\ pipeline: a modulated ultrasonic signal excites the IMU gyroscope into resonance, creating a quasi-steady rate bias that PX4’s cascaded controller converts into controlled repositioning.}
    \label{fig:sys_fig}
\end{figure*}

\subsection{UAV Flight-stack Overview}

Modern autopilots share a common structure: high-rate inertial sensing feeds a state estimator, which in turn drives a cascaded position$\to$velocity$\to$attitude$\to$rate controller \cite{mahony2012multirotor,alexis2012model}. We use PX4~\cite{px4:cuav_v6x} as a representative, widely deployed stack whose estimator/controller structure makes small attitude errors consequential (Fig.~\ref{fig:sys_fig}).

PX4's \emph{sensor stack} samples the IMU, including the 3-axis gyroscope and accelerometer, at high rate and supplements it with slower drift-bounding sources such as GNSS, magnetometer, barometer, and optionally optical flow. Because the gyroscope provides the low-latency body-rate signal, short-term attitude propagation is dominated by the IMU. PX4's EKF (\texttt{EKF2}) fuses these streams to estimate position, velocity, attitude, and IMU biases. If a gyro bias is not flagged as a fault, the EKF can absorb it as a persistent roll/pitch tilt.

Downstream, the cascaded controller converts a position setpoint to velocity, acceleration/tilt, attitude, and finally body-rate commands; the inner rate loop tracks these commands using the gyro stream. Hence, a persistent EKF roll/pitch bias can be interpreted as unintended lateral acceleration, causing the outer loop to command a steady horizontal correction. This estimator--controller coupling is the core pathway exploited by \modelAbbr. Further details on PX4's EKF and cascaded controller are provided in App.~\ref{sec:px4_coupling}.

\subsection{Telemetry as Attack Input}
\label{sec:telemetry}

\modelAbbr uses telemetry as the feedback input for attack scheduling and attacker--victim geometry estimation. In PX4-based systems, telemetry is commonly exchanged between the UAV and a ground station via MAVLink~\cite{koubaa2019micro}, enabling real-time mission supervision, logging, and failsafe awareness for operators. 
We assume that the adversary can receive this telemetry stream over the wireless link, consistent with prior PX4- and MAVLink-related security studies~\cite{bithas2020uav,allouch2019mavsec,kim2019rvfuzzer,kim2024systematic}.

The required feedback can be obtained from \textit{standard MAVLink state messages} over low-rate radio telemetry. \texttt{ATTITUDE} (ID~30) provides roll, pitch, yaw, and body rates for alias-phase estimation, while \texttt{LOCAL\_POSITION\_NED} (ID~32) and \texttt{GLOBAL\_POSITION\_INT} (ID~33) provide position, velocity, altitude, and heading for geometry and drift monitoring. Raw inertial streams such as \texttt{HIGHRES\_IMU} (ID~105) or \texttt{RAW\_IMU} (ID~27) can be used when available, but are not required. Because our attack tracks a low-frequency alias, typically $1$--$10\,\mathrm{Hz}$, tens-of-Hz attitude telemetry 
%, and optional 50\,Hz IMU telemetry
over a standard radio link~\cite{meier2015px4} is sufficient for phase-gated scheduling; the attack does not rely on high-speed Ethernet or USB telemetry. In our PX4/Pixhawk~V6X setup, this rate is sufficient for the observed 2--2.5\,Hz aliased response.
% \textbf{Attack preparation via sweep-and-lock.}
% Rather than assuming a priori knowledge of the exact aliased frequency, e.g., $\approx$1\,Hz, the attacker sweeps a narrow ultrasonic band around the expected MEMS resonance and monitors the victim's low-frequency attitude response in real time. In our setting, this feedback is obtained from telemetry, while a telemetry-free deployment would require external tracking to recover the same phase and attitude cues. Due to sampling and aliasing, an excited ultrasonic carrier near resonance appears in the EKF band as a low-frequency bias-like term; empirically, a suitable low-frequency component is consistently observable during sweep (App.~\ref{sec:sweep_and_aliased}). Once a responsive carrier is found, the attacker \emph{locks} onto it and applies envelope modulation to shape the effective bias. This feedback-driven procedure tolerates sampling jitter and clock drift because the sweep and lock can be repeated or refreshed when conditions change.

\section{Threat Model}
\label{sec:threat_model}

\subsection{Victim Model}
\label{sec:victim}

\textbf{Victim platform.}
We assume the PX4-style multirotor flight stack described in Sec.~\ref{sec:preliminaries}: an EKF-based estimator and a cascaded position/velocity $\to$ attitude $\to$ rate controller. The victim uses a commodity MEMS IMU together with standard drift-bounding sources such as GNSS, magnetometer, barometer, and optical/visual flow. Our attack targets the gyroscope, whose resonances often lie in the 20--30\,kHz band and can be acoustically excited~\cite{son2015rocking,tu2018injected,trippel2017walnut}.

\textbf{Estimator conditions.}
The vulnerability arises when ultrasonic excitation induces a small, low-frequency \emph{bias-like} term in the measured body rates. Although the rate controller attenuates the ultrasonic carrier, the EKF may integrate the aliased component and retain it as a residual roll/pitch attitude error. This leakage is persistent under two estimator behaviors below that are common in UAV stacks and \textit{broadly used} to maintain stable attitude estimates under noisy inertial sensing:

\textbf{Setting~1
\phantomsection\label{asp:1}
(Accelerometer updates down-weighted).}
During vibration, gusts, or translational maneuvers, accelerometer updates may be gated, rejected, or assigned larger covariance, shifting attitude estimation toward gyro-dominant propagation and weakening tilt correction.

\textbf{Setting~2
\phantomsection\label{asp:2}
(Stiff gyro-bias model).}
Gyro-bias process noise may be tuned small to suppress jitter, slowing adaptation to slowly varying injected biases and leaving a mismatch between the true and estimated gyro bias.

\textbf{Closed-loop consequence.}
A persistent EKF roll/pitch error can be converted by the cascaded controller into lateral displacement. The inner attitude loop tracks the biased attitude estimate, while the outer position loop interprets the apparent tilt as horizontal acceleration and commands a compensating counter-tilt. Under normal position/velocity aiding, this produces a \emph{bounded} displaced equilibrium rather than an immediate crash (Sec.~\ref{sec:control}, App.~\ref{sec:px4_coupling}). Under weaker position correction (e.g., GNSS-denied operation), estimator saturation, or controller limits, the same bias--tilt pathway can produce larger excursions until constrained by saturation, safety limits, or failsafe transitions.

\subsection{Adversary Model}
\label{sec:adversary}

We assume an \emph{acoustic} adversary equipped with a directional ultrasonic projector, implemented as a single transducer or an array. The adversary may be stationary, such as a rooftop or mast-mounted emitter, or mobile, such as a chasing UAV. The adversary must maintain line-of-sight, a favorable acoustic incidence angle, and sufficient acoustic pressure ($>105\,dB$ in our cases) at the target IMU. Our evaluated setting focuses on hover or slow flight, where maintaining incidence and timing feedback is most reliable; faster moving-target deployment requires stronger tracking, pointing, and acoustic power management under changing geometry.

\textbf{Telemetry and feedback.}
We assume that the adversary can receive the victim UAV's standard MAVLink telemetry stream over radio as an observation channel, as elaborated in Sec.~\ref{sec:telemetry}, but does not inject commands or modify the mission. The attack does not require custom topics, private onboard data, or high-speed debug links; it uses \textit{ordinary state telemetry} exposed on the radio link. This is weaker than protocol takeover: even if command injection is blocked by flight-stack validation, MAVLink signing, or operator supervision, telemetry may remain observable on broadcast, weakly protected, or otherwise accessible monitoring links. In our attack, telemetry supports two functions: estimating the low-frequency aliased gyro phase for acoustic scheduling and maintaining attacker--victim geometry, such as rear/side or top-down acoustic incidence. This assumption lets us evaluate the core acoustic-injection and estimator--controller pathway without solving fully black-box phase recovery or target-relative positioning. Telemetry-free operation using cameras, LiDAR, radar, or other perception sensors is a deployment extension not demonstrated in this paper.

\textbf{IMU identification and sweep-and-lock.}
The attacker does not assume prior knowledge of the exact aliased frequency or arbitrary control over the victim's IMU output. Instead, the attacker first narrows the resonance search space by identifying the vehicle or IMU model family, using or inferring information such as drone make/model, flight-controller enclosure, public part lists, or common autopilot BOMs. The attacker then performs a short sweep around the expected MEMS resonance band and monitors the victim's low-frequency attitude response. Once a responsive carrier is found, the attacker locks onto it and applies bias-preserving modulation. This feedback-driven procedure accounts for manufacturing variation, installation effects, sampling jitter, and clock drift, and can be refreshed when conditions change.

\textbf{Operating range.}
The attacker operates within the projector's effective range, where sufficient acoustic pressure reaches the IMU to excite resonance. Our evaluation uses short standoff distances and safety-bounded conditions to demonstrate the mechanism, rather than to maximize range. Higher-power emitters or arrayed projectors can extend range and scale up coupling \cite{kumar2017long,qiu2022review,tu2018injected}, but the feasibility and safety of such range extension depend on acoustic propagation, incidence angle, sensor packaging, and environmental constraints \cite{gao2022kite,gao2023exploring}. To account for sensor diversity, we evaluate multiple representative IMUs/gyros in Sec.~\ref{sec:imu_generality}.

\subsection{Attack Surface Analysis}
\label{sec:attack-surface}

\textbf{Gyro resonance and target channel.}
The attack targets the gyroscope channel of commodity MEMS IMUs. Prior work shows that such sensors can exhibit narrow ultrasonic resonances that leak into inertial readings~\cite{tu2018injected,son2015rocking,trippel2017walnut}. Our prototype uses the MPU-6500~\cite{invensense_mpu6500} as a representative sensor, but the same pathway applies to other MEMS gyros that admit acoustic excitation. The effective carrier band, affected axis, and coupling strength may vary across sensor models and installations.

\textbf{Axis dependence.}
\modelAbbr focuses on roll/pitch-dominant coupling because residual roll/pitch bias directly maps to lateral acceleration under multirotor control. Some platforms may instead exhibit stronger yaw-axis coupling or different mounting-dependent mode shapes~\cite{gao2022kite,trippel2017walnut}. In those cases, the sweep-and-lock procedure can still identify responsive axes, but the closed-loop effect may differ, e.g., yaw drift rather than lateral displacement. We explicitly scope our evaluation to roll/pitch-dominant cases and treat yaw-dominant resonance as an important boundary condition for generality.

\textbf{Directional sensitivity.}
Attack strength also depends on acoustic incidence direction because of enclosure geometry, PCB modes, IMU mounting, and axis placement~\cite{gao2022kite}. In the main flight experiments, we use controlled top-down coupling to make incidence reproducible and reduce moving-target beam-control confounds. In Sec.~\ref{sec:imu_generality}, we further evaluate rear/side speaker-detached excitation in constrained tests. Fully mobile realization with continuously changing geometry remains outside the scope of this paper.

\subsection{Risk Assessment}
\label{sec:risk}

\textbf{Bounded displacement and practical impact.}
The induced displacement is intentionally \emph{bounded} under normal position-hold control: the UAV remains stable but settles at a shifted hover point. This is not a weak form of the attack, but the operating regime in which a controlled nudge is most useful. For example, biasing a hovering UAV away from a boundary can keep it outside a restricted area; shifting a drone during inspection can move its camera away from the intended crack, window, cable, or structural target; and perturbing final alignment can disrupt docking, landing, package drop-off, or close-proximity navigation without requiring a crash. In crowded or indoor airspace, a meter-scale offset can also create controlled separation from people, equipment, or other UAVs using a less destructive effect than forced failure. We make no claim that the demonstrated displacement or standoff range is universally sufficient; the effect depends on acoustic coupling, incidence angle, estimator tuning, controller gains, and environment. In weaker-aiding or saturated regimes, the same estimator--controller pathway can produce larger excursions, while our physical evaluation keeps the effect safety-bounded to validate the mechanism.

\textbf{Scope and limitations.}
We do not claim that the demonstrated standoff distances, acoustic power levels, or displacement magnitudes are universal. Attack impact depends on sensor resonance, estimator tuning, controller gains, and environmental conditions; fully detached, black-box moving-target deployment remains outside this paper's scope. Nevertheless, this paper uses acoustic excitation as an exemplar way to inject controlled gyro errors and study the broader estimator--controller coupling vulnerability. Ethical and collateral-risk considerations are discussed separately in the Ethical Considerations section.

\section{Gyro-Bias Signal Generation}
\label{sec:sig_gen}

This section describes how an attacker can generate a biased gyroscope signal that later propagates through the UAV estimator--controller stack. Although our implementation uses ultrasonic excitation of MEMS gyroscopes, the signal-generation logic is not limited to acoustic attacks. \emph{The same abstraction applies to any modality that can induce a phase-observable perturbation in the gyro measurement chain}, such as transduction-based~\cite{gao2022kite} or laser-induced acoustic interference~\cite{shamsi2024wip}, provided that the perturbation can be shaped into a low-frequency bias-like component. We first formulate how a high-frequency resonant excitation is aliased into a low-frequency gyroscope output, and then describe three bias-shaping mechanisms for producing a nonzero mean in the measured angular rate.

\subsection{Frequency Formulation of Resonant Excitation}
\label{sec:signal_formulation}
Ultrasonic excitation at physical frequency $F$ drives a gyroscope near resonance. After sampling by the IMU ADC at rate $F_S^{\mathrm{IMU}}$ and again by the PX4 estimator at $F_S^{\mathrm{PX4}}$, the EKF does not ``see" the carrier but a \emph{low-frequency alias} $f_d \in [0, \tfrac{1}{2}F_S^{\mathrm{PX4}}]$. Intuitively, tones above Nyquist rate fold during IMU sampling, and the subsequent PX4 rate conversion folds them again, yielding a slow, bias-like component in the gyro stream that can leak into roll/pitch. The exact aliasing relation can be found in \cite{tu2018injected}.

\textbf{MPU-6500 instantiation.} On a representative PX4 platform with an MPU-6500 (gyroscope resonance in the $\sim27$kHz band) sampled at $F_S^{\mathrm{IMU}}\!\approx\!1$\,kHz and an EKF rate $F_S^{\mathrm{PX4}}\!\approx\!250$\,Hz, we sweep within the band and select $F$ so that the EKF observes a convenient alias (e.g., $f_d\!\approx\!1$\,Hz). That slow component integrates to a small, quasi-steady attitude error under realistic EKF tuning, enabling our controlled “nudge.”

\subsection{Bias Injection in Gyroscope Readings}
\label{sec:amplitude}

\begin{figure}[!t]
    \centering
    % Row 1
    \begin{subfigure}[b]{0.23\textwidth}
        \centering
        \includegraphics[width=\textwidth,trim=20mm 0 0 0,clip]{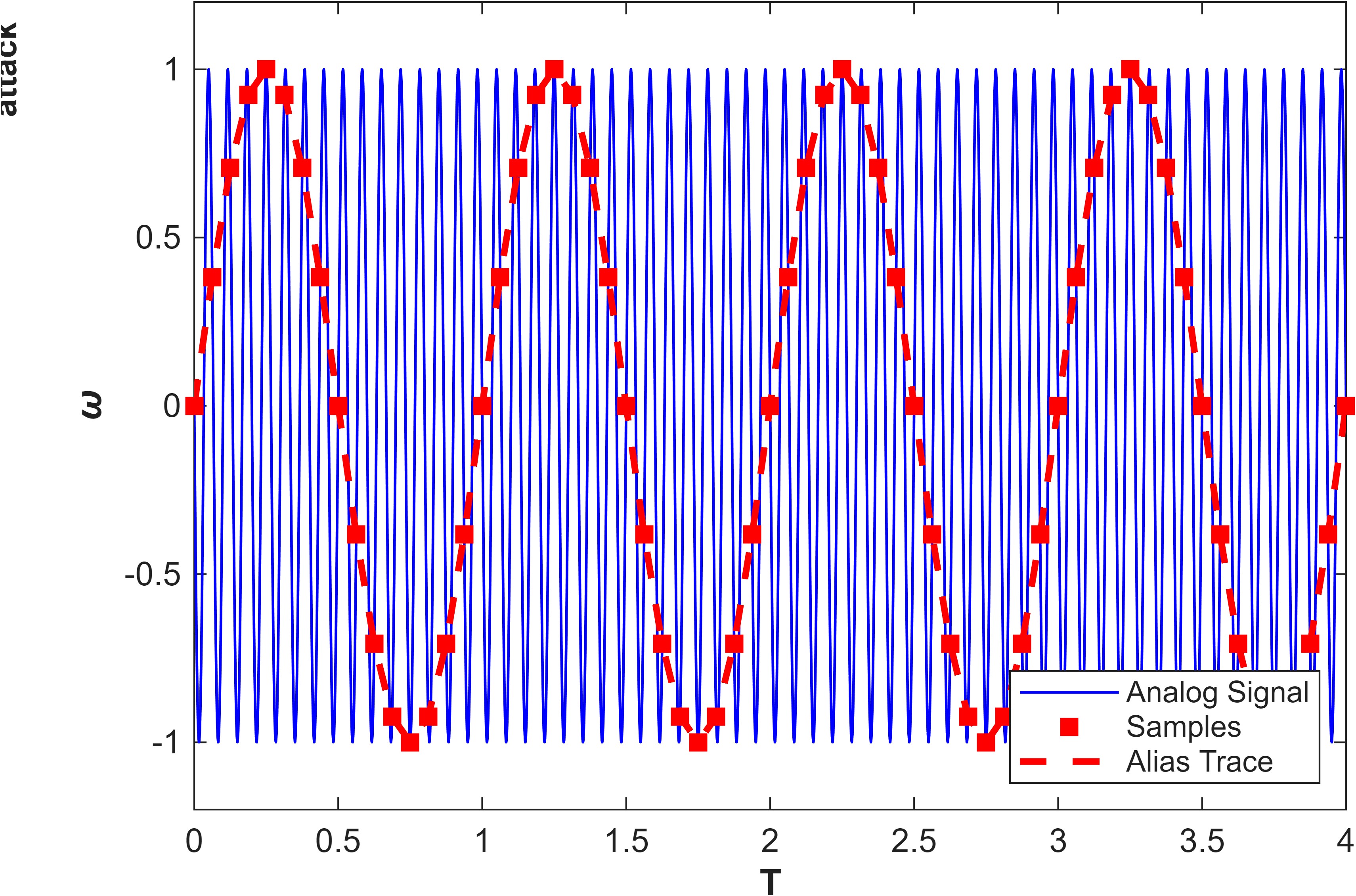}
        \caption{Non-biased waveform}
        \label{fig:no_bias}
    \end{subfigure}
    \hfill
    \begin{subfigure}[b]{0.23\textwidth}
        \centering
        \includegraphics[width=\textwidth,trim=20mm 0 0 0,clip]{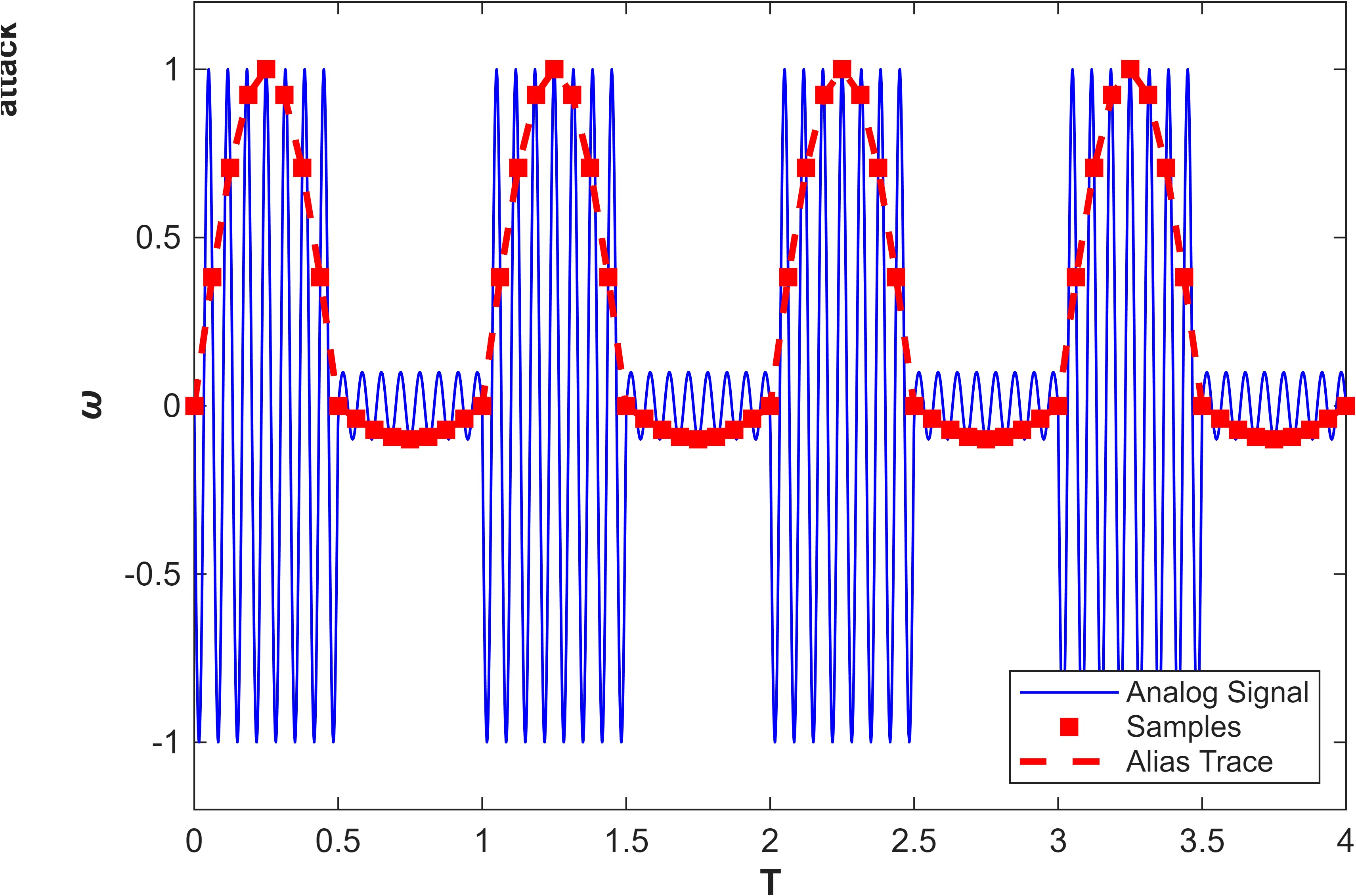}
        \caption{DAM}
        \label{fig:DAM}
    \end{subfigure}
    
    \vspace{1em} % space between rows
    
    % Row 2
    \begin{subfigure}[b]{0.23\textwidth}
        \centering
        \includegraphics[width=\textwidth,trim=5mm 2mm 15mm 0,clip]{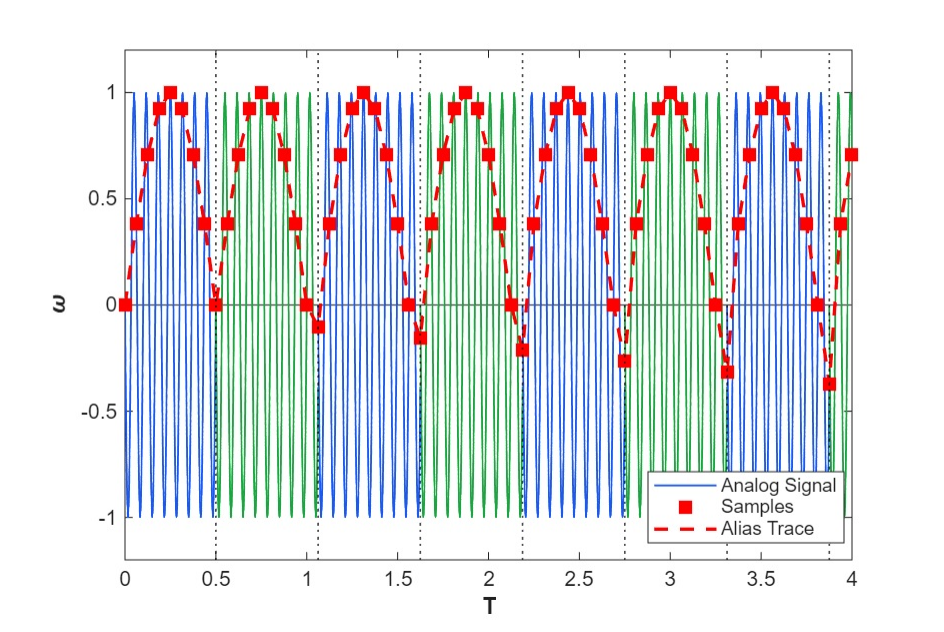}
        \caption{WCS}
        \label{fig:WCS}
    \end{subfigure}
    \hfill
    \begin{subfigure}[b]{0.23\textwidth}
        \centering
        \includegraphics[width=\textwidth,trim=20mm 0 0 0,clip]{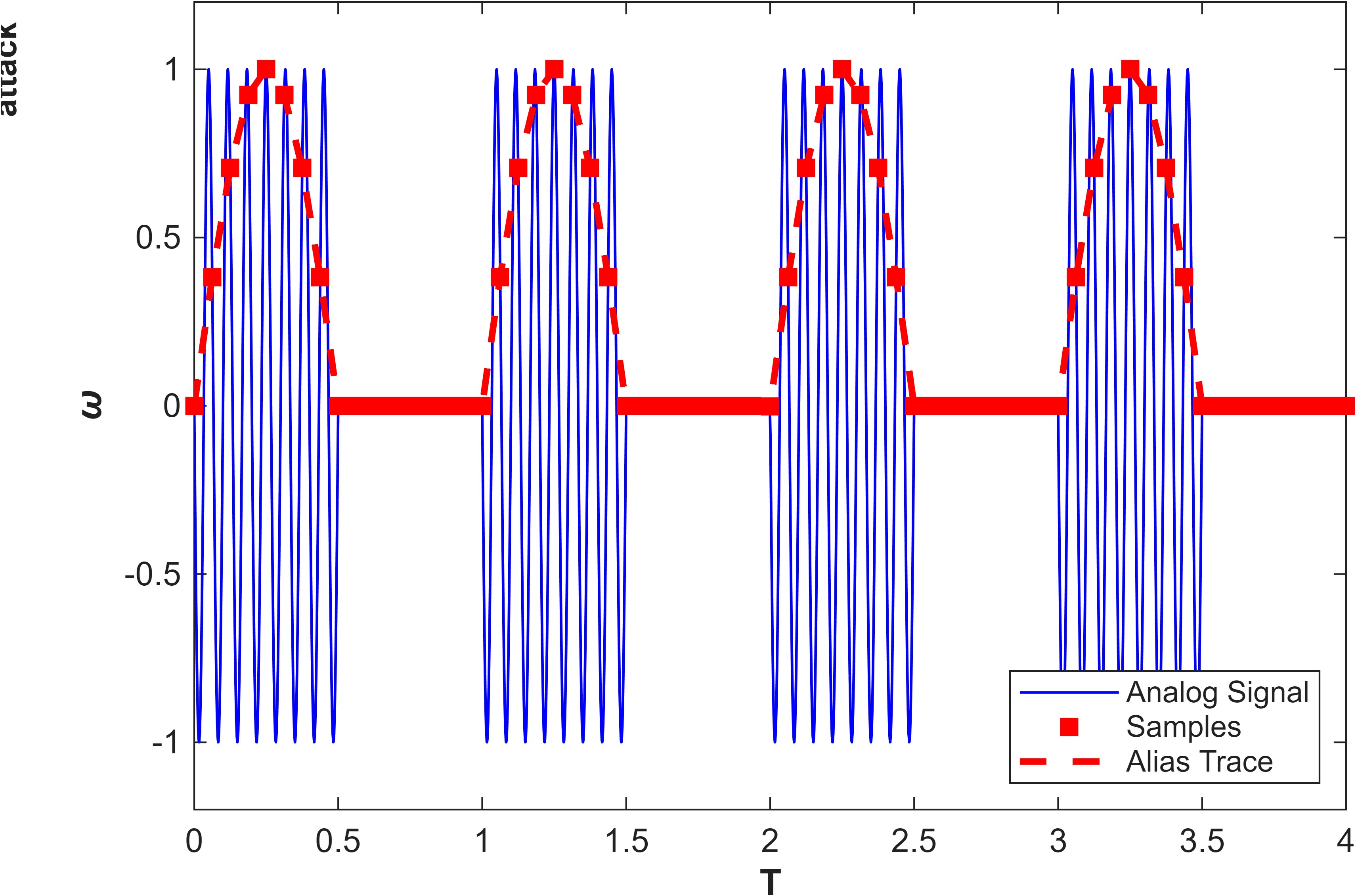}
        \caption{AGR}
        \label{fig:AGR}
    \end{subfigure}
    
    \caption{Resonance waveforms for gyro bias injection.}
    \label{fig:waves}
\end{figure}

Once the aliased digital gyroscope frequency $f_d$ is fixed (Sec.~\ref{sec:signal_formulation}), the remaining challenge is to introduce a \emph{net bias} (nonzero mean) in the measured rate.\footnote{The sign of the bias determines the drift direction but not the existence of repositioning. For clarity, we focus on a positive effective bias in the remainder of this section and \modelAbbr controls.} A direct DC offset at the signal source is ineffective: typical audio amplifiers and tweeters are AC-coupled and reject low-frequency and DC components. Instead, the attack must encode the bias in the \emph{envelope} or \emph{gating} of an otherwise AC-coupled resonant signal. We describe three feasible mechanisms.

For notational simplicity, we denote by $F_S$ the effective sampling rate at which the EKF (or flight stack) consumes the gyroscope stream (i.e., $F_S^{\mathrm{PX4}}$ in Sec.~\ref{sec:signal_formulation}), and by $F_{\mathrm{obs}}$ the aliased frequency seen at that rate.

\textbf{Digital Amplitude Modulation (DAM).}
In DAM, the resonant waveform is modulated by a slowly varying amplitude:
\begin{equation}
    s[i] = A[i] \cdot \sin\!\left(2\pi F_{\mathrm{obs}} \tfrac{i}{F_S} + \varphi_{0}\right),
    \qquad i \in \{0,1,2,\ldots\},
    \label{equ:modulated}
\end{equation}
where $A[0], A[1], \ldots$ is a sample-dependent envelope implemented by changing the drive strength of the acoustic source. A simple and practical choice is to use one gain $A_+\!\geq 0$ on the positive half-cycles and another $A_-\!\geq 0$ on the negative half-cycles, as illustrated in Fig.~\ref{fig:DAM}. In this case, the sampled mean (bias) over any integer number $N$ of periods satisfies
\begin{equation}
    \mu_{\mathrm{DAM}} \;=\; \frac{1}{N}\sum_{i=0}^{N-1} s[i]
    \;\approx\; \frac{A_+ - A_-}{\pi},
    \label{equ:dam_bias}
\end{equation}
so the maximum bias is $\mu_{\mathrm{DAM}}^{*} = A_+/\pi$ when $A_- = 0$.

DAM produces a relatively smooth, natural waveform and is therefore well suited for simulation studies (Sec.~\ref{sec:simulation}). In physical experiments, however, it requires amplitude-programmable excitation, which may not be supported by standard off-the-shelf function generators. 
% However, it is sensitive to relative phase \textcolor{red}{has issues}: if the modulation starts at an unfavorable phase of the aliased gyro signal, the downsampled sequence may exhibit reduced amplitude or even an effective bias with~the~opposite~sign.

\textbf{Waveform-Coherent Switching (WCS).}
WCS exploits the non-uniqueness of $F$ for a desired alias $f_d$. Consider two sinusoidal candidates $s_1(t) = A \sin(2\pi F_1 t + \varphi_1)$ and $s_2(t) = A \sin(2\pi F_2 t + \varphi_2)$,
where $F_1 \approx F_2$ and both produce the same low-frequency alias $f_d$ after the PX4 sampling pipeline.
To switch from $s_1$ to $s_2$ at time $t_s$ without introducing an amplitude discontinuity, we require the value continuity condition
$s_1(t_s) = s_2(t_s)$, 
which implies a phase alignment
$2\pi F_1 t_s + \varphi_1 \equiv 2\pi F_2 t_s + \varphi_2 \pmod{2\pi}.$
Equivalently,
\begin{equation}
    t_s = \frac{\varphi_2 - \varphi_1 + 2k\pi}{2\pi(F_1 - F_2)}, 
    \qquad k \in \mathbb{Z},
    \label{equ:switching_time}
\end{equation}
with $k$ chosen so that $t_s$ lies in the desired time window.

Discretizing at the EKF rate with $t_s = i_s/F_S$, a stylized digital model of WCS is
\begin{equation}
    s[i] =
    \begin{cases}
        A \cdot \sin\!\bigl(2\pi F_{\mathrm{obs},1}\tfrac{i}{F_S} + \varphi_{1}\bigr), & i \le i_s, \\[4pt]
        A \cdot \sin\!\bigl(2\pi F_{\mathrm{obs},2}\tfrac{i}{F_S} + \varphi_{2}\bigr), & i > i_s,
    \end{cases}
    \label{equ:phase}
\end{equation}
where $F_{\mathrm{obs},1}$ and $F_{\mathrm{obs},2}$ both alias to the same $f_d$. If we choose $\lvert \varphi_{1} - \varphi_{2} \rvert = \pi$ and perform the switch near a zero crossing, the resulting envelope resembles a half-wave-rectified sinusoid; the corresponding maximum bias is approximately $\mu_{\mathrm{WCS}}^{*} = 2A/\pi$ (Fig.~\ref{fig:WCS}).

Compared to the fixed-switching strategy in \cite{tu2018injected}, which is highly sensitive to sampling drift, the WCS conceptually enables phase-aware switching based on observed drone motion. In practice, however, reliably identifying and maintaining a pair of ``twin" excitation frequencies that yield the same alias but different phases is difficult in real time, which limits WCS as a primary mechanism in our physical experiments.

\textbf{Alias-Gated Resonance (AGR).}
To avoid the complexity of DAM envelopes and the twin-frequency search of WCS, we introduce an alias-gated approach, AGR. AGR does not track the ultrasonic carrier directly; it tracks a low-frequency phase proxy that is phase-correlated with the aliased gyro response. In simulation, MAVLink-style attitude/body-rate telemetry and the aliased gyro component are frequency-locked, which allows peak/trough timing to be inferred from telemetry. In physical experiments, we use MAVLink attitude/body-rate telemetry at tens of Hz as this practical proxy, sufficient for the observed 2--2.5\,Hz alias. If this phase relationship drifts or becomes unreliable, AGR can re-fit the alias phase over a rolling window before issuing new gates.

The attacker then gates the ultrasonic source so that only one half-cycle of the effective aliased response is excited. Conceptually, the digital signal seen by the EKF is
\begin{equation}
    s[i] =
    \begin{cases}
        0, \qquad\qquad \text{if } \sin\!\left(2\pi F_{\mathrm{obs}} \tfrac{i}{F_S} + \varphi_{0}\right) < 0,  \\[4pt]
        A \cdot \sin\!\left(2\pi F_{\mathrm{obs}} \tfrac{i}{F_S} + \varphi_{0}\right), \qquad  \text{otherwise},
    \end{cases}
    \label{equ:bias_AGR}
\end{equation}
which is equivalent to half-wave rectification of the aliased sinusoid. The resulting mean (bias) over one period is $\mu_{\mathrm{AGR}}^{*} = \frac{A}{\pi}$, and the waveform is shown in Fig.~\ref{fig:AGR}. 

AGR can be implemented in hardware by toggling the ultrasonic drive (e.g., via a gate or enable pin on the amplifier or function generator) whenever the estimated alias enters the ``undesired" half of its cycle. A deterministic trigger executes this gate, so no per-sample amplitude programming or twin-frequency handling is required. Phase-selective gating yields a net, sign-controlled bias-like component in the gyroscope reading and proved most robust in our physical experiments. The brief zero interval between half-waves in Fig.~\ref{fig:AGR} reflects gate dead time; reducing this dead time (faster switching) increases the on-duty cycle and thus the~long-term~average~bias.

\section{\modelAbbr Control Scheme}
\label{sec:control}

We analyze \modelAbbr in two phases. During the attack, the injected signal biases the gyroscope reading, the EKF converts this rate bias into a roll/pitch attitude error, and the cascaded controller maps the attitude error into lateral displacement. After the attack stops, the remaining displacement depends on how quickly the EKF bias state and attitude estimate recenter. In simulation, we use DAM (Sec.~\ref{sec:amplitude}) for smoothly controlled amplitude envelopes; in physical experiments, we use AGR (Sec.~\ref{sec:agr_algo}) because our hardware supports reliable on/off gating rather than continuous amplitude modulation.

\subsection{During the Attack}
\label{sec:init_attack}

We consider the standard roll--pitch EKF that fuses gyroscope and accelerometer tilt observations (details in App.~\ref{sec:ekf}). The EKF uses process noise $\mathbf Q_k$ (including gyro noise and bias random walk) and measurement noise $\mathbf R_k$ (accelerometer tilt), parameterized as
\begin{equation}
{%\footnotesize
\begin{aligned}
\mathbf{Q}_{\scriptstyle k} &= \mathrm{diag}\!\Bigl(
    \sigma_{gr,\phi}^2\,\Delta t^2,\;
    \sigma_{gr,\theta}^2\,\Delta t^2,\;
    \sigma_{b,p}^2\,\Delta t,\;
    \sigma_{b,q}^2\,\Delta t
\Bigr), \\
\mathbf{R}_{\scriptstyle k} &= \mathrm{diag}\!\Bigl(
    \sigma_{acc,\phi}^2,\;
    \sigma_{acc,\theta}^2
\Bigr).
\end{aligned}
}
\label{equ:QR}
\end{equation}
Let $K_{\phi,k}$ and $K_{b,p,k}$ denote the scalar gains, from the roll-angle innovation, that update $\hat\phi$ and $\hat b_p$, respectively.

\textbf{Injected rate.}
During the attack, the measured roll rate is
\begin{equation}
p_{\mathrm{meas},k}=p_k+p_{m,k},\qquad p_{m,k}=\mu_p+\tilde p_k,
\label{equ:bias_def_main}
\end{equation}
where $\mu_p$ is the per-cycle mean of the injected signal and $\tilde p_k$ is a zero-mean AC component.

\textbf{Why drift appears.}
The EKF prediction integrates $p_{\mathrm{meas},k}-\hat b_p$, so the injected mean $\mu_p$ directly enters the roll prediction. When accelerometer tilt updates are down-weighted or gated (large $\mathbf R_k$ $\Rightarrow$ small $K_{\phi,k}$; \hyperref[asp:1]{Setting~1}), this injected component is only weakly corrected and appears as a residual roll/pitch attitude error. Over the growth time scale of the attack,
\begin{equation}
\begin{aligned}
\varepsilon_{\phi,k}&\triangleq \hat\phi_{k|k}-\hat\phi^{\mathrm{free}}_{k|k}, \\
\varepsilon_{\phi,k}-\varepsilon_{\phi,k-1}
&=(\mu_p+\tilde p_k-\hat b_{p,k-1})\Delta t .
\label{equ:eps_phi_main}
\end{aligned}
\end{equation}
Thus, if \hyperref[asp:1]{Setting~1} holds, the EKF can retain a residual attitude error during excitation and the cascaded controller converts this error into lateral drift. The gyro-bias adaptation rate (\hyperref[asp:2]{Setting~2}) affects how much of the injected component is absorbed into $\hat b_p$, but it is not required for drift during the active attack. A step-by-step derivation and the corresponding pitch case are provided in App.~\ref{sec:ekf_bias_derivation}.

\subsection{The Aftermath When the Attack Stops}
\label{sec:aftermath}

Let the acoustic attack stop at time step $T$. For $k\ge T$, the injected component vanishes ($p_{m,k}=0$), so $p_{\mathrm{meas},k}=p_k$; in near-hover, $p_k\approx 0$. The EKF prediction becomes $\hat{\phi}_{k|k-1}
= \hat{\phi}_{k-1|k-1}
-\hat{b}_{p,k-1|k-1}\Delta t$,
whereas the attack-free prediction is approximately $\hat{\phi}^{\mathrm{free}}_{k|k-1}\approx \hat{\phi}_{k-1|k-1}$. Hence the post-attack roll-error update satisfies
\begin{equation}
\varepsilon_{\phi,k}-\varepsilon_{\phi,k-1}
\approx -\hat{b}_{p,k-1|k-1}\Delta t,
\qquad k\ge T .
\label{equ:aftermath}
\end{equation}

The post-attack behavior is therefore governed mainly by \hyperref[asp:2]{Setting~2}. If the gyro-bias model is stiff, i.e., the bias random-walk variance is small and $K_{b,p,k}$ is low, $\hat b_p$ relaxes slowly and the residual attitude offset persists; the UAV can remain near the nudged hover point even after excitation stops. If \hyperref[asp:2]{Setting~2} does not hold, the bias estimate and attitude error recenter more quickly, and the position controller pulls the UAV back toward the original hover point. Thus, \hyperref[asp:1]{Setting~1} explains drift during the attack, while \hyperref[asp:2]{Setting~2} explains post-attack persistence.

\subsection{\modelAbbr Control}
\label{sec:control}

Most multirotors use a cascaded position/velocity--attitude--rate controller, as in PX4, Paparazzi, and ArduPilot~\cite{mahony2012multirotor,meier2015px4,brisset2006paparazzi,ardupilot}. 
Our attack induces a persistent EKF tilt bias (roll shown; pitch is analogous), so the controller regulates using $\hat\phi=\phi+\varepsilon_\phi$ rather than $\phi$.

\textbf{Inner-loop tracking under biased state feedback.}
In hover, $\phi_{sp}\approx 0$, and the angle controller computes a rate setpoint
\begin{equation}
p_{sp}=K_{AC}(\phi_{sp}-\hat\phi)\approx -K_{AC}\varepsilon_\phi,
\label{equ:p_ctrl_main}
\end{equation}
so the fast rate/attitude loops drive the \emph{estimated} attitude toward the setpoint, i.e., $\hat\phi\approx \phi_{sp}$, which implies $\phi\approx \phi_{sp}-\varepsilon_\phi$.
Thus, a nonzero $\varepsilon_\phi$ initially creates a physical tilt and a lateral acceleration.

\textbf{Outer-loop compensation and bounded offset.}
The outer position/velocity loops react to the resulting drift by commanding a counter-tilt $\phi_{sp}$ until the net lateral acceleration is driven near zero, yielding a finite displaced equilibrium.
In the near-hover regime (small angles, negligible vertical coupling, and P-dominant outer loop),
\begin{equation}
\phi_{sp}\approx \frac{K_{vel,P}K_{pos,P}}{g}\,\Delta y,
\qquad
a_y \approx g(\phi_{sp}-\varepsilon_\phi),
\label{equ:outer_main}
\end{equation}
so at equilibrium $a_y\approx 0$ and the steady lateral offset satisfies
\begin{equation}
\Delta y \approx \frac{g\,\varepsilon_\phi}{K_{vel,P}K_{pos,P}}.
\label{equ:y_disp}
\end{equation}
Equation~\eqref{equ:y_disp} is used as a local equilibrium approximation; it explains why \modelAbbr yields a \emph{bounded, directionally programmable} displacement: larger residual tilt bias increases the offset, while stiffer outer-loop gains reduce it. Details and boundary conditions (integrator action, tilt limits, and saturation) are deferred to App.~\ref{app:control}.

\section{Simulation Experiments}
\label{sec:simulation}

\subsection{Baseline Attack Injection and Analysis}
\label{sec:baseline_simulation}

We conduct simulation experiments in MATLAB to validate the proposed \modelAbbr scheme. The simulated quadrotor is modeled as a rigid body with parameters representative of a small aerial platform \cite{mahony2012multirotor,khazraei2023stealthy}. The total mass is set to $1\,\mathrm{kg}$, and gravity is 
$g = 9.81\,\mathrm{m/s^2}$. The inertia matrix is assumed diagonal, i.e., 
$
\mathbf{I}_{sys} = \mathrm{diag}(0.016,\,0.016,\,0.0274)\,\mathrm{kg \cdot m^2},
$
reflecting approximate symmetry about the $x$- and $y$-axes and a slightly larger inertia about the $z$-axis. Rotor dynamics are included via the rotor inertia 
$I_m = 3.789\times 10^{-6}\,\mathrm{kg \cdot m^2}$, which influences motor response and gyroscopic effects. Together, these parameters provide a physically consistent baseline for simulating the vehicle’s translational and rotational dynamics. 

The sampling period is set to $\Delta t = 4\,\text{ms}$. The gyro measurement noise is $\sigma_{\mathrm{gr}} = 0.0087\,\text{rad/s}$, and the accelerometer tilt noise is modeled as $\sigma_\mathrm{acc} = 0.05\,\text{rad}$. The gyro-bias random-walk standard deviations are 
$\sigma_{\mathrm{b}} = [10^{-3},10^{-3}]^{\!\top}$. These parameters define the process noise covariance $\mathbf{Q}$ in (\ref{equ:QR}), and are chosen to be consistent with \hyperref[asp:1]{Setting~1} (down-weighted accelerometer corrections) and \hyperref[asp:2]{Setting~2} (slow bias adaptation).

The drone then takes off and hovers at the desired position $(p_x, p_y, p_z)_d=(0, 0, 1\, \text{m})$. After the system reaches a stable hover, an attack signal formulated using DAM (Sec.~\ref{sec:amplitude}) is injected at $t_0=10\, \text{s}$. The injected signal parameters are
\begin{equation}
    F_{\mathrm{obs}} = 1 \,\text{Hz}\footnote{We use $F_{\mathrm{obs}}=1~\mathrm{Hz}$ only as an illustrative example. In practice, the effective aliased component is determined by our sweep-and-lock approximation procedure (Sec.~\ref{sec:agr_algo}, App.~\ref{sec:sweep_and_aliased}).}, \,\, \varphi_0=0, \,\, A_+=0.04, \,\, A_-=0.004,
\label{equ:baseline_attack_sig}
\end{equation}
so that the induced bias in the gyroscope channel is 
$\mu_{DAM}=\frac{A_+ - A_-}{\pi} = \tfrac{0.036}{\pi}$. The signal is turned off at $t_1=30\, \text{s}$.

\begin{figure}[!t]
    \centering
    % Row 1
    \begin{subfigure}[b]{0.23\textwidth}
        \centering
        \includegraphics[width=\textwidth,trim=0mm 0 0 0,clip]{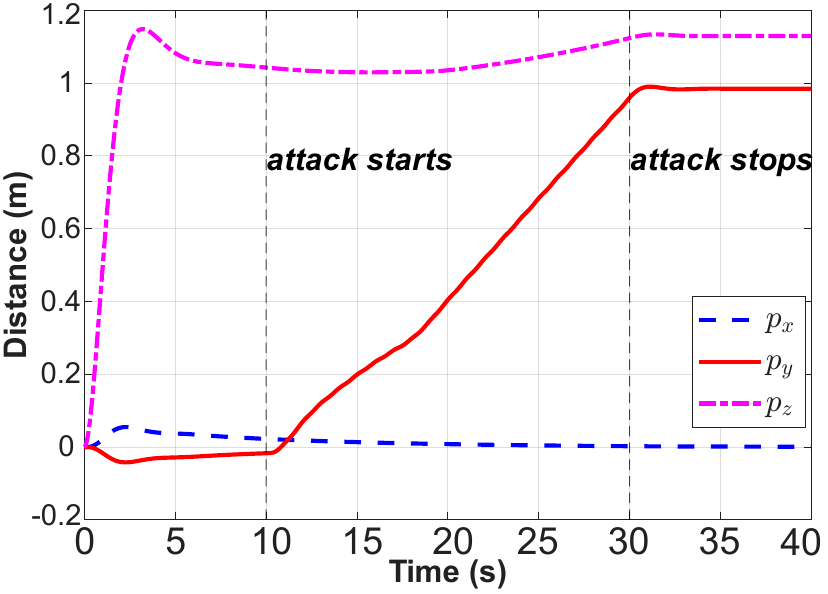}
        \caption{Drone horizontal position}
        \label{fig:pos}
    \end{subfigure}
    \hfill
    \begin{subfigure}[b]{0.23\textwidth}
        \centering
        \includegraphics[width=\textwidth,trim=0mm 0 0 0,clip]{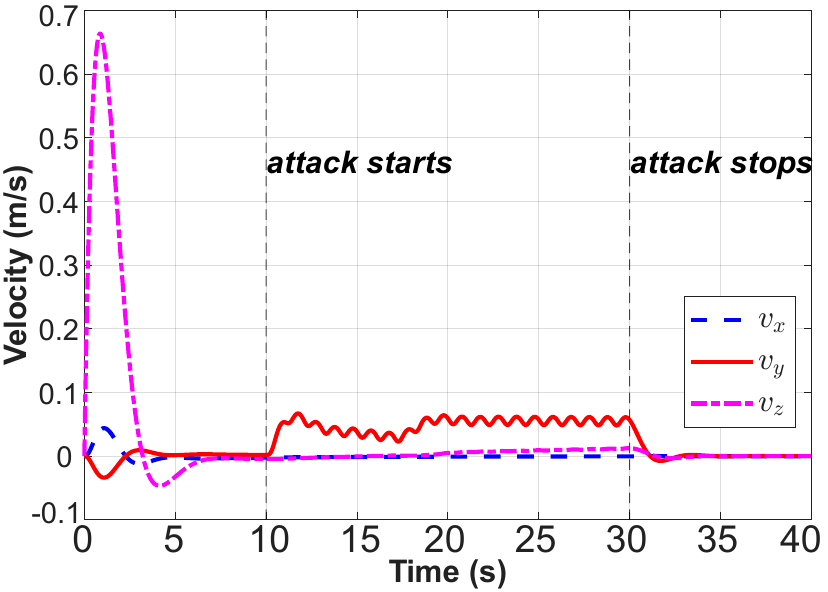}
        \caption{Drone horizontal velocity}
        \label{fig:vel}
    \end{subfigure}
    \caption{Time histories of single-axis position and velocity during the attack.}

    \label{fig:pos_vel}
\end{figure}
% \input{figures/angles}
% \input{figures/diff_A}
% In the preamble (if not already):
% \usepackage{graphicx}
% \usepackage{subcaption}

\begin{figure*}[!t]
    \centering
    \begin{subfigure}[t]{0.3\textwidth}
        \centering
        \includegraphics[width=\linewidth,trim=0mm 0 0 0,clip]{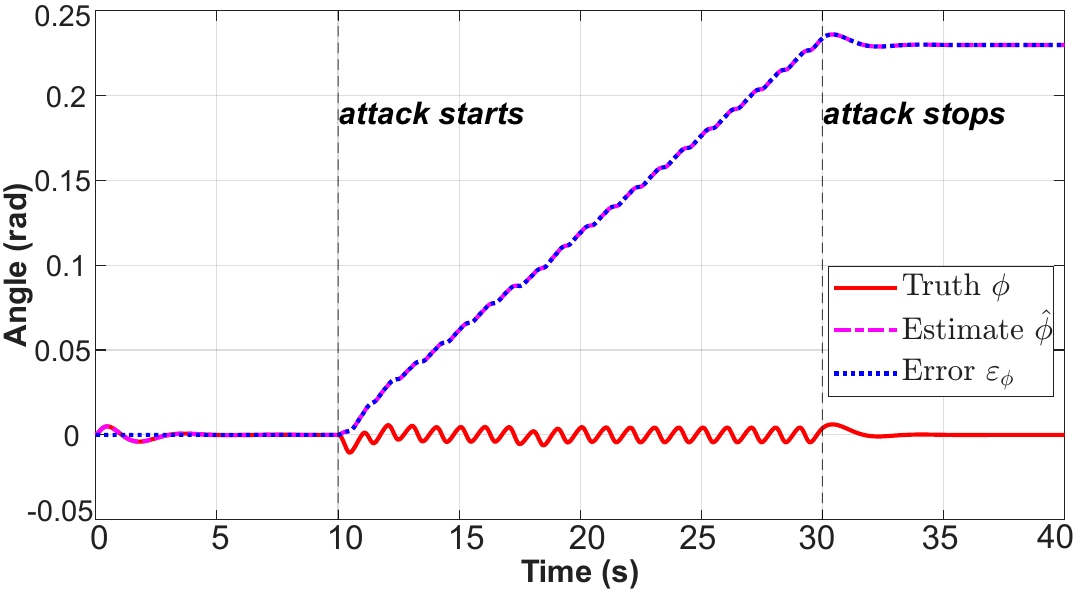}
        \caption{Roll under attack: EKF estimate $\hat\phi$, predicted error $\varepsilon_\phi$ (Eq.~\ref{equ:eps_phi_main}), and ground truth $\phi$.}
        \label{fig:angle}
    \end{subfigure}\hfill
    \begin{subfigure}[t]{0.31\textwidth}
        \centering
        \includegraphics[width=\linewidth,trim=0mm 0 0 0,clip]{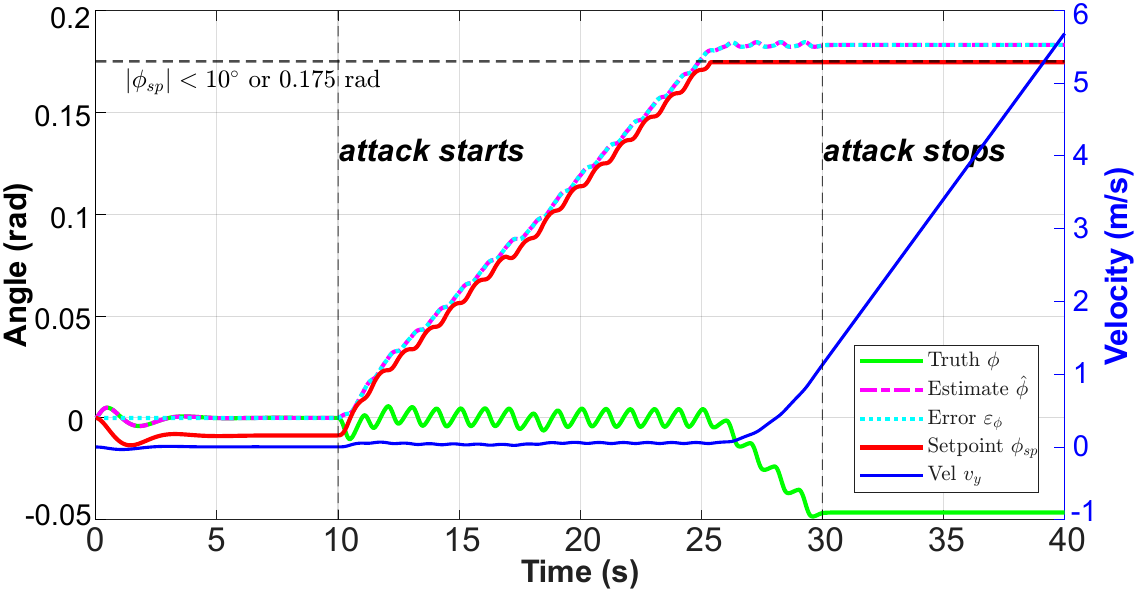}
        \caption{Bounded roll setpoint $|\phi_{sp}|\le 10^\circ$ in PX4 position control induces exponential lateral drift ($v_y\rightarrow\infty$) than the unbounded case until failsafe logic is triggered.}
        \label{fig:bounded_phi}
    \end{subfigure}\hfill
    \begin{subfigure}[t]{0.34\textwidth}
        \centering
        \includegraphics[width=\linewidth,trim=15mm 0 8mm 0,clip]{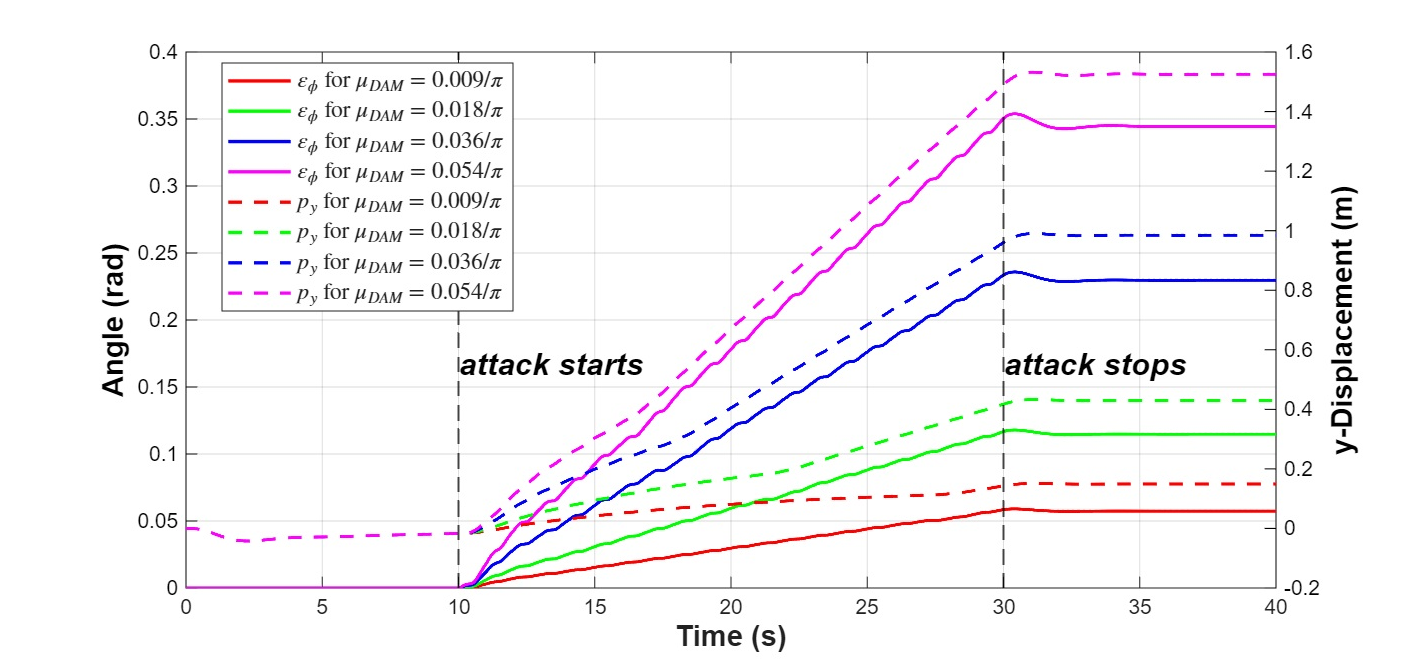}
        \caption{Estimated roll angles $\varepsilon_\phi$ and y-axis displacement $p_y$ under four different low-amplitude gyroscope biases.}
        \label{fig:diff_A}
    \end{subfigure}

    \caption{Attack-induced roll estimation error and lateral drift behavior in PX4.}
    \label{fig:roll_attack_triptych}
\end{figure*}

The position and velocity responses are shown in Fig.~\ref{fig:pos_vel}. During the first $\sim 6\,\text{s}$ the vehicle takes off and settles into a hover at $(p_x,p_y,p_z)_d$. Once the attack starts at $t_0$, the biased gyro reading produces a biased attitude estimate $\hat\phi$, which the cascaded controller interprets as a genuine tilt error. This causes a horizontal acceleration and the vehicle begins to drift laterally (here along the $y$-axis) at an average speed of about $0.05\,\text{m/s}$ while maintaining altitude (vertical deviation $<5\,\text{cm}$). Because the simulation allows us to inject the bias into a single Euler channel, the drift is effectively one-dimensional; in practice, acoustic excitation couples into both roll and pitch, as seen in Sec.~\ref{sec:evaluation}. As discussed in Sec.~\ref{sec:control}, an appropriate modulation and selection of the dominant axis can still realize a coarse drift direction from the set $\{\textit{go forward}, \textit{go backward}, \textit{go left}, \textit{go right}\}$. When the attack is active and the outer position loop has rebalanced the acceleration, the vehicle converges to a new, offset hover point, and the lateral velocity decays back to zero.

% Fig.~\ref{fig:rate} (and the associated rate plots) illustrates the relation between the injected attack signal $s[i]$, the measured angular rate $p_{meas}$, and the true rate $p$. 
The biased $s[i]$ yields a biased $p_{meas}$, which is integrated by the EKF into an offset Euler angle $\hat\phi$, as shown in Fig.~\ref{fig:angle}. The attitude controller then tilts the vehicle to cancel this (false) error, and the resulting physical tilt generates the lateral motion observed in Fig.~\ref{fig:pos_vel}. As long as the biased component of $s[i]$ persists and the EKF bias state adapts slowly (per \hyperref[asp:2]{Setting~2}), the estimation error $\varepsilon_\phi$ grows to a new quasi-steady value, and the drift increases until the outer loops reach a new equilibrium; the P-only position loop prevents runaway, yielding a moderate, bounded displacement rather than an unstable divergence.

To study the effect of attitude/tilt limits, we impose a conservative bound on the roll setpoint from the position controller, $\abs{\phi_{sp}}\leq 10^\circ$, consistent with the practical constraints discussed around (\ref{equ:y_max}). The resulting behavior is shown in Fig.~\ref{fig:bounded_phi}. Around $t\approx 25\,\text{s}$, the position loop saturates at $\phi_{sp}\approx 10^\circ$. Beyond this point, the controller can no longer fully counteract the estimator bias, so the true roll $\phi$ and the estimation error $\varepsilon_\phi$ approach their respective bounds, and the lateral acceleration remains biased. Consequently, the horizontal speed grows significantly and the vehicle exhibits rapidly increasing drift (approximately quadratic growth in displacement), which can persist for some time even after the attack is lifted.

Finally, we vary the injected bias from $\mu_{\text{DAM}} = 0.009/\pi$ up to $0.054/\pi$. As shown in Fig.~\ref{fig:diff_A}, the average drift speed and the corresponding steady-state error in the estimated roll angle $\varepsilon_\phi$  both increase approximately linearly with $\mu_{DAM}$, in line with the analytical relationship $\Delta y \propto \varepsilon_\phi$ derived in Sec.~\ref{sec:control}. All biases in these experiments are positive; injecting negative biases produces symmetric behavior, with the lateral drift occurring in the opposite direction.

This validates the gyro-bias-to-displacement mechanism and confirms the bounded equilibrium predicted by Equation (\ref{equ:y_disp}).

\subsection{Attack Sensitivity to Diverse EKF Weighting}
\label{sec:sensitivity_simulation}

\begin{figure}[!t]
    \centering
    % Row 1
    \begin{subfigure}[b]{0.235\textwidth}
        \centering
        \includegraphics[width=\textwidth,trim=0mm 0 0 0,clip]{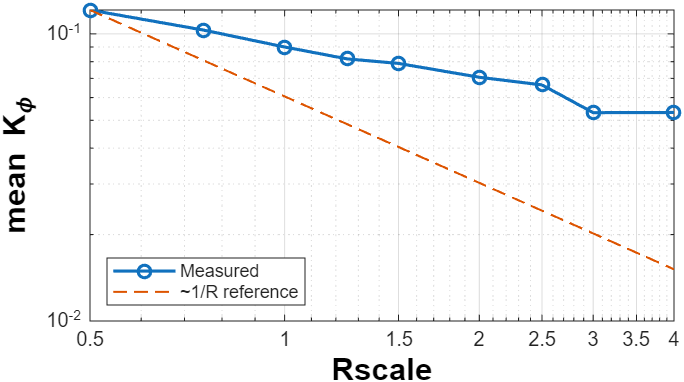}
        \caption{$K_\phi$ vs. Rscale}
        \label{fig:k_phi}
    \end{subfigure}
    \hfill
    \begin{subfigure}[b]{0.235\textwidth}
        \centering
        \includegraphics[width=\textwidth,trim=0mm 0 0 0,clip]{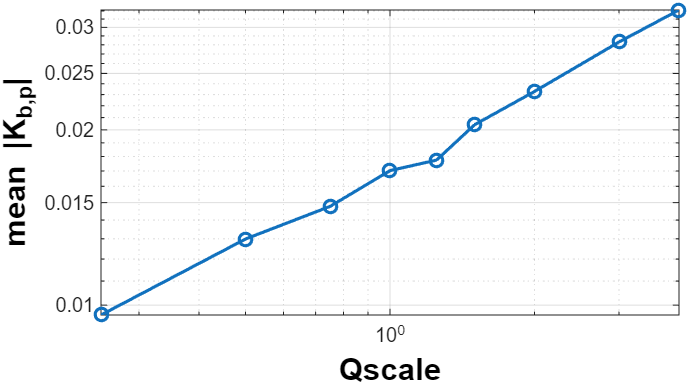}
        \caption{$|K_{b,p}|$ vs. Qscale}
        \label{fig:k_b_p}
    \end{subfigure}
    \caption{EKF gain sensitivity to estimator weighting under attack. (a) The mean value of $K_\phi$ decreases while Rscale increases at a fixed Qscale (=1). The dashed $\sim 1/R$ line shows the expected inverse gain--covariance trend. The measured $K_\phi$ decreases similarly, with small deviations because the state covariance varies under closed-loop dynamics. (b) The mean value of $|K_{b,p}|$ increases while Qscale increases at a fixed Rscale (=1).}

    \label{fig:sensitivity_K}
\end{figure}
\begin{figure}[!t]
    \centering
    \begin{subfigure}[b]{0.48\textwidth}
        \centering
        \includegraphics[width=\textwidth,trim=0mm 0 0mm 0,clip]{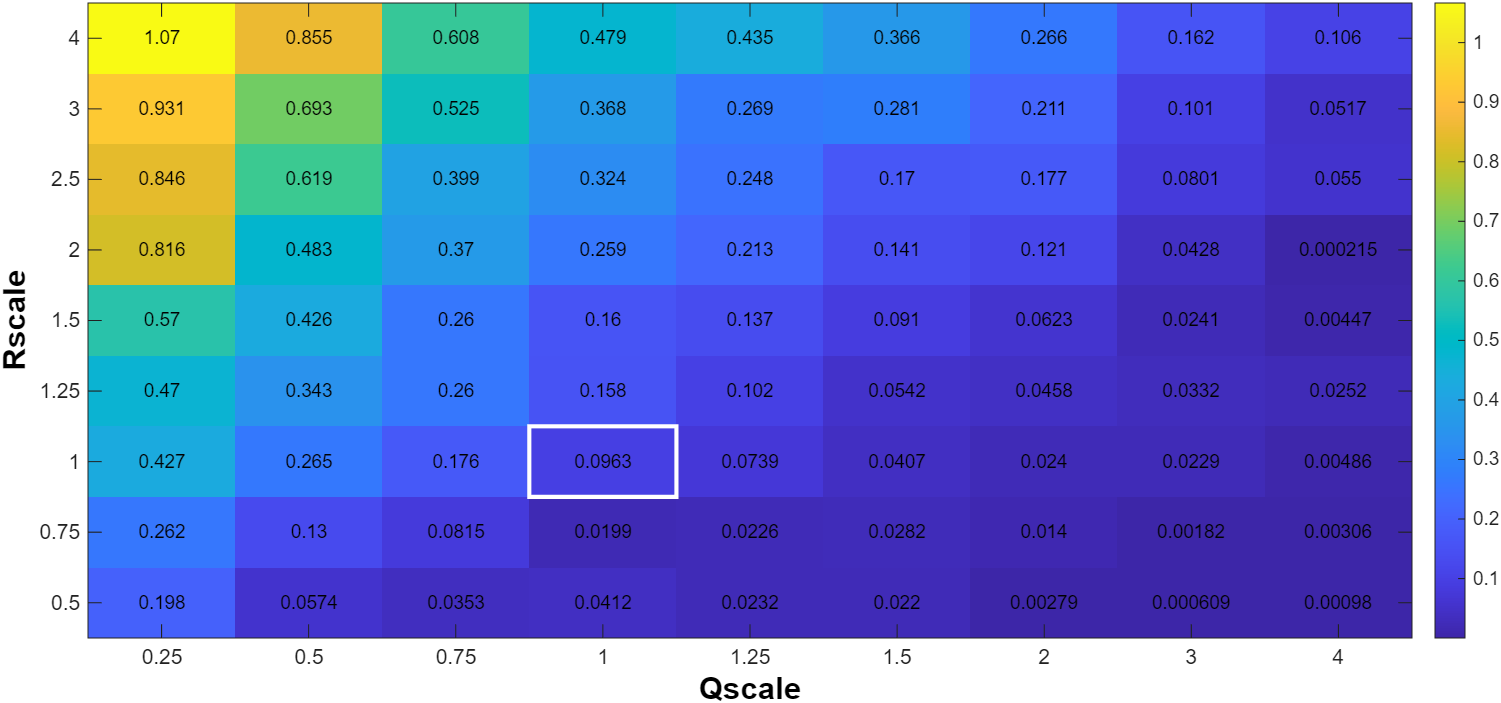}
    \end{subfigure}
    
    % \vspace{1em} % space between rows
        
    \caption{Steady attack impact under EKF reweighting. Heatmap shows the magnitude of the mean estimated roll bias $|\mathrm{mean}(\hat{\phi})|$ over $[13,23]$~s for a grid of (Rscale, Qscale) settings. Larger Rscale (weaker accelerometer correction) increases the residual attitude bias, while larger Qscale (faster bias adaptation) reduces it. The white box marks the baseline setting $(\mathrm{Rscale,Qscale})=(1,1)$ experimented in Sec.~\ref{sec:baseline_simulation}. }
    \label{fig:heatmap}
\end{figure}

We next examine whether the proposed gyro attack remains effective under a broad range of estimator tunings, as would be expected across different autopilot stacks and EKF configurations. To this end, we perform an EKF sensitivity sweep over the two dominant weighting terms in the estimator: the accelerometer tilt measurement covariance $\mathbf{R}_k$ and the gyro-bias random-walk covariance $\mathbf{Q}_k$. We retain the baseline EKF noise settings from Sec.~\ref{sec:baseline_simulation}, i.e., $\sigma_{\mathrm{gr,base}}=0.0087~\mathrm{rad/s}$, $\sigma_{\mathrm{b,base}}=[10^{-3},\,10^{-3}]^{\!\top}~\mathrm{rad/s}$, and $\sigma_{acc,\mathrm{base}}=0.05~\mathrm{rad}$, and introduce dimensionless scaling factors $\mathrm{Rscale}$ and $\mathrm{Qscale}$ such that
$
\sigma_{\mathrm{acc}} \leftarrow \sqrt{Rscale}\,\sigma_{\mathrm{acc,base}},\,
\sigma_{\mathrm{b}} \leftarrow \sqrt{Qscale}\,\sigma_{\mathrm{b,base}}, 
$
which is equivalent to scaling $\mathbf{R}_k \leftarrow Rscale\,\mathbf{R}_{k,\mathrm{base}}$ and $\mathbf{Q}_k \leftarrow Qscale\,\mathbf{Q}_{k,\mathrm{base}}$ for the corresponding blocks. To show a wider analysis scope, we choose $\mathrm{Rscale}=[0.5, 0.75, 1, 1.25, 1.5, 2, 2.5, 3, 4]$, and $\mathrm{Qscale}=[0.25, 0.5, 0.75, 1, 1.25, 1.5, 2, 3, 4]$. For each $(\mathrm{Rscale},\mathrm{Qscale})$ pair, we inject the same attack signal as (\ref{equ:baseline_attack_sig}), yielding a full grid of \textbf{81 experiments}. In addition to the estimated roll angle $\hat{\phi}$, we log the EKF Kalman gain $\mathbf{K}$ to explicitly connect estimator weighting to attack-induced attitude error, focusing on the accelerometer-to-roll correction gain $K_{\phi}$ and the bias-adaptation gain $|K_{b, p}|$.

The sweep results follow the expected sensitivity trends and provide a gain-level explanation for attack transferability across EKF settings. With $\mathrm{Qscale}=1$ held fixed, the mean $K_{\phi, k}$ decreases monotonically as $\mathrm{Rscale}$ increases (Fig.~\ref{fig:k_phi}), indicating reduced reliance on accelerometer-derived tilt and correspondingly weaker correction of the attitude estimate. Conversely, with $\mathrm{Rscale}=1$ held fixed, the mean $|K_{b_q}|$ increases with $\mathrm{Qscale}$ (Fig.~\ref{fig:k_b_p}), consistent with faster bias adaptation when the bias random-walk covariance is larger. These gain changes translate directly into steady attack impact: the heatmap of $|\mathrm{mean}(\hat{\phi})|$ (Fig.~\ref{fig:heatmap}) during 10~s of attack window ($t=[13,23]~\mathrm{s}$) shows that the residual roll bias grows with increasing $\mathrm{Rscale}$ (weaker accelerometer correction) and diminishes with increasing $\mathrm{Qscale}$ (stronger bias tracking). 

Taken together, these results indicate that the proposed gyro perturbation remains effective over a wide range of EKF weightings, with its steady-state influence on Euler-angle estimation varying predictably according to the estimator's measurement trust and bias adaptation rate. The EKF sweep also provides a \textbf{defense-side interpretation}: stronger bias adaptation and less conservative accelerometer correction reduce the residual attitude bias, suggesting estimator-level hardening levers.

\section{Physical Experiment Evaluation}
\label{sec:evaluation}
\begin{figure*}[t]
    \centering

    \begin{subfigure}[t]{0.62\textwidth}
        \centering
        \includegraphics[height=0.125\textheight,keepaspectratio]{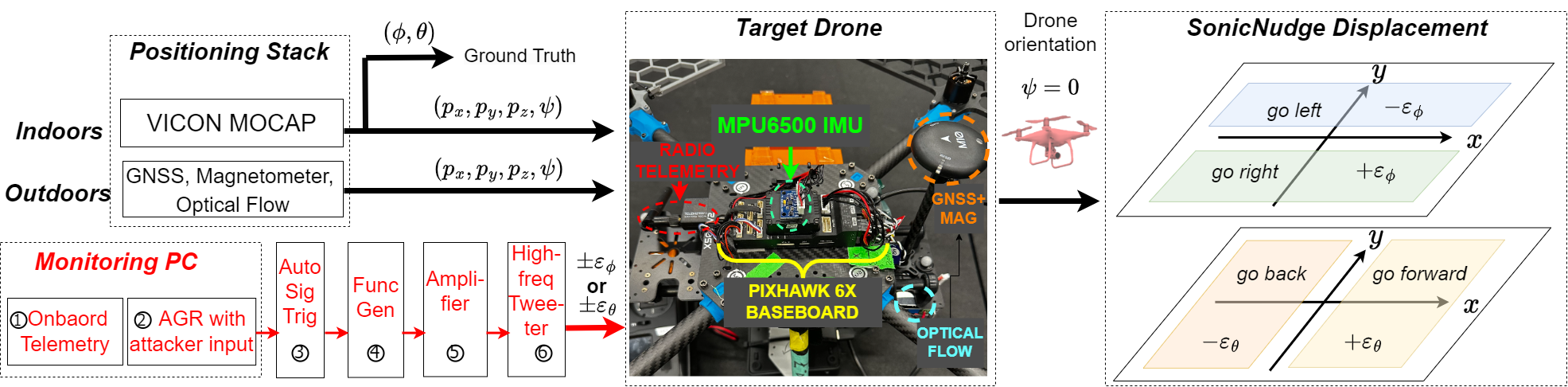}
        \caption{End-to-end \modelAbbr\ pipeline with the target UAV in the loop.}
        \label{fig:pipeline}
    \end{subfigure}
    \hfill
    \begin{subfigure}[t]{0.34\textwidth}
        \centering
        \includegraphics[height=0.12\textheight,keepaspectratio]{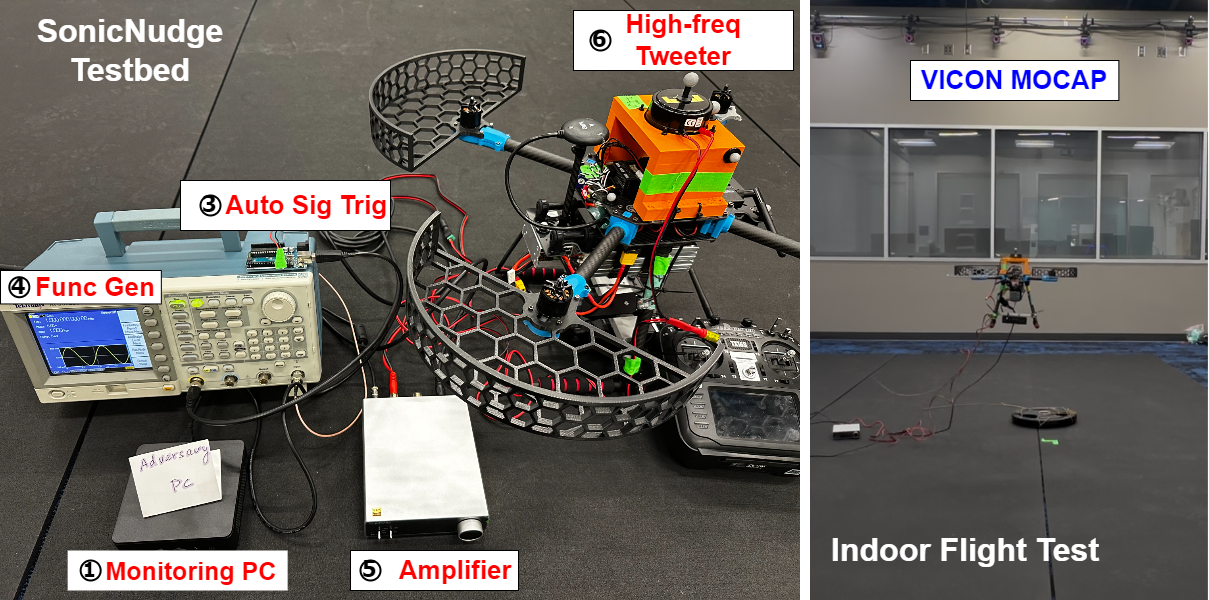}
        \caption{Adversary hardware and indoor test site.}
        \label{fig:drone}
    \end{subfigure}

    \caption{Experimental platform and physical attack setup. 
    (a) An MPU-6500 IMU on a Pixhawk V6X baseboard sits at the UAV's centroid. Directed ultrasound biases the gyro, inducing roll/pitch errors ($\varepsilon_\phi,\varepsilon_\theta$) that the controller converts into horizontal-plane displacement. 
    (b) Physical adversary module and indoor test environment.}
    \label{fig:exp_setup}
\end{figure*}

\subsection{Experimental Platform Design}
\label{sec:experimental_design}

We use a Holybro X500~V2 quadrotor as the victim platform, equipped with a Pixhawk~6X flight controller (Baseboard~v2A) configured to use an MPU6500 IMU as the primary inertial sensor. Indoor experiments are conducted in a motion-capture arena where a VICON MOCAP system provides global position and ground-truth Euler angles at 10~Hz; outdoor experiments use GNSS at 5~Hz. In both settings, the PX4 estimator fuses IMU, magnetometer, optical-flow, and external position measurements, matching a typical PX4-based multirotor sensing and control stack.

For attack scheduling and logging, the attacker-side monitoring PC receives the target UAV's radio telemetry: \texttt{ATTITUDE} and \texttt{LOCAL\_POSITION\_NED} at 10\,Hz indoors, and \texttt{ATTITUDE} at 10\,Hz with \texttt{GLOBAL\_POSITION\_INT} at 5\,Hz outdoors. These streams support AGR scheduling and attacker--victim geometry monitoring, including favorable acoustic incidence (Sec.~\ref{sec:telemetry}). Fig.~\ref{fig:pipeline} shows the platform overview.

The monitoring PC runs the AGR-based phase-gating algorithm (Alg.~\ref{alg:phase_gating}) to choose an attack axis $a\in\{x,y\}$ and sign $s\in\{+,-\}$, thereby biasing EKF pitch or roll toward the desired drift direction. The resulting gated waveform schedule drives a Tektronix AFG3022C function generator through an Arduino~UNO microcontroller; the analog signal is amplified and emitted by a Fostex FT17 high-frequency tweeter~\cite{fostex_ft17h} at an SPL of $\sim$105~dB measured at the sensor side. SPL at the sensor side primarily affects the injected gyro-rate bias $\mu_p$, i.e., the growth rate of the Euler-angle bias $\epsilon_p$, but the realized bias also depends on resonance alignment, phase-lock quality, acoustic coupling, and estimator adaptation.

For the main flight validation, we use a controlled acoustic-coupling setup by mounting the tweeter on the victim drone along the $-z$ direction of the sensor (Fig.~\ref{fig:drone}), with at least 5~cm separation from the IMU and silicone padding around the tweeter housing. This setup keeps acoustic incidence reproducible and reduces beam-alignment variability, while the padding minimizes mechanical transfer to the flight controller and avoids pure transduction-style PCB or mounting-substrate attacks as in~\cite{gao2023exploring}. Thus, the observed displacement can be attributed primarily to airborne ultrasonic excitation of the IMU and its propagation through the estimator--controller stack. 

Speaker-detached tests are reported in Sec.~\ref{sec:imu_generality}, demonstrating sensor susceptibility up to 1.52~m using a single speaker without range-optimized beamforming or power delivery. Fully detached, black-box, moving-target free-flight deployment remains future work.

\subsection{AGR-based Angular Bias Injection}
\label{sec:agr_algo}

Our goal is to inject a small but consistently \emph{signed} angular bias by phase-gating Alias-Gated Resonance (AGR) at specific points of the target's \emph{low-frequency aliased} motion. The key observation is that the aliased gyro component is typically in the 1--10~Hz range, so the scheduler only needs to gate slow peak/trough events rather than track the ultrasonic carrier directly. For near-sinusoidal motion, gyro-rate zero-crossings align with peaks/troughs of the corresponding Euler angles $(\phi,\theta,\psi)$ because angle and rate are in quadrature.

AGR shapes the effective gyro bias by switching between two attack waveforms with a $180^\circ$ phase offset: \emph{Type-A} (in-phase, rising at onset) and \emph{Type-B} (anti-phase, falling at onset). As detailed in App.~\ref{sec:alg_app} and Alg.~\ref{alg:phase_gating}, the scheduler takes a gyro stream $(t,g_x,g_y,g_z)$, an axis $a\in\{x,y\}$, a desired sign $s\in\{+,-\}$, and fixed offsets $\Delta t_{\mathrm{on}},\Delta t_{\mathrm{off}}$ that compensate end-to-end latency (ROS\,2 $\rightarrow$ gate $\rightarrow$ function generator). We restrict $a$ to $\{x,y\}$ so that biasing $g_x$ mainly perturbs EKF pitch $\theta$ (forward/backward drift), while biasing $g_y$ perturbs EKF roll $\phi$ (left/right drift); the sign $s$ selects the drift direction.

% At each sample, \textsc{FitSineOnAxis} estimates the aliased frequency and phase $(\hat f,\hat\phi_{\rm abs})$ from a rolling window on the chosen axis. After a short preparation period, the phase is locked, and \textsc{PhaseGateAttack} predicts the next peak and trough using wall-clock time. For $s=+$, AGR enables \emph{Type-A} over the peak$\rightarrow$trough half-cycle; for $s=-$, it enables \emph{Type-B} over the trough$\rightarrow$peak half-cycle. Because the required timing precision is governed by the aliased period, typically hundreds of milliseconds, millisecond-scale sampling jitter and gate latency are tolerated. Over repeated cycles, this asymmetric half-cycle gating accumulates a net signed bias in the aliased gyro rate, yielding the controlled drift analyzed in Sec.~\ref{sec:control}.

At each sample, a rolling window of length $T_w$ is maintained on the chosen axis, and \textsc{FitSineOnAxis} in Alg.~\ref{alg:phase_gating} performs a lightweight least-squares fit to obtain $(\hat f,\hat\phi_{\rm abs})$ for the aliased component. This phase estimate requires only a short \emph{preparation time} (a few cycles) and is then \emph{locked} to avoid noisy refits; the subsequent schedule uses wall-clock time and the locked $(\hat f,\hat\phi_{\rm abs})$ to predict the next peak and trough. \textsc{PhaseGateAttack} then selects the attack type via the gating window: for $s=+$ it enables \emph{Type-A} over the peak$\rightarrow$trough half-cycle (rising onset, reinforcing a positive bias), whereas for $s=-$ it enables \emph{Type-B} over the trough$\rightarrow$peak half-cycle (falling onset, reinforcing a negative bias). Importantly, the required timing precision is set by the aliased period (hundreds of milliseconds), so millisecond-scale sampling jitter and gate latency are tolerated; $\Delta t_{\mathrm{on}},\Delta t_{\mathrm{off}}$ absorb fixed delays, and locking mitigates short-term jitter. Clock drift is negligible over our attack horizon; if needed, the same fitting step can be periodically re-invoked to re-synchronize the phase. Over many cycles, this asymmetric half-cycle gating produces a net signed bias in the aliased gyro rate (pitch if $a=x$, roll if $a=y$), yielding the controlled drift analyzed in Sec.~\ref{sec:control}.

\subsection{Real Experiment Analysis}
\label{sec:real_exp}
\begin{table}[t]
  \centering
  \footnotesize
  \caption{Phase-gated trigger schedule for the indoor run, including trigger times and target bias signs (+/-). The attack signal was manually interrupted after T4, T7, T10, T14, and T16, with each new attack segment launched shortly after signal reactivation.}

  \label{tab:trigger_indoor}
  \begin{tabular}{c|cccccccc}
    \hline
    Index    & T1   & T2   & T3   & T4   & T5   & T6   & T7   & T8   \\
    \hline
    Time (s) & 53.8 & 55.3 & 56.6 & 57.6 & 58.8 & 59.8 & 60.3 & 68.9 \\
    Sign     & +    & +    & +    & +    & --   & --   & --   & --   \\
    \hline
  \end{tabular}

  \vspace{4pt}

  \begin{tabular}{c|cccccccc}
    \hline
    Index    & T9   & T10  & T11  & T12  & T13  & T14  & T15  & T16  \\
    \hline
    Time (s) & 69.8 & 70.5 & 76.1 & 80.0 & 81.3 & 82.4 & 83.6 & 84.9 \\
    Sign     & --   & --   & +    & +    & +    & +    & --   & --   \\
    \hline
  \end{tabular}
\end{table}

\begin{figure}[!t]
    \centering
    \begin{subfigure}{0.49\textwidth}
    \centering        \includegraphics[width=0.99\textwidth,trim=0mm 3mm 0 0,clip]{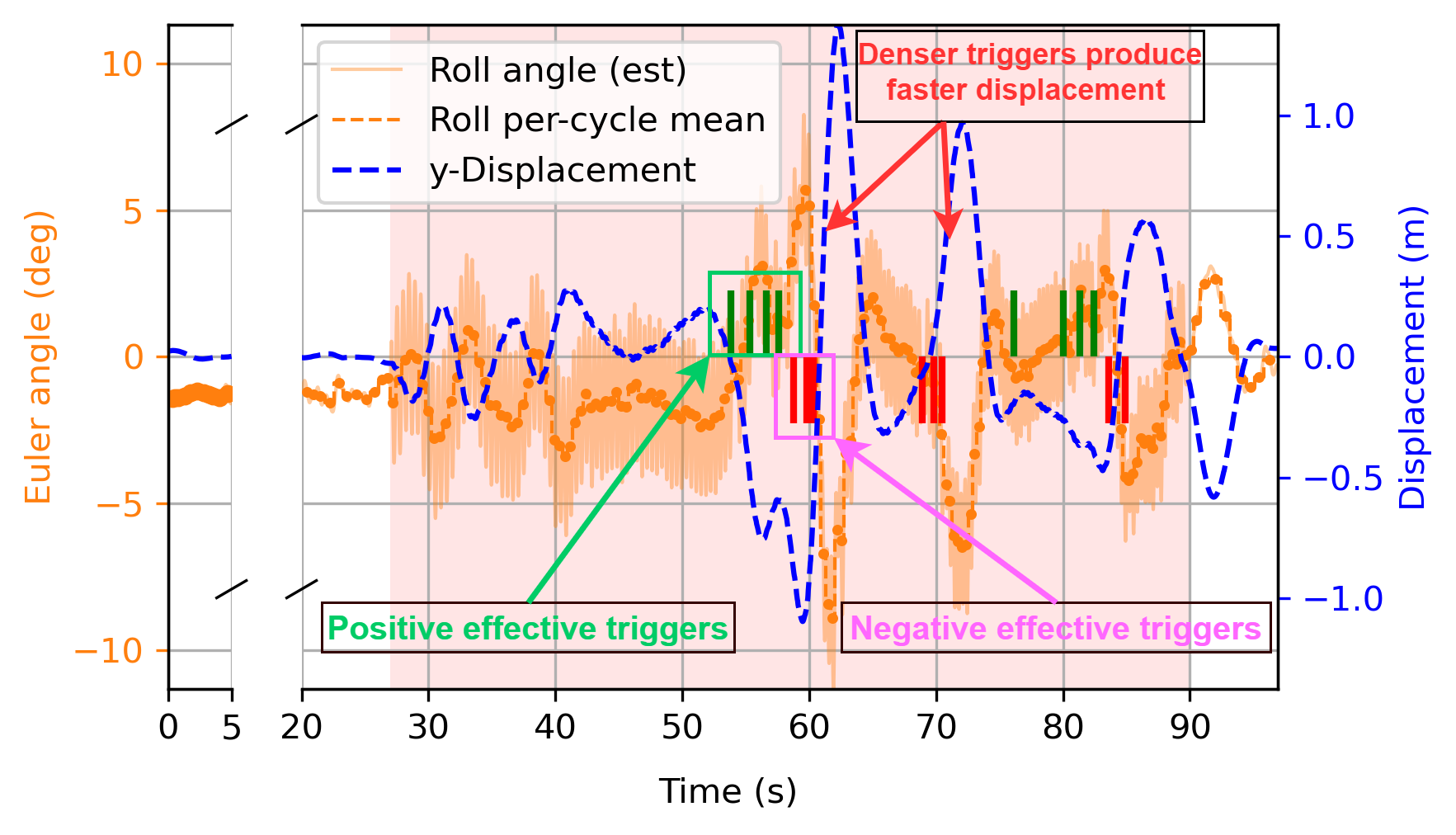}
        % \caption{}
    \end{subfigure}
    \caption{Relationship between biased roll angle and lateral $y$-displacement; red shading marks the attack window.}
    \label{fig:roll_est_y}
\end{figure}

\begin{figure}[!t]
    \centering
    \begin{subfigure}{0.49\textwidth}
    \centering        \includegraphics[width=0.99\textwidth,trim=0mm 3mm 0 0,clip]{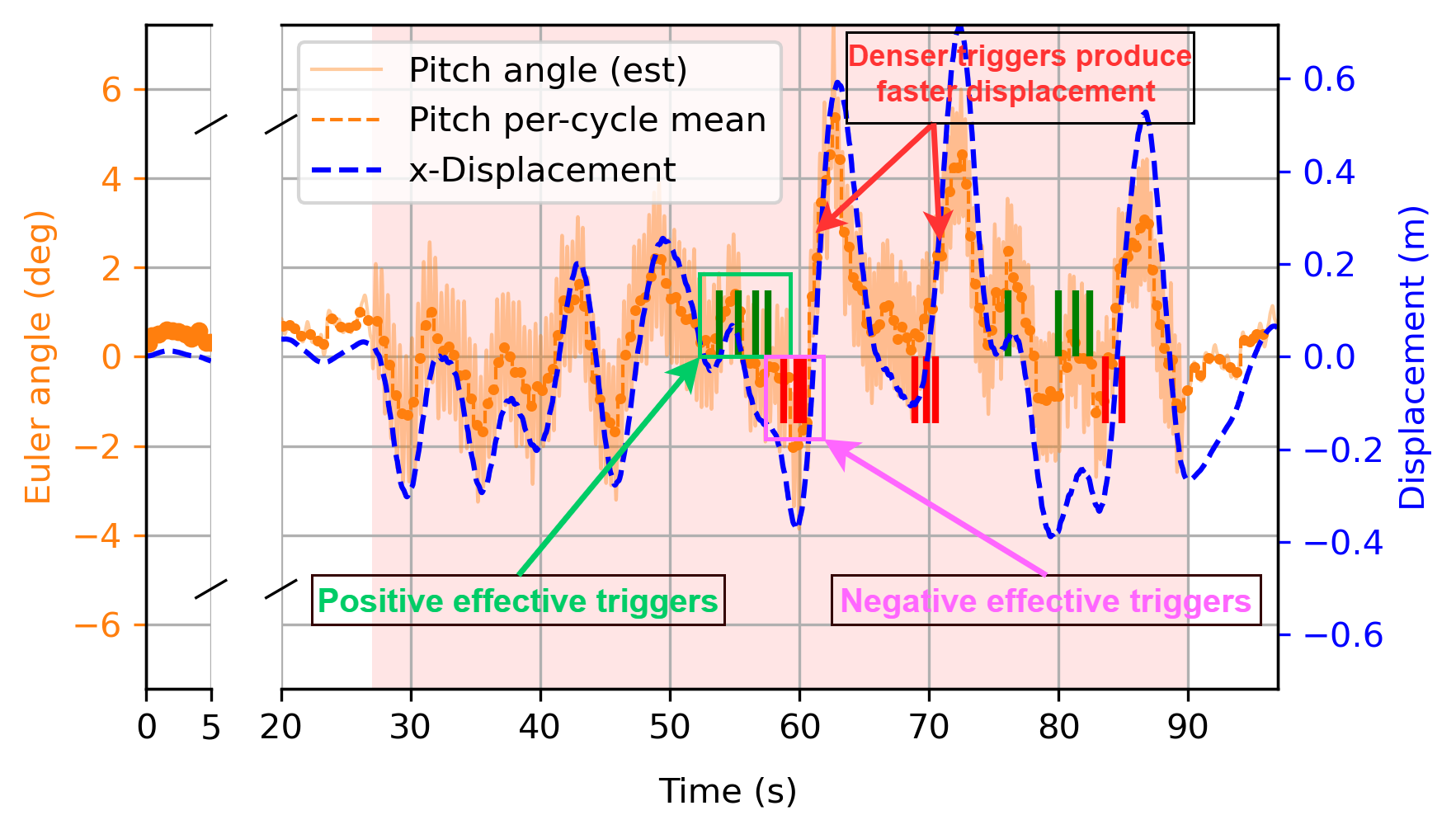}
        % \caption{}
    \end{subfigure}
    \caption{Relationship between biased pitch angle and lateral $x$-displacement; red shading marks the attack window.}
    \label{fig:pitch_est_x}
\end{figure}

\begin{figure}[!t]
    \centering
    \begin{subfigure}{0.49\textwidth}
    \centering        \includegraphics[width=0.99\textwidth,trim=0mm 3mm 0 0,clip]{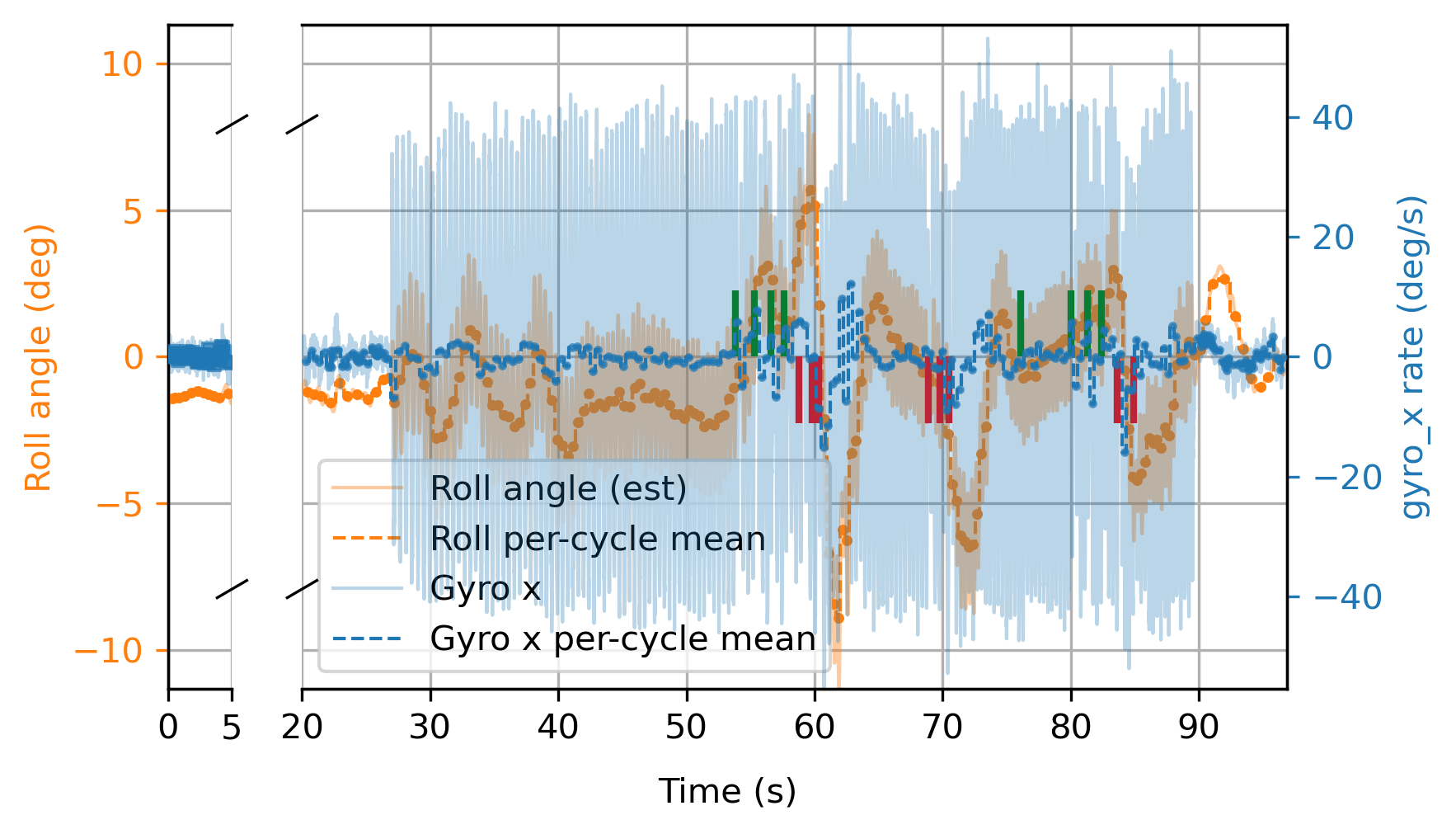}
        % \caption{}
    \end{subfigure}
    \caption{Relationship between x-axis gyroscope rate ($p$) and roll angle ($\phi$). The blue shade shows attack-caused resonance. }
    \label{fig:gyro_roll_est}
\end{figure}

\begin{figure}[!t]
    \centering
    \begin{subfigure}{0.49\textwidth}
    \centering       \includegraphics[width=0.99\textwidth,trim=0mm 3mm 0 0,clip]{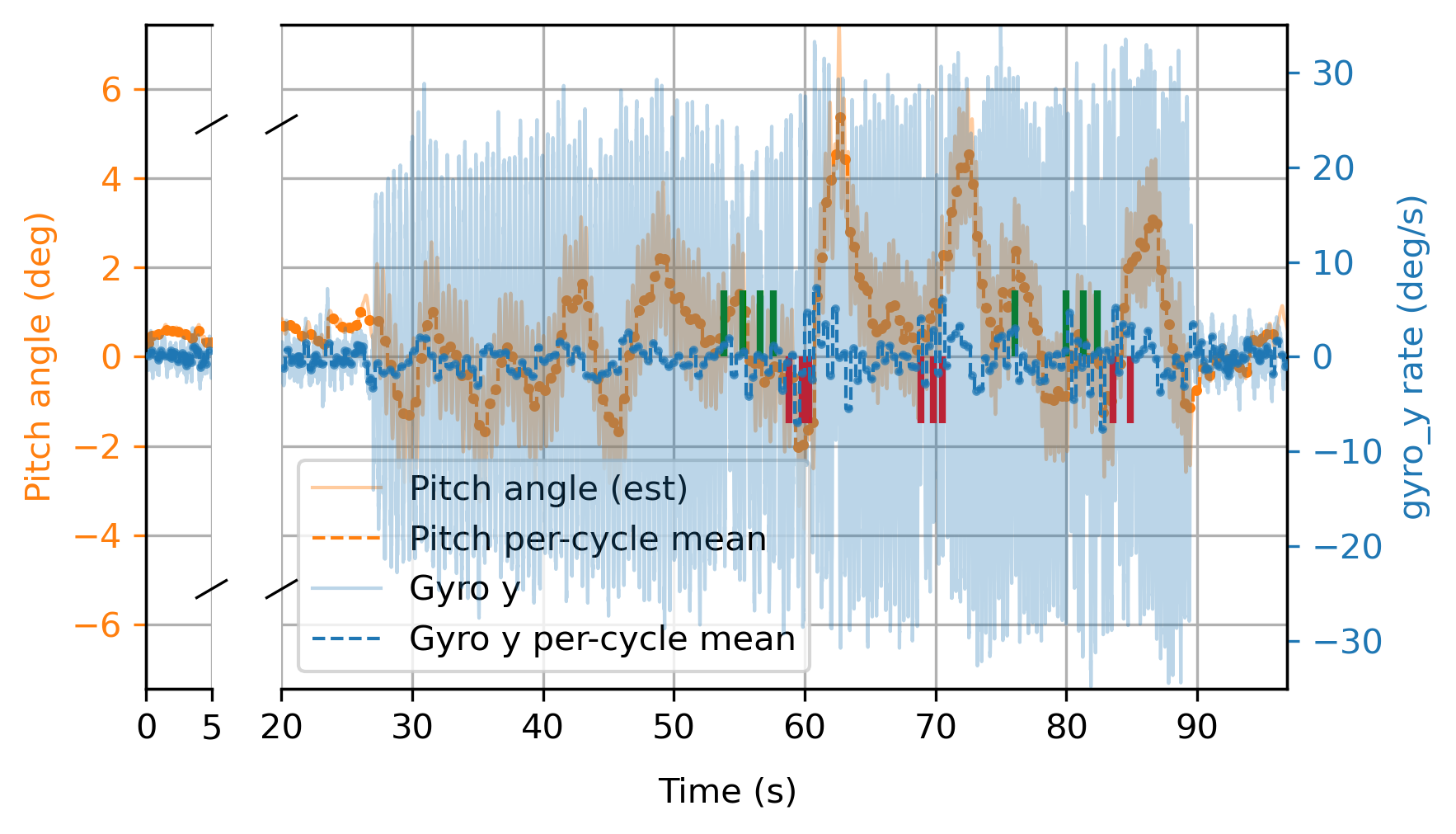}
        % \caption{}
    \end{subfigure}
    \caption{Relationship between y-axis gyroscope rate ($q$) and pitch angle ($\theta$). The blue shade shows attack-caused resonance. }
    \label{fig:gyro_pitch_est}
\end{figure}

We first perform indoor experiments in a VICON-equipped arena (safe zone of 3~m~$\times$~3~m) and enforce conservative attitude bounds $\varepsilon_\phi,\varepsilon_\theta \in (-10^\circ,10^\circ)$ for safety. In the run shown in Fig.~\ref{fig:roll_est_y} (duration $\sim\!100$~s), we attack the roll angle $\phi$ to induce lateral motion along $+y$ or $-y$ while PX4 runs with $K_{vel,P}=0.6~\mathrm{s^{-1}}$ and $K_{pos,P}=1.5~\mathrm{s^{-1}}$. The observed aliased gyro oscillation is approximately 2--2.5~Hz. The AGR modulation is enabled from $t_0=27$~s to $t_1=90$~s (red shaded region). Within this window, Alg.~\ref{alg:phase_gating} schedules multiple trigger windows; the 16 effective\footnote{Effective triggers mean phase-aligned triggers; this excludes delayed/phase-misaligned or falsely fired triggers.} triggers are listed in Tbl.~\ref{tab:trigger_indoor}; green and red markers in Fig.~\ref{fig:roll_est_y} indicate positive and negative bias injections, respectively. The triggers are scheduled near EKF roll peaks/troughs, i.e., when the underlying gyro rate is close to zero, so that each short on/off burst maximally contributes to a signed bias.

The coupling between gyro measurements and EKF attitude is illustrated in Fig.~\ref{fig:gyro_roll_est} and Fig.~\ref{fig:gyro_pitch_est}. Under resonance, the $x$-axis gyro amplitude reaches roughly $\pm 40$~deg/s, yet the raw rate remains visually centered. Small calibration offsets are visible in the EKF roll before the attack, but the dominant growth in $\hat\phi$ and $\hat\theta$ occurs during the biased phase triggers. In contrast, the incremental change in the gyro signal per trigger is subtle. The same pattern holds for pitch.

Figure~\ref{fig:roll_est_y} also shows how the injected tilt bias maps into $y$-displacement, consistent with~(\ref{equ:y_disp}). Under our body-frame sign convention, positive roll-bias triggers correspond to negative $y$ displacement. Once AGR locks to the underlying oscillation, the first four positive-sign triggers (starting at $T_1=53.8$~s) drive the UAV toward $-y$, reaching a peak roll bias of $\approx -5.5^\circ$ and a displacement of $\sim\!-1.1$~m, matching the bounded-drift behavior predicted by~(\ref{equ:y_disp}). The displacement is intentionally kept meter-scale by \textbf{manually turning off the signal} to avoid wall collisions in the 3\,m$\times$3\,m arena; in less constrained settings, a larger residual tilt bias or weaker outer-loop correction could yield a larger equilibrium offset according to~(\ref{equ:y_disp}). To demonstrate bidirectional control, subsequent triggers alternate sign, steering the UAV between nearby $+y$ and $-y$ equilibria. Around $t\approx 60$~s, we observe a brief fast return toward $+y$: the vehicle starts from a displaced $-y$ state and AGR flips the bias sign, so the outer-loop position controller commands thrust that aligns with the induced tilt, temporarily amplifying lateral drive. As sine-fitting becomes less reliable under growing uncertainty and incurs nontrivial compute cost, triggers become sparse; during pauses, the EKF bias decays and the UAV drifts back toward the origin with residual fluctuations. After resonant excitation is disabled ($t>90$~s), both attitude and position converge back to hover because we use moderate bias random-walk variance (i.e., \hyperref[asp:2]{Setting~2} is not enforced).

From the indoor experiments in Fig.~\ref{fig:roll_est_y}, we adopt an empirical phase-alignment criterion to distinguish effective from mismatched triggers. Specifically, a trigger is counted as \emph{phase-aligned} if its scheduled on/off window is centered within $\Delta t_{\max}=20$\,ms (corresponding to approximately $14.4^\circ$--$18^\circ$ phase error for the observed 2--2.5~Hz aliased oscillation) of the predicted peak (or trough) of the fitted gyro oscillation for at least three consecutive cycles.\footnote{This criterion is chosen to reflect practical scheduling latency and timing jitter in the on-board pipeline; it is sufficient to explain the observed directionality and drift-rate differences in our runs.} Phase-aligned triggers reinforce the signed bias injection and lead to faster drift, whereas phase mismatch primarily attenuates the effective injected bias and results in slower (and occasionally inconsistent) displacement. This trend is evident in Fig.~\ref{fig:roll_est_y}: denser sequences of effective triggers produce noticeably faster displacement growth (e.g., the first two batches of red markers), while sparser sequences correlate with weaker drift. Over the full attack window in this run, the fraction of phase-aligned triggers is $47\%$ (16 phase-aligned triggers in scheduled 34 trigger windows). 
Across 10 repeated indoor trials under the same setup, the phase-aligned-trigger rate had a standard deviation of 5.8 percentage points.

Finally, Fig.~\ref{fig:pitch_est_x} shows the $x$-axis motion when we attack roll and primarily target $y$ displacement. In this particular run, the phase gap between roll and pitch happens to be small, so the $x$ trajectory partially mirrors $y$ and the motion is mildly two-dimensional. However, the excited gyro bias on the $x$-axis is only about 60\% of that on $y$, leading to smaller $x$ drift. This asymmetry---stemming from sensor placement and resonance characteristics---is typical and is discussed further as a limitation in Sec.~\ref{sec:limitation}.

% From VICON ground truth during the attack window, the roll deviation (\(\phi\)) is \(2.5^\circ\), versus the EKF-estimated \(2.9^\circ\). This implies a \(\approx 14\%\) attenuation of the actual attitude relative to the estimator, consistent with \modelAbbr's position loop limiting induced tilt and lateral drift. We observe the same trend for \(\theta\). As resonance grows more severe, this attenuation provides a safety margin, reducing peak values and lowering the risk of loss of control.

We also carried out \textbf{outdoor flights} using the GNSS-based positioning stack in Fig.~\ref{fig:pipeline}. Because their behavior closely mirrors the indoor results, we defer plots and detailed analysis to App.~\ref{app:outdoor}. 
The indoor and outdoor experiments validate the signed, safety-bounded displacement in a real PX4 stack.

The bounded displacement observed in our physical flight experiments should be interpreted as a consequence of normal position-hold control, not as a lack of attack impact. Under closed-loop correction, the UAV remains stable but settles at a shifted hover point; after excitation stops in our physical setting, the gyro-bias estimate can recover under moderate bias random-walk variance, allowing the vehicle to return to hover. This behavior remains security-relevant for tasks in which meter-scale position errors can be operationally meaningful (detailed in Sec.~\ref{sec:risk}). The measured displacement is also capped by safety-zone geometry and conservative attitude limits, so it validates the gyro-to-position pathway rather than defining its maximum achievable effect. The following subsection complements the full-flight study with speaker-detached tests on constrained UAV setups. These tests characterize IMU generality, affected axes, and standoff behavior under detached acoustic excitation, while fully black-box, moving-target free flight remains future work because it requires robust tracking, beam steering, and acoustic power management under changing geometry.

\subsection{IMU/Gyro Generality Analysis}
\label{sec:imu_generality}

% Preamble:
% \usepackage{graphicx} % for \scalebox

\begin{table*}[t]
\centering
\footnotesize
\caption{Results of \modelAbbr on drone system with PX4 EKF in-the-loop.}
\label{tbl:imu_generality}
\renewcommand{\arraystretch}{1.15}
\setlength{\tabcolsep}{4.5pt}

\begin{tabular}{|c|c|c|c|c||c|c|c|c|c|c|}
\hline
\textbf{Sensor Model$^{\dagger}$} &
\textbf{\shortstack{Resonant\\Freq. (kHz)}} &
\textbf{\shortstack{Affected\\Axes}} &
\textbf{\shortstack{$D_{max}$\\(m)}} &
\textbf{\shortstack{$T_{min}$\\(s)}} &
\textbf{Sensor Model$^{\dagger}$} &
\textbf{\shortstack{Resonant\\Freq. (kHz)}} &
\textbf{\shortstack{Affected\\Axes}} &
\textbf{\shortstack{$D_{max}$\\(m)}} &
\textbf{\shortstack{$T_{min}$\\(s)}} \\
\hline

IS MPU6000      & 30.0$\pm$3        & {\scriptsize \scalebox{0.85}[1]{$z\!:\!53.56$}}   & 0.45 & NA$^{\ddagger}$   &
IS MPU6500      & 27.0$\pm$2        & {\scriptsize \scalebox{0.85}[1]{$x\!:\!253.95,\;y\!:\!114.3,\;z\!:\!26.36$}} & 0.46 & 45 \\
\hline

IS MPU9250      & 26.1$\pm$2      & {\scriptsize \scalebox{0.85}[1]{$x\!:\!14.2,\;y\!:\!10.1$}}  & 0.18 & 57  &
ST LSM6DS3      & 19.6$\pm$1               & {\scriptsize \scalebox{0.85}[1]{$x\!:\!27.7,\;y\!:\!73,\;z\!:\!13.11$}}      & 0.67   & 15 \\
\hline

ST LSM6DSO      & 20.1$\pm$1               & {\scriptsize \scalebox{0.85}[1]{$x\!:\!29.22,\;y\!:\!10.98$}} & 0.42   & 12 &
BS BMI088       & 24.30$\sim$24.37     & {\scriptsize \scalebox{0.85}[1]{$x\!:\!26.1,\;y\!:\!24.0$}} & 1.52   & 4 \\
\hline

ST LSM9DS1      & 19.44$\sim$20.1 & {\scriptsize \scalebox{0.85}[1]{$x\!:\!32.5,\;y\!:\!42.25$}} & 0.82 & 91  &
ST ISM330DHCX   & 19.9$\sim$20.1  & {\scriptsize \scalebox{0.85}[1]{$x\!:\!51.36,\;y\!:\!389.25$}} & 1.35   & 9 \\
\hline

IS ICM20948     & 23.6$\pm$1      & {\scriptsize \scalebox{0.85}[1]{$x\!:\!1.40,\;z\!:\!1.20$}} & 0.08 & 209 &
BS BMI270     & 27.8$\sim$28.1  &  {\scriptsize \scalebox{0.85}[1]{$x\!:\!43.45,\;y\!:\!19.26; z\!:\!39.74$}}        & 1.12   & 13 \\
\hline

\end{tabular}

\vspace{2pt}
{\footnotesize $^{\dagger}$ IS: InvenSense,\; ST: STMicroelectronics,\; BS: Bosch.\; All sensors have both gyro and accelerometer.\; $^{\ddagger}$ NA: Not applicable to \modelAbbr.}
\end{table*}

To assess whether \modelAbbr depends on a specific IMU/gyro model, we evaluate ten drone-grade inertial sensors, including six not covered by prior acoustic-injection datasets~\cite{son2015rocking,tu2018injected,mohanimufuzzer}. Following prior studies~\cite{son2015rocking,tu2018injected}, this experiment focuses on sensor susceptibility under acoustic excitation. 
We conduct the measurements on a PX4 EKF-in-the-loop, speaker-detached testbed (detailed in App.~\ref{app:imu_benchmark}) using sensors commercially available, widely deployed in the drone ecosystem, or representative of modern flight-controller designs. Because MEMS responses depend on acoustic incidence angle~\cite{gao2022kite,trippel2017walnut}, we use a fixed geometry across all sensors: the Fostex FT17 ultrasonic projector is placed behind the Holybro X500, i.e., along the body-frame $+x$ direction, while the drone is constrained on a 3-DoF UAV testing platform~\cite{sharma2023open,kim2023development,ahmad2024attitude} with fixed yaw, corresponding to $\psi=0$ in Fig.~\ref{fig:pipeline}. This setup keeps the attacker--victim geometry consistent so that differences in response mainly reflect sensor and installation properties.

For each IMU/gyro, we measure $D_{\max}$, the largest standoff distance where inertial disturbance exceeds $0.1$~rad/s, and $T_{\min}$, the time to reach at least $5^\circ$ attitude-estimation error at $D_{\max}$. Measurements use a Fostex FT17H tweeter rated at 96~dB SPL at 1~W/1~m~\cite{fostex_ft17h}; this single-tweeter setup is not range- or power-optimized, so $D_{\max}$ should not be interpreted as a fundamental acoustic limit. Higher-power or arrayed ultrasonic projectors may support longer range~\cite{tu2018injected}. We also report affected axes using gyro perturbation amplitude at $0.1$~m, normalized to the no-attack baseline as in~\cite{son2015rocking}; axes below 1.1 
% in the metrics
are treated as unaffected. Results are summarized in Tbl.~\ref{tbl:imu_generality}, with logs and scripts released in~\cite{sonic_nudge}.

Overall, several sensors exhibit clear susceptibility under speaker-detached excitation, with the strongest coupling usually appearing on roll/pitch axes ($x/y$) rather than yaw ($z$). This supports our focus on tilt-induced displacement while clarifying important boundary conditions: acoustic coupling depends on sensor model, installation geometry, affected axis, and attacker--victim relative position.

\section{Discussion}
\label{sec:discussion}

\subsection{Countermeasures}

A direct countermeasure against acoustic or electromagnetic signal injection is to harden the IMU itself, e.g., through rigid shielding of the sensor module~\cite{jang2023paralyzing} or meta-material acoustic resonators that attenuate narrow frequency bands at the MEMS gyroscope~\cite{churgin2025mitigating}. However, such protections add cost, weight, calibration burden, and design complexity, and are unlikely to be broadly deployed on consumer- and prosumer-grade drones. We therefore focus on firmware- and estimator-level mitigations that can be applied to existing platforms.

Our analysis shows that the induced attitude bias is the predictable outcome of two reasonable estimator choices: down-weighting accelerometer updates during dynamic flight and using small gyro-bias random-walk variance to reduce jitter. Mitigation therefore follows from adjusting these choices without simply making the estimator overly reactive. Accelerometer updates should remain robust but not overly conservative, e.g., by moderating the tilt-update covariance $R_{\text{tilt}}$ and using principled gating based on $\|\mathbf{a}\|-g$, so that accelerometer information is not disabled for long periods. Gyro-bias states should also retain sufficient agility by increasing the bias random-walk variances $\sigma_{bw}^2$ within stability margins, allowing $\hat b_p$ and $\hat b_q$ to track low-frequency content instead of lagging behind it. In addition, prefiltering or detecting asymmetric low-frequency gyro components, e.g., through moving-mean tests, offset monitors, or consistency checks between gyro-integrated attitude and independent tilt cues, can prevent quasi-DC and $\mathcal{O}(1\,\mathrm{Hz})$ components from accumulating into systematic attitude error.

Another mitigation is to bound the EKF tilt estimate or the position-loop tilt command. As discussed in Sec.~\ref{sec:control}, limiting $\lvert \hat\phi \rvert$ or $\lvert \phi_{sp} \rvert$ directly bounds the maximum drift via~(\ref{equ:y_max}). This defense is attractive because it constrains the downstream control consequence even if the sensor perturbation is not fully detected. However, it introduces a control tradeoff: overly restrictive tilt limits reduce lateral authority for trajectory tracking, gust rejection, and emergency avoidance, and may be unsuitable for aggressive maneuvers or windy environments.

Adding integral action to the outer position loop could also reduce steady-state displacement by driving position error toward zero. However, the thrust/tilt-to-position path is effectively a double integrator, and a naive PI upgrade can reduce phase margin, induce oscillations, slow recovery, or cause windup under actuator saturation. It may also integrate estimator or frame discontinuities, such as EKF origin or yaw resets, into long-lived position biases. Thus, outer-loop integral action is not a straightforward hardening measure for safety-critical UAVs; it would require anti-windup logic, reset handling, and careful gain scheduling.

\subsection{Limitations and Generalizability}
\label{sec:limitation}

\modelAbbr is effective in our hover setting, but its practicality is constrained by sensor, controller, and triggering conditions.

\textbf{Axis asymmetry and direction control.}
For a given IMU, resonance strength and phase can differ across the $x$-, $y$-, and $z$-axes. Roll and pitch gyros may therefore require different acoustic amplitudes to achieve comparable drift speeds. A nominally ``pure'' global direction, such as straight forward, can also become forward-left or forward-right depending on the phase relation among roll, pitch, and yaw. If the UAV yaws during the attack, the same body-frame bias maps to a different world-frame direction, requiring continuous geometry and yaw compensation for precise directional control.

\textbf{Controller tuning dependence.}
Beyond resonance identification, displacement depends on the outer-loop gains $K_{vel,P}$ and $K_{pos,P}$ via~\eqref{equ:y_disp}. These gains are platform- and mission-specific and are not directly observable, making precise drift prediction on an unknown target difficult. We use representative tunings in simulation and experiments; more aggressive or more conservative settings could reduce or amplify the effect. This dependence also means that \modelAbbr should be interpreted as an estimator--controller pathway whose magnitude is shaped by the closed-loop stack rather than a sensor effect.

\textbf{Phase-locked triggering robustness.}
AGR requires reliable fitting of the aliased gyro response to predict the next peak or trough. Sampling-rate drift and aliasing effects~\cite{tu2018injected} can shift the effective frequency after locking, reducing trigger accuracy. Nonlinear fitting also introduces computational delay; if fitting and prediction lag the incoming stream, triggers may miss the desired phase window, weakening the induced bias or reversing its sign. These constraints limit how densely and responsively AGR can fire triggers, especially on resource-constrained attacker hardware or under aggressive motions.

% Our attack is not trajectory-stealthy: global and local position sensors (e.g., GNSS, optical flow) do observe the induced drift. The challenge is not \emph{detecting} displacement but \emph{interpreting} it: from the controller's perspective, a persistent lateral offset may equally arise from external disturbances, miscalibration, or inertial bias, and there is no obvious baseline that cleanly separates these cases online. A promising direction is \emph{state-dependent} outer-loop design, where the controller selectively enables integral action or adjusts PID gains based on flight context. For example, the system can favor a robust P-only position loop under aggressive or uncertain conditions (strong gusts, transient sensor anomalies), and enable PI controller only during calm hover near protected regions to reduce steady-state drift. Such conditional scheduling preserves most of the maneuverability benefits of cascaded control in nominal flight, while narrowing the operational window in which a quasi-steady gyro bias turns into a sustained, exploitable drift.

Subject to these limitations, the underlying mechanism is not specific to multirotors. Any IMU-driven cyber--physical system that uses a resonance-susceptible MEMS IMU and closes a feedback loop around attitude estimates may exhibit related estimator--controller coupling. Although our experiments focus on UAVs as a concrete, safety-relevant case, this analysis applies to broader IMU-based platforms, including but not limited to camera gimbals, head-mounted displays, and self-balancing personal transporters.

\section{Related Work}
\label{sec:related_work}

\subsection{Counter--UAV and Displacement Techniques}

Counter--UAV techniques are commonly divided into destructive neutralization and non-destructive diversion. Destructive methods include RF jamming of command-and-control or telemetry links~\cite{abunada2020design,multerer2017low}, platform takeover~\cite{dulo2015unmanned,kang2025multi}, GNSS denial or jamming~\cite{noh2019tractor,ceccato2020spatial,steiner2024drone}, net-based capture~\cite{yu2022design}, and kinetic interception~\cite{fishman2021dynamic,fedoseev2021dronetrap}. While effective in some settings, these approaches may introduce legal, safety, and collateral-interference risks, especially near people, infrastructure, or benign RF/GNSS users. They may also be less reliable against platforms with encrypted links, inertial/vision fusion, and robust failsafe behaviors.

Non-destructive approaches seek to keep the UAV airworthy while moving it away from a protected region. Geofencing~\cite{gurriet2016towards} and airspace-management policies mainly affect compliant drones, while physical barriers are static and infrastructure-heavy. Existing displacement-oriented attacks typically manipulate navigation or translational-motion inputs: GNSS spoofing biases perceived global position~\cite{khazraei2024black,shen2020drift, zhang2025ghost}, and optical-flow spoofing creates fictitious ground-plane motion that the controller compensates for~\cite{davidson2016controlling}. \modelAbbr introduces a different displacement primitive: it perturbs angular-rate sensing and studies bounded displacement as an estimator--controller outcome, rather than as a direct navigation-spoofing or destructive neutralization objective.

\subsection{Acoustic Attacks on Inertial Sensors} 

Prior work shows that MEMS inertial sensors can be affected by ultrasonic excitation near mechanical resonance. Acoustic inputs can alias through the sensor sampling path and induce structured errors in accelerometers or gyroscopes~\cite{son2015rocking,tu2018injected,trippel2017walnut}. Son et al.~\cite{son2015rocking} demonstrated drone destabilization through acoustic gyroscope interference, and later studies examined out-of-band injection, feedback-guided signal generation, and coupling through PCB or housing modes~\cite{tu2018injected,mohanimufuzzer,gao2022kite,gao2023exploring}. These works establish that acoustic or transduction-based perturbations can corrupt inertial measurements and, in some cases, mislead higher-level systems such as mobile devices, robots, and drones~\cite{son2015rocking,meng2025mars}. 

However, most prior acoustic IMU attacks focus on sensor corruption, denial-of-service, imbalance, or crash. They typically treat the estimator and controller as a black box, rather than analyzing how a quasi-steady gyroscope bias is absorbed by an EKF and then acted on by cascaded UAV control loops. In contrast, \modelAbbr uses acoustic excitation as an exemplar way to inject signed low-level inertial errors, and studies when these errors become bounded, directionally programmable lateral displacement.

% Prior work shows that MEMS inertial sensors can be affected by ultrasonic excitation near mechanical resonance. Acoustic inputs can alias through the sensor sampling path and induce structured errors in accelerometers or gyroscopes~\cite{son2015rocking,tu2018injected,trippel2017walnut}. Son et al.~\cite{son2015rocking} demonstrated drone destabilization through acoustic gyroscope interference, and later studies examined out-of-band injection, feedback-guided signal generation, and coupling through PCB or housing modes~\cite{tu2018injected,mohanimufuzzer,gao2022kite,gao2023exploring}. These works establish that acoustic or transduction-based perturbations can corrupt inertial measurements and mislead higher-level systems such as mobile devices, robots, and drones~\cite{son2015rocking,meng2025mars}. \modelAbbr differs by using acoustic excitation as an exemplar way to inject signed low-level gyroscope errors and by analyzing how a quasi-steady bias propagates through the EKF and cascaded UAV controller to produce bounded, directionally programmable lateral displacement rather than imbalance or crash.

\subsection{Estimation--Control-Path Attacks on UAVs}

A broader line of CPS security research studies how crafted sensor errors can manipulate estimators and feedback controllers. False-data injection attacks show that adversaries can bias Kalman-filtered state estimates, evade residual-based detectors, excite weakly damped modes, or steer systems toward unsafe equilibria~\cite{khazraei2024black,shen2020drift,chen2024adversary,khazraei2023stealthy,duo2022survey}. In UAVs, related work has shown that falsified GNSS, IMU, or visual inputs can mislead navigation stacks through control-aware attacks rather than simple jamming~\cite{davidson2016controlling,khazraei2024black}. 

\modelAbbr extends this perspective to a lower-level inertial pathway, showing that residual roll/pitch bias in angular-rate sensing can be retained by the estimator and converted by the normal cascaded controller into a shifted hover equilibrium without directly spoofing position or motion inputs.

\section{Conclusion}
\label{sec:conclusion}

This paper shows that UAV displacement can emerge from estimator--controller coupling rather than direct manipulation of navigation inputs. Through analysis, simulation, and controlled flight experiments, we demonstrate that small gyroscope perturbations can produce repeatable lateral displacement while preserving stable flight. The results show that the outcome depends on estimator tuning, bias persistence, position-correction strength, and sensor-specific resonance properties. These findings suggest that UAV defenses should account not only for anomalous sensor readings, but also for how residual sensing errors propagate through estimation and control. More broadly, \modelAbbr motivates treating the estimator--controller stack as an attack surface for UAVs and other IMU-based cyber--physical systems.

% \cleardoublepage
\section*{Acknowledgment}
\label{app:ack}

This work is sponsored in part by the ONR under agreement N00014-23-1-2206, AFOSR under the award number FA9550-19-1-0169, and by the NSF under NAIAD Award 2332744 as well as the National AI Institute for Edge Computing Leveraging Next Generation Wireless Networks, Grant CNS-2112562. Moreover,  the research was sponsored by the Army Research Office and was accomplished under Cooperative Agreement Number W911NF-26-2-A165. The views and conclusions contained in this document are those of the authors and should not be interpreted as representing the official policies, either expressed or implied, of the Army Research Office or the U.S. Government. The U.S. Government is authorized to reproduce and distribute reprints for Government purposes notwithstanding any copyright notation herein. %remove for submission

\bibliographystyle{plainurl}
\bibliography{reference}

@inproceedings{tu2018injected,
  title={Injected and delivered: Fabricating implicit control over actuation systems by spoofing inertial sensors},
  author={Tu, Yazhou and Lin, Zhiqiang and Lee, Insup and Hei, Xiali},
  booktitle={27th USENIX security symposium (USENIX Security 18)},
  pages={1545--1562},
  year={2018}
}

@inproceedings{son2015rocking,
  title={Rocking drones with intentional sound noise on gyroscopic sensors},
  author={Son, Yunmok and Shin, Hocheol and Kim, Dongkwan and Park, Youngseok and Noh, Juhwan and Choi, Kibum and Choi, Jungwoo and Kim, Yongdae},
  booktitle={24th USENIX security symposium (USENIX Security 15)},
  pages={881--896},
  year={2015}
}

@inproceedings{trippel2017walnut,
  title={WALNUT: Waging doubt on the integrity of MEMS accelerometers with acoustic injection attacks},
  author={Trippel, Timothy and Weisse, Ofir and Xu, Wenyuan and Honeyman, Peter and Fu, Kevin},
  booktitle={2017 IEEE European symposium on security and privacy (EuroS\&P)},
  pages={3--18},
  year={2017},
  organization={IEEE}
}

@inproceedings{gao2022kite,
  title={KITE: Exploring the practical threat from acoustic transduction attacks on inertial sensors},
  author={Gao, Ming and Zhang, Lingfeng and Shen, Leming and Zou, Xiang and Han, Jinsong and Lin, Feng and Ren, Kui},
  booktitle={Proceedings of the 20th ACM conference on embedded networked sensor systems},
  pages={696--709},
  year={2022}
}

@misc{invensense_mpu6500,
  title        = {MPU-6500 6-Axis MotionTracking Device},
  author       = {{InvenSense, Inc.}},
  year         = {2013},
  note         = {Datasheet},
  howpublished = {\url{https://invensense.tdk.com/download-pdf/mpu-6500-datasheet/}},
}

@misc{sonic_nudge,
  title        = {Sonic{N}udge Anonymous Open Source Website},
  author       = {Sonic{N}udge},
  year         = {2026},
  howpublished = {\url{https://tinyurl.com/Sonic-Nudge/}},
}

@article{meng2025mars,
  title={MARS: Defending Unmanned Aerial Vehicles From Attacks on Inertial Sensors with Model-based Anomaly Detection and Recovery},
  author={Meng, Haocheng and Luo, Shaocheng and Liang, Zhenyuan and Huang, Qing and Khazraei, Amir and Pajic, Miroslav},
  journal={arXiv preprint arXiv:2505.00924},
  year={2025}
}

@misc{ardupilot,
  title        = {ArduPilot: Open Source Autopilot Software Suite},
  author       = {{ArduPilot Community}},
  year         = {2025},
  note         = {Project Website},
  howpublished = {\url{https://ardupilot.org/}},
}

@online{px4:cuav_v6x,
  title   = {CUAV Pixhawk V6X},
  author  = {{PX4 Autopilot Documentation Team}},
  year    = {2025},
  url     = {https://docs.px4.io/main/en/flight_controller/cuav_pixhawk_v6x},
  note    = {PX4 User Guide},
  urldate = {2025-10-24}
}

@inproceedings{abunada2020design,
  title={Design and implementation of a RF based anti-drone system},
  author={Abunada, Abdularahman Hasan and Osman, Ahmed Yousif and Khandakar, Amith and Chowdhury, Muhammad Enamul Hoque and Khattab, Tamer and Touati, Farid},
  booktitle={2020 IEEE International Conference on Informatics, IoT, and Enabling Technologies (ICIoT)},
  pages={35--42},
  year={2020},
  organization={IEEE}
}

@inproceedings{multerer2017low,
  title={Low-cost jamming system against small drones using a 3D MIMO radar based tracking},
  author={Multerer, Thomas and Ganis, Alexander and Prechtel, Ulrich and Miralles, Enric and Meusling, Askold and Mietzner, Jan and Vossiek, Martin and Loghi, Mirko and Ziegler, Volker},
  booktitle={2017 European radar conference (EURAD)},
  pages={299--302},
  year={2017},
  organization={IEEE}
}

@article{noh2019tractor,
  title={Tractor beam: Safe-hijacking of consumer drones with adaptive GPS spoofing},
  author={Noh, Juhwan and Kwon, Yujin and Son, Yunmok and Shin, Hocheol and Kim, Dohyun and Choi, Jaeyeong and Kim, Yongdae},
  journal={ACM Transactions on Privacy and Security (TOPS)},
  volume={22},
  number={2},
  pages={1--26},
  year={2019},
  publisher={ACM New York, NY, USA}
}

@inproceedings{yu2022design,
  title={Design and testing of a net-launch device for drone capture},
  author={Yu, Derek and Judasz, Andrea and Zheng, Minghui and Botta, Eleonora M},
  booktitle={AIAA Scitech 2022 Forum},
  pages={0273},
  year={2022}
}

@inproceedings{fishman2021dynamic,
  title={Dynamic grasping with a" soft" drone: From theory to practice},
  author={Fishman, Joshua and Ubellacker, Samuel and Hughes, Nathan and Carlone, Luca},
  booktitle={2021 IEEE/RSJ International Conference on Intelligent Robots and Systems (IROS)},
  pages={4214--4221},
  year={2021},
  organization={IEEE}
}

@inproceedings{brisset2006paparazzi,
  title={The paparazzi solution},
  author={Brisset, Pascal and Drouin, Antoine and Gorraz, Michel and Huard, Pierre-Selim and Tyler, Jeremy},
  booktitle={MAV 2006, 2nd US-European competition and workshop on micro air vehicles},
  pages={1--15},
  year={2006}
}

@article{ceccato2020spatial,
  title={Spatial GNSS spoofing against drone swarms with multiple antennas and Wiener filter},
  author={Ceccato, Marco and Formaggio, Francesco and Tomasin, Stefano},
  journal={IEEE Transactions on Signal Processing},
  volume={68},
  pages={5782--5794},
  year={2020},
  publisher={IEEE}
}

@inproceedings{steiner2024drone,
  title={Drone e-ID Vulnerability to GNSS Interference},
  author={Steiner, Jakub and {\v{C}}ern{\`y}, Michal},
  booktitle={2024 New Trends in Civil Aviation (NTCA)},
  pages={207--212},
  year={2024},
  organization={IEEE}
}

@inproceedings{fedoseev2021dronetrap,
  title={Dronetrap: Drone catching in midair by soft robotic hand with color-based force detection and hand gesture recognition},
  author={Fedoseev, Aleksey and Serpiva, Valerii and Karmanova, Ekaterina and Cabrera, Miguel Altamirano and Shirokun, Vladimir and Vasilev, Iakov and Savushkin, Stanislav and Tsetserukou, Dzmitry},
  booktitle={2021 IEEE 4th International Conference on Soft Robotics (RoboSoft)},
  pages={261--266},
  year={2021},
  organization={IEEE}
}

@inproceedings{gurriet2016towards,
  title={Towards a generic and modular geofencing strategy for civilian UAVs},
  author={Gurriet, Thomas and Ciarletta, Laurent},
  booktitle={2016 International Conference on Unmanned Aircraft Systems (ICUAS)},
  pages={540--549},
  year={2016},
  organization={IEEE}
}

@article{dulo2015unmanned,
  title={Unmanned aircraft: the rising risk of hostile takeover [leading edge]},
  author={Dulo, Donna A},
  journal={IEEE Technology and Society Magazine},
  volume={34},
  number={3},
  pages={17--19},
  year={2015},
  publisher={IEEE}
}

@article{kang2025multi,
  title={Multi-modem-based FHSS-drone takeover with precision spoofing},
  author={Kang, Jae Gu and Choi, Byeong Cheol},
  journal={ETRI Journal},
  year={2025},
  publisher={Wiley Online Library}
}

@article{gao2023exploring,
  title={Exploring practical acoustic transduction attacks on inertial sensors in MDOF systems},
  author={Gao, Ming and Zhang, Lingfeng and Shen, Leming and Zou, Xiang and Han, Jinsong and Lin, Feng and Ren, Kui},
  journal={IEEE Transactions on Mobile Computing},
  volume={23},
  number={5},
  pages={3539--3557},
  year={2023},
  publisher={IEEE}
}

@inproceedings{mohanimufuzzer,
  title        = {{IMUFUZZER}: Resilience-based Discovery of Signal Injection Attacks on Robotic Aerial Vehicles},
  author       = {Mohan, Sudharssan and Yang, Kyeongseok and Kong, Zelun and Kwon, Yonghwi and Rhee, Junghwan and Summers, Tyler and Choi, Hongjun and Lee, Heejo and Kim, Chung Hwan},
  booktitle    = {Proceedings of the 40th IEEE/ACM International Conference on Automated Software Engineering (ASE~2025)},
  year         = {2025},
  organization = {IEEE/ACM}
}

@inproceedings{shen2020drift,
  title={Drift with devil: Security of $\{$Multi-Sensor$\}$ fusion based localization in $\{$High-Level$\}$ autonomous driving under $\{$GPS$\}$ spoofing},
  author={Shen, Junjie and Won, Jun Yeon and Chen, Zeyuan and Chen, Qi Alfred},
  booktitle={29th USENIX security symposium (USENIX Security 20)},
  pages={931--948},
  year={2020}
}

@mastersthesis{churgin2025mitigating,
  title={Mitigating Ultrasonic Attack on Drones Using Meta-Material Acoustic Resonators},
  author={Churgin, Corey},
  year={2025},
  school={Rowan University}
}

@inproceedings{jang2023paralyzing,
  title={Paralyzing Drones via EMI Signal Injection on Sensory Communication Channels.},
  author={Jang, Joon-Ha and Cho, Mangi and Kim, Jaehoon and Kim, Dongkwan and Kim, Yongdae},
  booktitle={NDSS},
  year={2023}
}

@inproceedings{chen2024adversary,
  title={Adversary is on the road: Attacks on visual $\{$SLAM$\}$ using unnoticeable adversarial patch},
  author={Chen, Baodong and Wang, Wei and Sikorski, Pascal and Zhu, Ting},
  booktitle={33rd USENIX Security Symposium (USENIX Security 24)},
  pages={6345--6362},
  year={2024}
}

@inproceedings{khazraei2024black,
  title={Black-box Stealthy GPS Attacks on Unmanned Aerial Vehicles},
  author={Khazraei, Amir and Meng, Haocheng and Pajic, Miroslav},
  booktitle={2024 IEEE 63rd Conference on Decision and Control (CDC)},
  pages={2857--2862},
  year={2024},
  organization={IEEE}
}

@article{khazraei2023stealthy,
  title={Stealthy perception-based attacks on unmanned aerial vehicles},
  author={Khazraei, Amir and Meng, Haocheng and Pajic, Miroslav},
  journal={arXiv preprint arXiv:2303.02112},
  year={2023}
}

@article{duo2022survey,
  title={A survey of cyber attacks on cyber physical systems: Recent advances and challenges},
  author={Duo, Wenli and Zhou, MengChu and Abusorrah, Abdullah},
  journal={IEEE/CAA Journal of Automatica Sinica},
  volume={9},
  number={5},
  pages={784--800},
  year={2022},
  publisher={IEEE}
}

@inproceedings{davidson2016controlling,
  title={Controlling $\{$UAVs$\}$ with sensor input spoofing attacks},
  author={Davidson, Drew and Wu, Hao and Jellinek, Rob and Singh, Vikas and Ristenpart, Thomas},
  booktitle={10th USENIX workshop on offensive technologies (WOOT 16)},
  year={2016}
}

@techreport{DOT-VNTSC-OST-05-01,
  title        = {Global Positioning System (GPS) Civil Monitoring Performance Specification},
  institution  = {U.S. Department of Transportation, Volpe National Transportation Systems Center},
  type         = {Technical Report},
  number       = {DOT-VNTSC-OST-05-01},
  address      = {Washington, DC},
  year         = {2005},
  month        = dec,
  url          = {https://www.gps.gov/sites/default/files/2025-07/2005-civil-monitoring-performance-specification.pdf},
  urldate      = {2025-11-13},
  note         = {Approved for public release}
}

@inproceedings{tibaldo2025gnss,
  title={$\{$GNSS-WASP$\}$:$\{$GNSS$\}$ Wide Area $\{$SPoofing$\}$},
  author={Tibaldo, Christopher and Sathaye, Harshad and Camurati, Giovanni and Capkun, Srdjan},
  booktitle={34th USENIX Security Symposium (USENIX Security 25)},
  pages={7039--7058},
  year={2025}
}

@inproceedings{sathaye2022experimental,
  title={An experimental study of $\{$GPS$\}$ spoofing and takeover attacks on $\{$UAVs$\}$},
  author={Sathaye, Harshad and Strohmeier, Martin and Lenders, Vincent and Ranganathan, Aanjhan},
  booktitle={31st USENIX security symposium (USENIX security 22)},
  pages={3503--3520},
  year={2022}
}

@inproceedings{shamsi2024wip,
  title={WIP: threat modeling laser-induced acoustic interference in computer vision-assisted vehicles},
  author={Shamsi, Nina and Chandrasekar, Kaeshav and Long, Yan and Limbach, Christopher and Rebello, Keith and Fu, Kevin},
  booktitle={Symposium on Vehicles Security and Privacy (VehicleSec)},
  year={2024}
}

@article{sharma2023open,
  title={Open gimbal: A 3 degrees of freedom open source sensing and testing platform for nano-and micro-uavs},
  author={Sharma, Suryansh and Dijkstra, Tristan and Prasad, Ranga Venkatesha},
  journal={IEEE Sensors Letters},
  volume={7},
  number={9},
  pages={1--4},
  year={2023},
  publisher={IEEE}
}

@article{kim2023development,
  title={Development and verification of a ros-based multi-dof flight test system for unmanned aerial vehicles},
  author={Kim, Shinhyung and Philip, Opayemi Alaba and Tullu, Abera and Jung, Sunghun},
  journal={Ieee Access},
  volume={11},
  pages={37068--37081},
  year={2023},
  publisher={IEEE}
}

@article{ahmad2024attitude,
  title={Attitude modelling and real-time robust control of a 3-DoF quadcopter UAV test bench},
  author={Ahmad, SM and Fareed, S},
  journal={The Aeronautical Journal},
  volume={128},
  number={1326},
  pages={1767--1788},
  year={2024},
  publisher={Cambridge University Press}
}

@inproceedings{allouch2019mavsec,
  title={MAVSec: Securing the MAVLink protocol for ardupilot/PX4 unmanned aerial systems},
  author={Allouch, Azza and Cheikhrouhou, Omar and Koub{\^a}a, Anis and Khalgui, Mohamed and Abbes, Tarek},
  booktitle={2019 15th International Wireless Communications \& Mobile Computing Conference (IWCMC)},
  pages={621--628},
  year={2019},
  organization={IEEE}
}

@inproceedings{kim2019rvfuzzer,
  title={$\{$RVFuzzer$\}$: Finding input validation bugs in robotic vehicles through $\{$Control-Guided$\}$ testing},
  author={Kim, Taegyu and Kim, Chung Hwan and Rhee, Junghwan and Fei, Fan and Tu, Zhan and Walkup, Gregory and Zhang, Xiangyu and Deng, Xinyan and Xu, Dongyan},
  booktitle={28th USENIX Security Symposium (USENIX Security 19)},
  pages={425--442},
  year={2019}
}

@article{koubaa2019micro,
  title={Micro air vehicle link (mavlink) in a nutshell: A survey},
  author={Koub{\^a}a, Anis and Allouch, Azza and Alajlan, Maram and Javed, Yasir and Belghith, Abdelfettah and Khalgui, Mohamed},
  journal={IEEE Access},
  volume={7},
  pages={87658--87680},
  year={2019},
  publisher={IEEE}
}

@article{bithas2020uav,
  title={UAV-to-ground communications: Channel modeling and UAV selection},
  author={Bithas, Petros S and Nikolaidis, Viktor and Kanatas, Athanasios G and Karagiannidis, George K},
  journal={IEEE Transactions on Communications},
  volume={68},
  number={8},
  pages={5135--5144},
  year={2020},
  publisher={IEEE}
}

@inproceedings{zhang2025ghost,
  title={The Ghost Navigator: Revisiting the Hidden Vulnerability of Localization in Autonomous Driving},
  author={Zhang, Junqi and Cheng, Shaoyin and Hu, Linqing and Zhang, Jie and Shi, Chengyu and Han, Xingshuo and Zhang, Tianwei and Cheng, Yueqiang and Zhang, Weiming},
  booktitle={34th USENIX Security Symposium (USENIX Security 25)},
  pages={3979--3998},
  year={2025}
}

@inproceedings{kim2024systematic,
  title={A systematic study of physical sensor attack hardness},
  author={Kim, Hyungsub and Bandyopadhyay, Rwitam and Ozmen, Muslum Ozgur and Celik, Z Berkay and Bianchi, Antonio and Kim, Yongdae and Xu, Dongyan},
  booktitle={2024 IEEE Symposium on Security and Privacy (SP)},
  pages={2328--2347},
  year={2024},
  organization={IEEE}
}

@article{qiu2022review,
  title={Review of ultrasonic ranging methods and their current challenges},
  author={Qiu, Zurong and Lu, Yaohuan and Qiu, Zhen},
  journal={Micromachines},
  volume={13},
  number={4},
  pages={520},
  year={2022},
  publisher={MDPI}
}

@article{kumar2017long,
  title={Long-range measurement system using ultrasonic range sensor with high-power transmitter array in air},
  author={Kumar, Sahdev and Furuhashi, Hideo},
  journal={Ultrasonics},
  volume={74},
  pages={186--195},
  year={2017},
  publisher={Elsevier}
}

@article{mahony2012multirotor,
  title={Multirotor aerial vehicles: Modeling, estimation, and control of quadrotor},
  author={Mahony, Robert and Kumar, Vijay and Corke, Peter},
  journal={IEEE robotics \& automation magazine},
  volume={19},
  number={3},
  pages={20--32},
  year={2012},
  publisher={IEEE}
}

@article{alexis2012model,
  title={Model predictive quadrotor control: attitude, altitude and position experimental studies},
  author={Alexis, Kostas and Nikolakopoulos, George and Tzes, Anthony},
  journal={IET Control Theory \& Applications},
  volume={6},
  number={12},
  pages={1812--1827},
  year={2012},
  publisher={IET}
}

@inproceedings{meier2015px4,
  title={PX4: A node-based multithreaded open source robotics framework for deeply embedded platforms},
  author={Meier, Lorenz and Honegger, Dominik and Pollefeys, Marc},
  booktitle={2015 IEEE international conference on robotics and automation (ICRA)},
  pages={6235--6240},
  year={2015},
  organization={IEEE}
}

@misc{fostex_ft17h,
  title        = {{Fostex FT17H Horn Super Tweeter}},
  author       = {{Madisound Speaker Components}},
  howpublished = {\url{https://www.madisoundspeakerstore.com/bullet-tweeters/fostex-ft17h-horn-super-tweeter/}},
  note         = {Accessed: 2026-08-19}
}
\appendix
\section{Appendix}

\subsection{PX4 Estimator–Controller Coupling Exploited by \modelAbbr}
\label{sec:px4_coupling}

PX4 ingests high-rate inertial data and fuses it with slower absolute sensors. A 3-axis gyroscope and accelerometer stream at $\sim\!1\,\mathrm{kHz}$ and dominate short-horizon attitude information, while GNSS, magnetometer, barometer, and optional vision/optical-flow provide drift-bounding updates at lower rates. PX4’s \texttt{EKF2} fuses these streams to produce a NED-frame state (position, velocity, attitude, and IMU-bias estimates) at a few hundred hertz. Because only the gyroscope supplies low-latency angular-rate cues, roll/pitch estimates are gyro-led between absolute corrections.

Downstream, PX4 uses a cascaded controller structurally equivalent to Fig.~\ref{fig:sys_fig}. At about $50\,\mathrm{Hz}$, the outer position/velocity loop (P for position, PID for velocity) takes the commanded position $X_{sp}$ and the EKF state, producing an inertial-frame acceleration/tilt setpoint $A_{sp}$ and a yaw setpoint. This is mapped to an attitude setpoint $q_{sp}$. At about $250\,\mathrm{Hz}$, the attitude controller (P) tracks $q_{sp}$ and generates desired body rates $\Omega_{sp}$. Finally, at about $1\,\mathrm{kHz}$, the rate controller (PID) tracks $\Omega_{sp}$ directly from the \emph{raw gyroscope} and outputs actuator commands. In effect, whatever attitude the estimator reports is the attitude the controller tries to hold, while the fast inner loop stabilizes body dynamics largely independently of slow spatial objectives.

\modelAbbr exploits this estimator–controller handshake with a directional ultrasonic nudge. Ultrasound excites a MEMS-gyro resonance and, with bias-preserving modulation, imprints a low-frequency, bias-like component in the measured angular rate. The rate loop largely ignores the ultrasonic carrier due to low-pass filtering, but the same gyro stream feeds \texttt{EKF2}; integrating a residual rate bias yields a persistent roll/pitch error. In realistic flight, accelerometer corrections are often down-weighted (e.g., during thrusting, gusts, or vibration) and gyro-bias states are tuned with small process noise to suppress jitter, so the filter adapts slowly—leaving a mismatch between true and estimated bias that sustains the attitude error.

The biased attitude then propagates through the cascade. The inner attitude loop faithfully tracks the biased tilt; the outer position/velocity loop interprets the resulting apparent lateral acceleration and commands a counter-tilt, but because the estimate itself is biased it cannot drive both attitude and position errors to zero simultaneously. The UAV converges to a displaced equilibrium with a finite, directionally programmable horizontal offset—a gentle, bounded repulsion rather than an aggressive runaway—precisely the behavior \modelAbbr leverages.

\subsection{Initial Attack Formulation}
\label{sec:ekf}

\begin{table}[h]
\centering
\begin{minipage}{0.48\textwidth}
\centering
\caption{Notation for \modelAbbr control under acoustic attacks.}
\setlength{\tabcolsep}{2pt}       % tighter column spacing
\renewcommand{\arraystretch}{0.95}% tighter row spacing
\footnotesize                     % smaller font
\begin{tabular}{@{}p{0.23\linewidth} p{0.57\linewidth} p{0.18\linewidth}@{}}
\hline
\textbf{Symbol} & \textbf{Meaning} & \textbf{Units} \\
\hline
$t,\;k$ & Continuous time / discrete index & s / -- \\
$T_s$ & Sample period ($T_s = 1/f_s$) & s \\
\hline
$\phi(t)$ & True roll angle & rad \\
$\theta(t)$ & True pitch angle & rad \\
$\psi(t)$ & True yaw angle & rad \\
$p(t)=\dot{\phi}(t)$ & True roll rate & rad/s \\
$q(t)=\dot{\theta}(t)$ & True pitch rate & rad/s \\
$r(t)=\dot{\psi}(t)$ & True yaw rate & rad/s \\
$\phi_{\mathrm{sp}}$ & Commanded roll from outer loops & rad \\
$\phi_{\mathrm{sp}}$ & Commanded roll rate from attitude P & rad/s \\
\hline
$\hat{\phi},\;\hat{b}$ & EKF roll estimate and gyro-bias estimate & rad,\;rad/s \\
$z$ & Tilt (accelerometer) measurement & rad \\
$[K_\phi\;\;K_{\phi,b}]^\top$ & EKF Kalman gain (angle/bias) & -- \\
$\mathbf{P},\;\mathbf{Q},\;\mathbf{R}$ & Covariance, process noise, measurement noise & -- \\
$H$ & Measurement matrix ($z=H[\theta\;b]^\top+v$) & -- \\
$\varepsilon_{\cross}$ & Residual EKF angle bias (post-fusion) & rad \\
\hline
$p_{meas}$ & Measured/attacked gyro rate & rad/s \\
$p_m(t)$ & Injected gyro attack signal & rad/s \\
$\mu_p$ & Mean (DC) of $p_m$ over a cycle & rad/s \\
$\tilde p(t)$ & Zero-mean AC part of $p_m$ & rad/s \\
$A,\;F_{obs},\;\varphi$ & Attack amplitude, frequency, phase & rad/s,\;Hz,\;rad \\
\hline
$u$ & Control torque command (rate PI output) & N$\cdot$m \\
$K_{AC}$ & Attitude P gain (angle\,$\to$\,rate sp) & -- \\
$K_p,\;K_i,\;K_d$ & Rate-loop PID gains & --,\;--,\;-- \\
$I$ & Rate-loop integrator state & rad \\
$J$ & Body roll inertia & kg·m$^2$ \\
\hline
$g$ & Gravity magnitude (ENU with $+Z$ up) & m/s$^2$ \\
$a_y$ & Lateral accel $\approx g\,\phi$ (small-angle) & m/s$^2$ \\
$v_y$ & Lateral velocity (integral of $a_y$) & m/s \\
\hline
\end{tabular}
\end{minipage}
\end{table}

Following \cite{meng2025mars}, we consider an EKF that estimates the drone's roll and pitch by fusing gyroscope and accelerometer data. Let the state and input be
\begin{equation}
    \mathbf{x} = [\phi, \theta, b_p, b_q]^{\!\top}, \qquad
    \mathbf{u} = [p, q]^{\!\top},
\end{equation}
where $\phi$ and $\theta$ denote roll and pitch, respectively, and $b_p$, $b_q$ are the gyroscope bias states in the roll and pitch rate channels.

The EKF prediction model is
\begin{equation}
    \mathbf{x}_k = \mathbf{F}_k \, \mathbf{x}_{k-1} + \mathbf{B}_k \, \mathbf{u}_k + \mathbf{w}_k,
\end{equation}
where $\mathbf{F}_k$ is the state transition matrix, $\mathbf{B}_k$ the input matrix, and
$\mathbf{w}_k = [w_\phi, w_\theta, w_{b,p}, w_{b,q}]^{\!\top}$ is zero-mean process noise
with covariance $\mathbf{Q}_k$:
$\mathbf{w}_k \sim \mathcal{N}(0, \mathbf{Q}_k)$.
With sampling time $\Delta t$,
\[
\mathbf{F}_k =
\begin{bmatrix}
1 & 0 & -\Delta t & 0 \\
0 & 1 & 0 & -\Delta t \\
0 & 0 & 1 & 0 \\
0 & 0 & 0 & 1
\end{bmatrix},
\qquad
\mathbf{B}_k =
\begin{bmatrix}
\Delta t & 0 \\
0 & \Delta t \\
0 & 0 \\
0 & 0
\end{bmatrix}.
\]

The measurement model is
\begin{equation}
    \mathbf{z}_k = \mathbf{H}_k \, \mathbf{x}_k + \mathbf{v}_k,
\end{equation}
where $\mathbf{z}_k = [z_{\phi,k}, z_{\theta,k}]^{\!\top}$ is the accelerometer-derived tilt,
\[
\mathbf{H}_k =
\begin{bmatrix}
1 & 0 & 0 & 0 \\
0 & 1 & 0 & 0
\end{bmatrix},
\]
and $\mathbf{v}_k = [v_{\phi,k}, v_{\theta,k}]^{\!\top}$ is zero-mean measurement noise with covariance $\mathbf{R}_k$:
$\mathbf{v}_k \sim \mathcal{N}(0, \mathbf{R}_k)$.

The standard EKF prediction step is
\begin{equation}
\hat{\mathbf{x}}_{k|k-1}
= \mathbf{F}_k \, \hat{\mathbf{x}}_{k-1|k-1} + \mathbf{B}_k \, \mathbf{u}_k,
\label{equ:state_predict}
\end{equation}
\begin{equation}
\mathbf{P}_{k|k-1}
= \mathbf{F}_k \, \mathbf{P}_{k-1|k-1} \, \mathbf{F}_k^\top + \mathbf{Q}_k,
\label{equ:cov_predict}
\end{equation}
and the update step is
\begin{equation}
\mathbf{K}_k
= \mathbf{P}_{k|k-1} \, \mathbf{H}_k^\top
\left( \mathbf{H}_k \, \mathbf{P}_{k|k-1} \, \mathbf{H}_k^\top + \mathbf{R}_k \right)^{-1},
\label{equ:kalman_gain}
\end{equation}
\begin{equation}
\hat{\mathbf{x}}_{k|k}
= \hat{\mathbf{x}}_{k|k-1}
+ \mathbf{K}_k \bigl( \mathbf{z}_k - \mathbf{H}_k \, \hat{\mathbf{x}}_{k|k-1} \bigr),
\label{equ:state_update}
\end{equation}
\begin{equation}
\mathbf{P}_{k|k}
= \bigl( \mathbf{I} - \mathbf{K}_k \mathbf{H}_k \bigr) \, \mathbf{P}_{k|k-1},
\label{equ:cov_update}
\end{equation}
where $\mathbf{P}_{k|k}$ is the state covariance and $\mathbf{I}$ is the $4\times 4$ identity.
\subsection{EKF Bias Propagation Under Gyro Injection}
\label{sec:ekf_bias_derivation}

Following \cite{meng2025mars}, we consider an EKF that estimates roll $\phi$ and pitch $\theta$ by fusing gyroscope and accelerometer tilt. We denote by $K_{\phi,k}$ and $K_{b,p,k}$ the scalar Kalman gains multiplying the roll-angle innovation for the $\hat\phi$ and $\hat b_p$ updates, respectively. To make the noise structure explicit, we parameterize
\begin{equation}
\begin{aligned}
\mathbf{Q}_k &= \mathrm{diag}\!\Bigl(
    \sigma_{gr,\phi}^2\,\Delta t^2,\;
    \sigma_{gr,\theta}^2\,\Delta t^2,\;
    \sigma_{b,p}^2\,\Delta t,\;
    \sigma_{b,q}^2\,\Delta t
\Bigr), \\
\mathbf{R}_k &= \mathrm{diag}\!\Bigl(
    \sigma_{acc,\phi}^2,\;
    \sigma_{acc,\theta}^2
\Bigr),
\end{aligned}
\label{equ:QR_app}
\end{equation}
where $\sigma_{gr,\cdot}$ denotes gyroscope noise, $\sigma_{b,\cdot}$ the random-walk variance of the corresponding bias state, and $\sigma_{acc,\cdot}$ the accelerometer noise. For simplicity and reflecting typical tuning symmetry, we assume
$\sigma_{gr,\phi}=\sigma_{gr,\theta}=\sigma_{gr}$ and
$\sigma_{acc,\phi}=\sigma_{acc,\theta}=\sigma_{acc}$.

\paragraph{Injected rate and roll prediction.}
Within one period of the attack signal, consider the roll channel. The measured roll rate under attack is
\begin{equation}
p_{\mathrm{meas},k}=p_k+p_{m,k},\qquad p_{m,k}=\mu_p+\tilde p_k,
\label{equ:bias_def_app}
\end{equation}
where $p_k$ is the true rate, $\mu_p$ is the per-cycle mean (DC) of the injected signal, and $\tilde p_k$ is its zero-mean AC part (cf. the rectified asymmetric wave in Sec.~\ref{sec:amplitude}). The pitch case is analogous.

In EKF prediction,
\begin{equation}
\hat{\phi}_{k|k-1}
= \hat{\phi}_{k-1|k-1}
+ \bigl(p_{\mathrm{meas},k}-\hat{b}_{p,k-1|k-1}\bigr)\Delta t.
\label{equ:phi_predict_app}
\end{equation}
Substituting~\eqref{equ:bias_def_app} yields
\begin{equation}
\hat{\phi}_{k|k-1}
= \hat{\phi}_{k-1|k-1}
+ \bigl(p_k+\tilde p_k+\mu_p-\hat{b}_{p,k-1|k-1}\bigr)\Delta t.
\label{equ:phi_predict_sub_app}
\end{equation}
In the attack-free case,
\[
\hat{\phi}^{\mathrm{free}}_{k|k-1}
= \hat{\phi}_{k-1|k-1}+p_k\Delta t.
\]

\paragraph{Weak accelerometer correction.}
The accelerometer-based tilt update is
\begin{equation}
\hat{\phi}_{k|k}
= \hat{\phi}_{k|k-1}
+K_{\phi,k}\bigl(z_{\phi,k}-\hat{\phi}_{k|k-1}\bigr).
\label{equ:phi_update_app}
\end{equation}
As in \hyperref[asp:1]{Setting~1}, when accelerometer updates are down-weighted or gated (inflated $\mathbf R_k$), $K_{\phi,k}$ becomes small and, to first order,
\[
\hat{\phi}_{k|k}\approx \hat{\phi}_{k|k-1}.
\]
Define the roll estimation error relative to attack-free as
\begin{equation}
\Delta \varepsilon_{\phi,k}\triangleq
\hat{\phi}_{k|k}-\hat{\phi}^{\mathrm{free}}_{k|k}.
\label{equ:eps_def_app}
\end{equation}
Combining~\eqref{equ:phi_predict_sub_app} with the weak-correction approximation gives
\begin{equation}
\Delta \varepsilon_{\phi,k}\approx
\bigl(\tilde p_k+\mu_p-\hat b_{p,k-1|k-1}\bigr)\Delta t.
\label{equ:bias_final_app}
\end{equation}

\paragraph{Slow bias adaptation.}
The EKF bias-state update is
\begin{equation}
\hat b_{p,k|k}
=\hat b_{p,k|k-1}
+K_{b,p,k}\bigl(z_{\phi,k}-\hat{\phi}_{k|k-1}\bigr).
\label{equ:bias_update_app}
\end{equation}
Under \hyperref[asp:2]{Setting~2}, small $\sigma_{b,p}$ makes the bias model stiff and yields a small $K_{b,p,k}$, so $\hat b_p$ adapts slowly to low-frequency changes. Over the time scale on which $\varepsilon_{\phi,k}$ grows, we approximate
\[
\hat b_{p,k|k}\approx \hat b_{p,k-1|k-1}.
\]
Substituting into~\eqref{equ:bias_final_app} yields
\begin{equation}
\frac{\Delta \varepsilon_{\phi,k}}{\Delta t}\approx \mu_p+\tilde p_k,
\label{equ:conclusion_app}
\end{equation}
i.e., the injected DC component $\mu_p$ appears as a persistent roll (or pitch) bias, while $\tilde p_k$ contributes only zero-mean fluctuations.

In summary, in the initial attack phase, an asymmetric rectified periodic injection yields a per-cycle mean rate $\mu_p$ and a zero-mean component $\tilde p_k$. When accelerometer corrections are weakened and the bias state adapts slowly, the EKF carries $\mu_p$ through as a persistent roll/pitch bias, while $\tilde p_k$ appears as small zero-mean fluctuations.

\subsection{Control Derivation and Boundary Conditions}
\label{app:control}

\subsubsection{From cascaded loops to \eqref{equ:y_disp}}
PX4-like position control computes a lateral velocity setpoint
\[
v_{y,sp}=K_{pos,P}(y_{des}-y_{cur})=K_{pos,P}\Delta y,
\]
and a desired lateral acceleration (P term shown; integral omitted for clarity in the operating regime)
\begin{equation}
\begin{aligned}
    a_{y,des}&\approx K_{vel,P}(v_{y,sp}-v_{y,cur}) \\
    &\approx K_{vel,P}v_{y,sp} \\
    &=K_{vel,P}K_{pos,P}\Delta y.
\end{aligned}    
\end{equation}
For small angles, the attitude mapping gives $\phi_{sp}\approx a_{y,des}/g$, hence
\[
\phi_{sp}\approx \frac{K_{vel,P}K_{pos,P}}{g}\Delta y.
\]
With biased EKF feedback $\hat\phi=\phi+\varepsilon_\phi$ and tight attitude tracking $\hat\phi\approx \phi_{sp}$, we have
$\phi\approx \phi_{sp}-\varepsilon_\phi$ and therefore
$a_y \approx g\tan\phi \approx g(\phi_{sp}-\varepsilon_\phi)$, which yields
$\Delta y \approx g\varepsilon_\phi/(K_{vel,P}K_{pos,P})$ when $a_y\to 0$.

\subsubsection{Practical limits and larger-excursion regimes}
\textbf{Tilt bounds.} Autopilots impose explicit tilt limits (e.g., PX4 \texttt{MPC\_TILTMAX\_AIR})~\cite{meier2015px4}, implying
$|\varepsilon_\phi|\le \Phi^*$ and
\begin{equation}
|\Delta y_{\max}| \approx \frac{g\,\Phi^*}{K_{vel,P}K_{pos,P}}.
\label{equ:y_max}
\end{equation}

\textbf{Concrete scaling.} With representative tunings $K_{vel,P}\approx 0.3~\mathrm{s^{-1}}$ and $K_{pos,P}\approx 1.0~\mathrm{s^{-1}}$, a $1^\circ$ residual tilt ($0.01745$~rad) yields
$|\Delta y|\approx 9.81\cdot 0.01745/(0.3\cdot 1.0)\approx 0.6$~m per degree of bias.

\textbf{Setpoint saturation / weak absolute-position correction.}
If the outer loop cannot command sufficient counter-tilt (e.g., $|\phi_{sp}|\le \phi_{sp}^{\max}$, integrator clamping, or degraded GNSS/vision updates), then $\phi_{sp}$ may not reach $\varepsilon_\phi$, leaving $a_y\approx g(\phi_{sp}-\varepsilon_\phi)\neq 0$ and producing continued drift until other limits (tilt/velocity bounds or failsafes) intervene. This is the regime in which excursions can grow substantially beyond the equilibrium offset predicted by \eqref{equ:y_disp}.

\subsection{Aliasing-Based Low-Frequency Components and Why Sweeps Work}
\label{sec:sweep_and_aliased}

When a MEMS gyroscope is driven near a mechanical resonance, its sensing element oscillates at (or close to) the acoustic excitation frequency. A standard idealized model for the resulting analog signal at the sensor output is
\begin{equation}
    s(t)=A \sin(2\pi F t+\varphi_0),
    \label{equ:analog}
\end{equation}
where $F$ is the physical excitation frequency and $A$ is the effective amplitude, determined by the source level and propagation/structural coupling.

\paragraph{Stage-1 sampling (IMU).}
The IMU samples this analog waveform at rate $F_S^{\mathrm{IMU}}$ (ADC and/or internal digital processing), producing
\begin{equation}
    s[i]=A \sin\!\left(2\pi F \tfrac{i}{F_S^{\mathrm{IMU}}}+\varphi_0\right), \quad i=0,1,2,\ldots .
    \label{equ:digitized}
\end{equation}
Because commodity inertial sensors often have resonant frequencies well above their estimator-rate bandwidth (gyros commonly in the $\sim$20--30\,kHz band) \cite{son2015rocking,tu2018injected,trippel2017walnut,gao2022kite}, while usable estimation rates are typically a few hundred hertz, the condition $F>\tfrac{1}{2}F_S^{\mathrm{IMU}}$ is easy to satisfy. In that regime, aliasing is unavoidable: the discrete-time sequence in (\ref{equ:digitized}) is indistinguishable from a lower-frequency sinusoid.

\paragraph{Stage-2 rate conversion (flight stack).}
Autopilot stacks such as PX4 do not necessarily consume the raw IMU stream at the ADC/internal rate. Instead, the estimator consumes an internal sensor/estimator update stream at rate $F_S^{\mathrm{PX4}}$ (e.g., 250\,Hz), obtained via downsampling and filtering. This introduces an additional rate conversion stage that can further fold frequencies into the EKF band.

\begin{lemma}
\label{lem:two_stage_alias}
\textbf{Two-stage aliasing (IMU $\rightarrow$ EKF stream).}
Let an analog sinusoid of frequency $F$ be sampled by the IMU at rate $F_S^{\mathrm{IMU}}$ and then consumed by the estimator at rate $F_S^{\mathrm{PX4}}$ after downsampling/rate conversion. The effective observed frequency at the estimator, $F_{\mathrm{obs}}$, can be expressed as
\begin{equation}
    F_{\mathrm{obs}}
    = \Bigl|\, \bigl|F - n F_S^{\mathrm{IMU}}\bigr| - m F_S^{\mathrm{PX4}} \Bigr|,
    \qquad n,m \in \mathbb{Z},
    \label{equ:alias}
\end{equation}
with $0 \le F_{\mathrm{obs}} \le \tfrac{1}{2}F_S^{\mathrm{PX4}}$.
\end{lemma}

\noindent\textit{Sketch.}
Sampling at $F_S^{\mathrm{IMU}}$ maps $F$ to an equivalent discrete-time frequency
$F_{\mathrm{IMU}}=\lvert F-nF_S^{\mathrm{IMU}}\rvert$ for some $n\in\mathbb{Z}$.
The subsequent rate conversion to an EKF stream at $F_S^{\mathrm{PX4}}$ yields
$F_{\mathrm{obs}}=\lvert F_{\mathrm{IMU}}-mF_S^{\mathrm{PX4}}\rvert$ for some $m\in\mathbb{Z}$.
Substitution gives (\ref{equ:alias}). \hfill$\square$

\paragraph{Consequence: low-frequency components are generically produced.}
By Lemma~\ref{lem:two_stage_alias}, the estimator sees a sinusoid
\begin{equation}
    s[k] = A \sin\!\left(2\pi F_{\mathrm{obs}} \tfrac{k}{F_S^{\mathrm{PX4}}} + \varphi_0\right),
    \qquad k=0,1,2,\ldots,
    \label{equ:new_digitized}
\end{equation}
where $F_{\mathrm{obs}}$ lies in the EKF band $[0,\,\tfrac{1}{2}F_S^{\mathrm{PX4}}]$ even when $F$ is ultrasonic. For a representative MPU-6500 + PX4 configuration, $F_S^{\mathrm{IMU}}\approx 1$\,kHz and $F_S^{\mathrm{PX4}}\approx 250$\,Hz \cite{invensense_mpu6500}.

\paragraph{Why sweep-and-lock is practical under uncertainty.}
In practice, an attacker does \emph{not} need to know the exact $(n,m)$ in (\ref{equ:alias}) or the exact resonant frequency of a specific sensor unit a priori; both vary with sensor model, manufacturing, mounting, and the autopilot’s sampling pipeline. Instead, the key observation is that the mapping $F \mapsto F_{\mathrm{obs}}$ is \emph{many-to-one}:

\begin{corollary}
\label{cor:preimage}
\textbf{Non-uniqueness (many excitation tones yield EKF-band aliases).}
Fix $F_S^{\mathrm{IMU}}$ and $F_S^{\mathrm{PX4}}$.
For any desired EKF-band frequency $f_d \in [0,\tfrac{1}{2}F_S^{\mathrm{PX4}}]$, the set of excitation frequencies that produce $F_{\mathrm{obs}}=f_d$ is
\[
\mathcal{S}(f_d)=\{\, nF_S^{\mathrm{IMU}} \pm mF_S^{\mathrm{PX4}} \pm f_d \;:\; n,m\in\mathbb{Z}\,\}.
\]
In particular, around each lattice point $L=nF_S^{\mathrm{IMU}} \pm mF_S^{\mathrm{PX4}}$ there are two solutions $F=L\pm f_d$ separated by $2f_d$.
\end{corollary}

Corollary~\ref{cor:preimage} formalizes why a continuous sweep is effective: there are many candidate ultrasonic frequencies that fold into the EKF band, and small perturbations in $F$ (or in effective sampling rates) typically move the system between nearby elements of $\mathcal{S}(f_d)$ rather than eliminating the alias phenomenon. Therefore, a feedback-driven sweep can empirically identify a carrier/modulation pair that yields a stable low-frequency component in the estimator band (as shown in Fig.~\ref{fig:sys_fig}), without requiring the attacker to deterministically ``set'' a specific $f_d$ across platforms.

\paragraph{Instantiation used in this paper.}
For clarity of analysis and repeatability, we often illustrate a representative low-frequency component (e.g., $\sim$1\,Hz). This choice is \emph{not} essential: the same mechanism applies for other $f_d$ within the EKF band, and in practice $f_d$ is selected by sweep-and-lock based on the strongest and most stable observed estimator response.

\subsection{Gyro Bias Scheduling Algorithm}
\label{sec:alg_app}
\begin{algorithm}[!ht]
\footnotesize
\caption{Phase-Gated Gyro Bias Scheduling for EKF Tilt Injection}
\label{alg:phase_gating}
\begin{algorithmic}[1]

\State \textbf{Inputs:} gyro stream $(t,g_x,g_y,g_z)$
\State \textbf{Params:} window length $T_w$, axis $a\in\{x,y\}$;
sign $s\in\{+,-\}$; number of attack periods $P$
\State \textbf{Attack Types:} TYPE-A (in-phase rising sine), TYPE-B (anti-phase falling sine, $180^\circ$ shifted)
\State \textbf{State:}
buffer $B$,
lock flag $\texttt{locked}\gets\text{False}$,
locked sine $(\hat f,\hat\phi_{\rm abs})$,
attack phase $E\in\{\text{TYPE-A},\text{TYPE-B}\}$,
attack counter $k\gets 0$

\vspace{0.3em}
\Procedure{OnNewSample}{$t,g_x,g_y,g_z$}
  \State Append $(t,g_x,g_y,g_z)$ to $B$
  \State Remove all samples in $B$ with time $< t - T_w$
  \If{\Call{EnoughSamples}{$B$} = \textbf{false}} \Return \EndIf
  \If{\textbf{not} \texttt{locked}}
    \State \Call{FitSineOnAxis}{$B,a,N_c$}
  \EndIf
\EndProcedure

\vspace{0.3em}
\Procedure{FitSineOnAxis}{$B,a,N_c$}
  \State Extract timestamps $(t_i)$ and axis data $(g_i^a)$ from $B$
  \State Fit $g_i^a \approx A\sin(2\pi \hat f t_i + \hat\phi)$ by least squares
  \If{fit quality not poor}
    \State Compute absolute phase $\hat\phi_{\rm abs}$
    \State \Call{PhaseGateAttack}{$t,\hat f,\hat\phi_{\rm abs},a,s$}
  \EndIf
\EndProcedure

\vspace{0.3em}
\Procedure{PhaseGateAttack}{$t,\hat f,\hat\phi_{\rm abs},a,s$}
  \If{$k \ge P$}
    \State publish \texttt{/attack\_enable}$\gets 1$
  \EndIf

  \State Predict next peak $t_{\rm peak}$ and trough $t_{\rm trough}$
  \State $\texttt{locked} \gets \text{True}$

  \If{$s = +$}
    \Comment TYPE-A: in-phase, rising; inject during peak$\rightarrow$trough
    \State $t_{\mathrm{on}} \gets t_{\rm peak}$
    \State $t_{\mathrm{off}} \gets t_{\rm trough}$
    \State $E \gets \text{TYPE-A}$
  \Else
    \Comment TYPE-B: anti-phase ($180^\circ$), falling; inject during trough$\rightarrow$peak
    \State $t_{\mathrm{on}} \gets t_{\rm trough}$
    \State $t_{\mathrm{off}} \gets t_{\rm peak}$
    \State $E \gets \text{TYPE-B}$
  \EndIf

  \If{$t \ge t_{\mathrm{on}}$ and $t < t_{\mathrm{off}}$}
    \State publish \texttt{/attack\_enable}$\gets 1$
  \Else
    \State publish \texttt{/attack\_enable}$\gets 0$
  \EndIf

  \If{$t \ge t_{\mathrm{on}}$}
    \State $k \gets k + 1$
  \EndIf
\EndProcedure

\end{algorithmic}
\end{algorithm}

\begin{figure}[!t]
    \centering
    \begin{subfigure}[b]{0.48\textwidth}
        \centering
        \includegraphics[width=\textwidth,trim=0mm 0 0mm 0,clip]{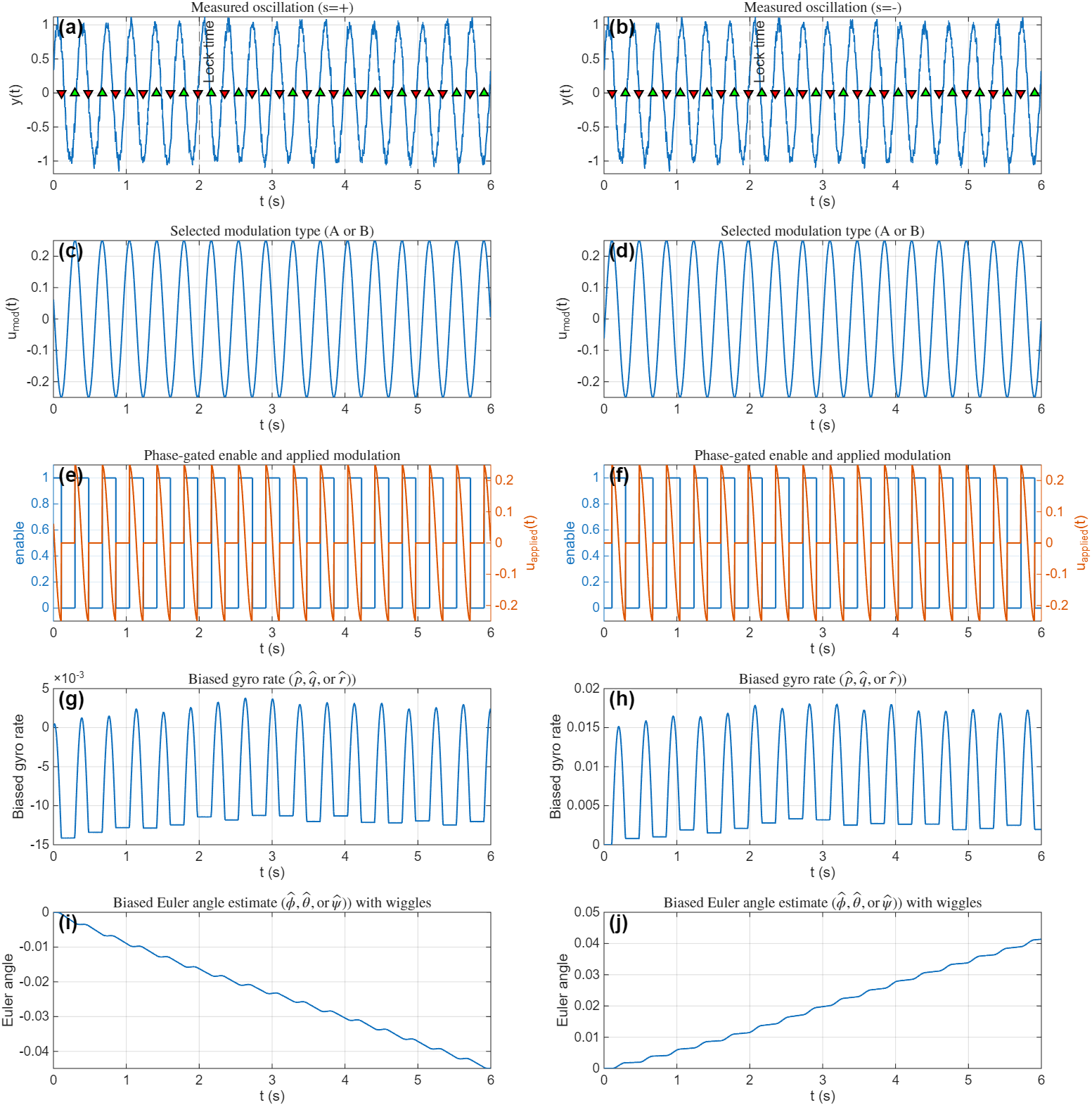}
    \end{subfigure}
    
    % \vspace{1em} % space between rows
        
    \caption{Demonstration of Alg.~\ref{alg:phase_gating}: phase locking, peak/trough-gated enabling, and Type-A/Type-B ($180^\circ$) selection produce a net signed accumulated effect. Left: $s=+$ (Type-A, peak-aligned gating). Right: $s=-$ (Type-B, trough-aligned gating).}
 
    \label{fig:AGR_demo}
\end{figure}

The proposed phase-aware AGR triggering procedure is summarized in Alg.~\ref{alg:phase_gating}. Its core mechanism—\emph{phase locking}, \emph{peak/trough gating}, and \emph{Type-A/Type-B} ($180^\circ$) selection—is visualized in Fig.~\ref{fig:AGR_demo}.

Figure~\ref{fig:AGR_demo} walks through the end-to-end scheduling logic of Alg.~\ref{alg:phase_gating} in a low-frequency (few-Hz) oscillation regime. The left column targets a positive signed effect ($s=+$), whereas the right column targets a negative signed effect ($s=-$). Subplots (a)--(j) follow the pipeline from gyro stream observation, to locking a sinusoidal phase model, to deterministic phase-gated enabling, and finally to the accumulated signed effect.

In Fig.~\ref{fig:AGR_demo}, panels (a)--(b) show the measured oscillation on the chosen gyro axis (illustrated with $g_y(t)$) used by \textsc{OnNewSample} and \textsc{FitSineOnAxis}. The vertical dashed line marks the \emph{lock time}, after which the fitted sinusoid parameters $(\hat f,\hat\phi_{\rm abs})$ are held fixed (\texttt{locked}$\leftarrow$True) as in Alg.~\ref{alg:phase_gating}. With the locked phase, \textsc{PhaseGateAttack} predicts upcoming extrema of $\sin(2\pi \hat f t+\hat\phi_{\rm abs})$: green upward triangles denote predicted peaks $t_{\rm peak}$ and red downward triangles denote predicted troughs $t_{\rm trough}$. These extrema act as phase anchors for gating, eliminating repeated refits and improving tolerance to sampling jitter.

Panels (c)--(d) show the selected injection waveform determined by the sign parameter. The attack provides two sinusoidal types separated by $180^\circ$: Type-A (in-phase) and Type-B (anti-phase). When $s=+$, the scheduler selects Type-A; when $s=-$, it selects Type-B. This choice enforces the intended local slope at the activation instant (rising vs.\ falling) so that repeated, phase-consistent activation yields a consistently signed cumulative effect.

Panels (e)--(f) correspond to \textsc{PhaseGateAttack}. The blue square wave is the published enable signal (\texttt{/attack\_enable}); the orange trace is the applied modulation $u_{\rm applied}(t)$ (the selected Type-A/Type-B waveform multiplied by the gate). The gate is aligned to the appropriate half-cycle of the locked oscillation: for $s=+$, enabling is scheduled around peak$\rightarrow$trough; for $s=-$, enabling is scheduled around trough$\rightarrow$peak. Thus, the algorithm converts a continuous carrier into discrete, phase-consistent injection windows, while allowing timing offsets (e.g., $\Delta t_{\rm on},\Delta t_{\rm off}$) to compensate end-to-end actuation latency.

Panels (g)--(h) plot the biased gyro reading after modulation, $(\hat p, \hat q, \mathrm{and }\,\hat r)$, illustrating cycle-by-cycle accumulation under repeated phase-gated injection (i.e., a signed gyro-rate bias). Panels (i)--(j) integrate the biased gyro and generate an Euler angle estimate (i.e.,  $\hat\phi, \hat\theta, \mathrm{and}\,\hat\psi)$), yielding a slowly drifting line with small oscillatory ``wiggles'' inherited from the gated waveform (i.e., biased gyro rates). Importantly, these oscillations are the same structure exploited by \textsc{FitSineOnAxis} to estimate and lock $(\hat f,\hat\phi_{\rm abs})$, closing the scheduling loop. The drift direction flips between the two columns, confirming that $s$ (implemented via Type-A/Type-B selection and peak/trough-gated enabling) controls the long-term sign of the accumulated Euler-angle bias, as intended by Alg.~\ref{alg:phase_gating}.

\subsection{Speaker-Detached IMU Benchmark Testbed}
\label{app:imu_benchmark}

\begin{figure}[!t]
    \centering
    \begin{subfigure}{0.45\textwidth}
    \centering
        \includegraphics[width=0.70\textwidth,trim=0 0 0 0,clip]{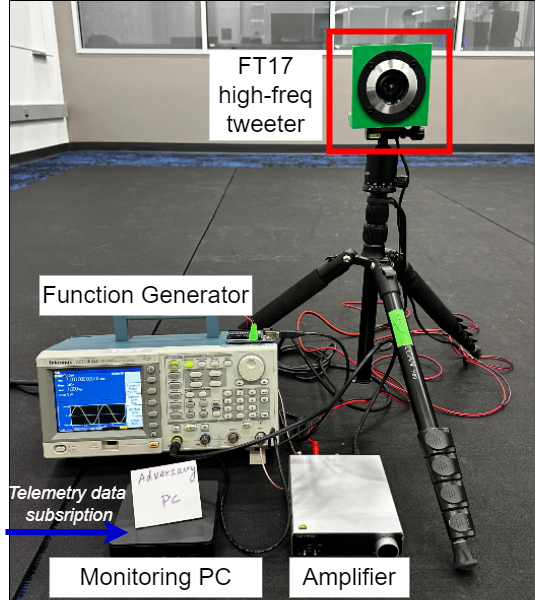}
        % \caption{}
    \end{subfigure}
    \caption{The acoustic testbed for different IMU/gyro sensor benchmarking. }
    \label{fig:imu_benchmark}
\end{figure}

We implement a speaker-detached IMU/gyro benchmark testbed, as shown in Fig.~\ref{fig:imu_benchmark}, to characterize sensor susceptibility under controlled acoustic excitation. Unlike the main flight experiments, where the tweeter is mounted on the UAV to ensure reproducible acoustic coupling, this benchmark places the ultrasonic projector physically separate from the UAV. The projector is positioned behind the sensor under evaluation, corresponding to the body-frame $+x$ direction, so that all IMUs are tested under a consistent attacker--victim geometry.

The UAV is constrained on a \textit{3-DoF testing platform}~\cite{sharma2023open,kim2023development,ahmad2024attitude}, while running the PX4 estimator in the loop. This setup allows the IMU, estimator, and attitude response to remain active while avoiding uncontrolled free-flight motion during high-intensity acoustic probing. For each sensor, we sweep or lock to the responsive resonant band and record the induced gyro disturbance, affected axes, and resulting attitude-estimation error. The same Fostex FT17 ultrasonic projector is used across all measurements to make results comparable across IMU/gyro models.

This benchmark is designed to evaluate IMU generality and standoff behavior rather than to demonstrate a fully detached moving-target attack. In particular, it isolates how acoustic coupling varies with sensor model, mounting, affected axis, and incidence direction. The measured quantities, including $D_{\max}$ and $T_{\min}$ in Tbl.~\ref{tbl:imu_generality}, therefore characterize whether a given IMU can support the gyro-bias pathway under detached excitation. Fully detached free-flight deployment, which would additionally require target detection, tracking, beam steering, and online geometry control, is left to future work.

\subsection{Outdoor Experiment Results}
\label{app:outdoor}

\begin{table}[t]
  \centering
  \footnotesize
  \caption{Phase-gated trigger schedule for the outdoor run: marker times and target sign (+/-). The attack signal was manually interrupted after T2, T6, T11, and T14, with each new attack segment launched shortly after signal reactivation.}
  \label{tab:trigger_outdoor}
  
  % ----------- First 8 -----------
  \begin{tabular}{c|cccccccc}
    \hline
    Index    & T1    & T2   & T3   & T4   & T5    & T6    & T7    & T8    \\
    \hline
    Time (s) & 21.5  & 24   & 26.0 & 28   & 32.5  & 33.5  & 35.5  & 36.5  \\
    Sign     & +     & --   & +    & +    & +     & --    & +     & +     \\
    \hline
  \end{tabular}

  \vspace{4pt}

  % ----------- Remaining 6 ----------
  \begin{tabular}{c|cccccc}
    \hline
    Index    & T9   & T10  & T11  & T12  & T13   & T14   \\
    \hline
    Time (s) & 39.5 & 41   & 49   & 51   & 53.5  & 54.5  \\
    Sign     & +    & +    & --   & +    & +     & --    \\
    \hline
  \end{tabular}
\end{table}

\begin{figure}[!t]
    \centering
    \begin{subfigure}{0.42\textwidth}
    \centering
        \includegraphics[width=0.99\linewidth]{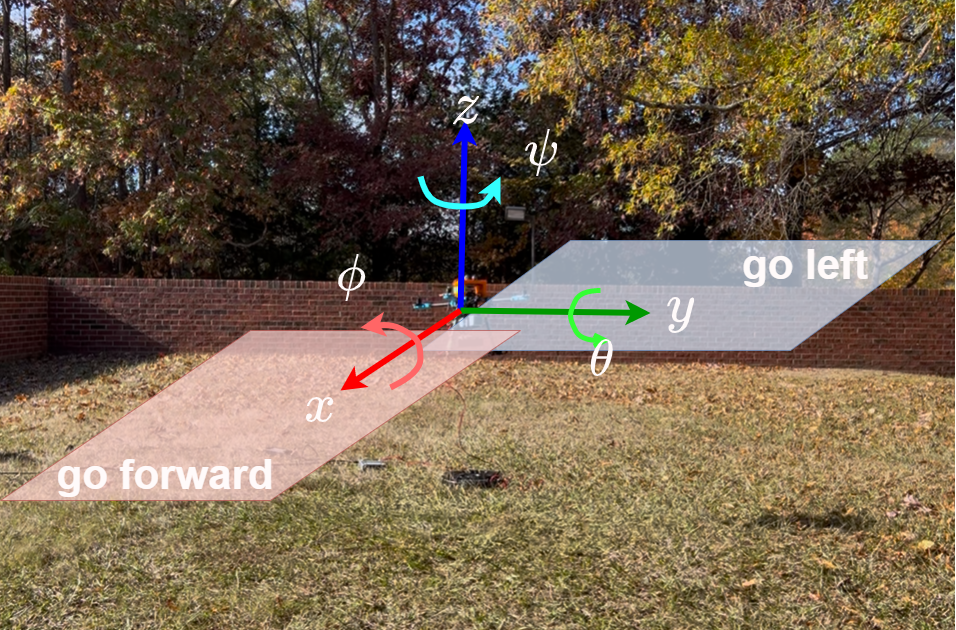}
        % \caption{}
    \end{subfigure}
    \caption{The outdoor experiment setting. }
    \label{fig:outdoor_setting}
\end{figure}

\begin{figure}[!t]
    \centering
    \begin{subfigure}{0.49\textwidth}
    \centering
        \includegraphics[width=0.99\linewidth]{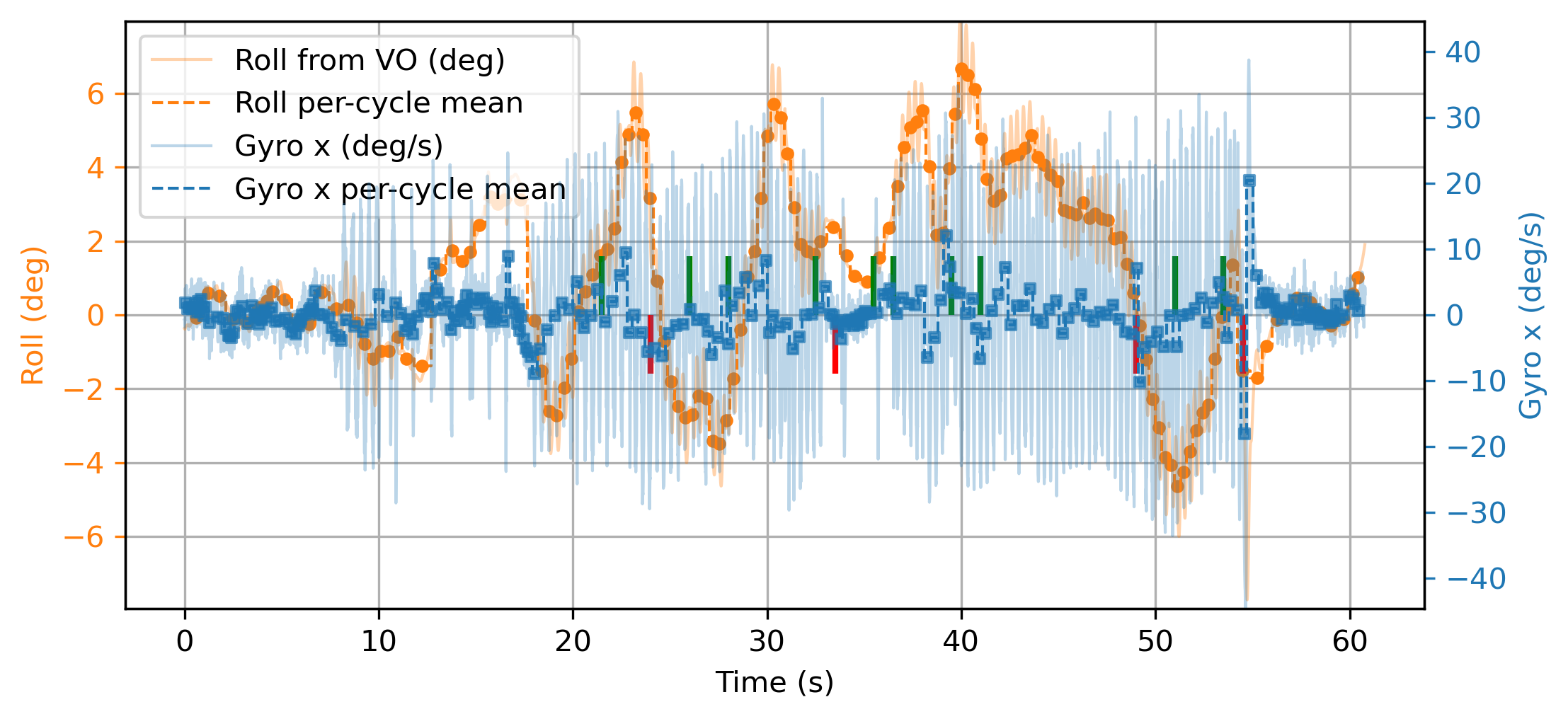}
        % \caption{}
    \end{subfigure}
    \caption{Relationship between x-axis gyroscope rate ($p$) and roll angle ($\phi$).}
    \label{fig:outdoor_gyro_x_vo_roll}
\end{figure}

\begin{figure}[!t]
    \centering
    \begin{subfigure}{0.49\textwidth}
    \centering
        \includegraphics[width=0.99\linewidth]{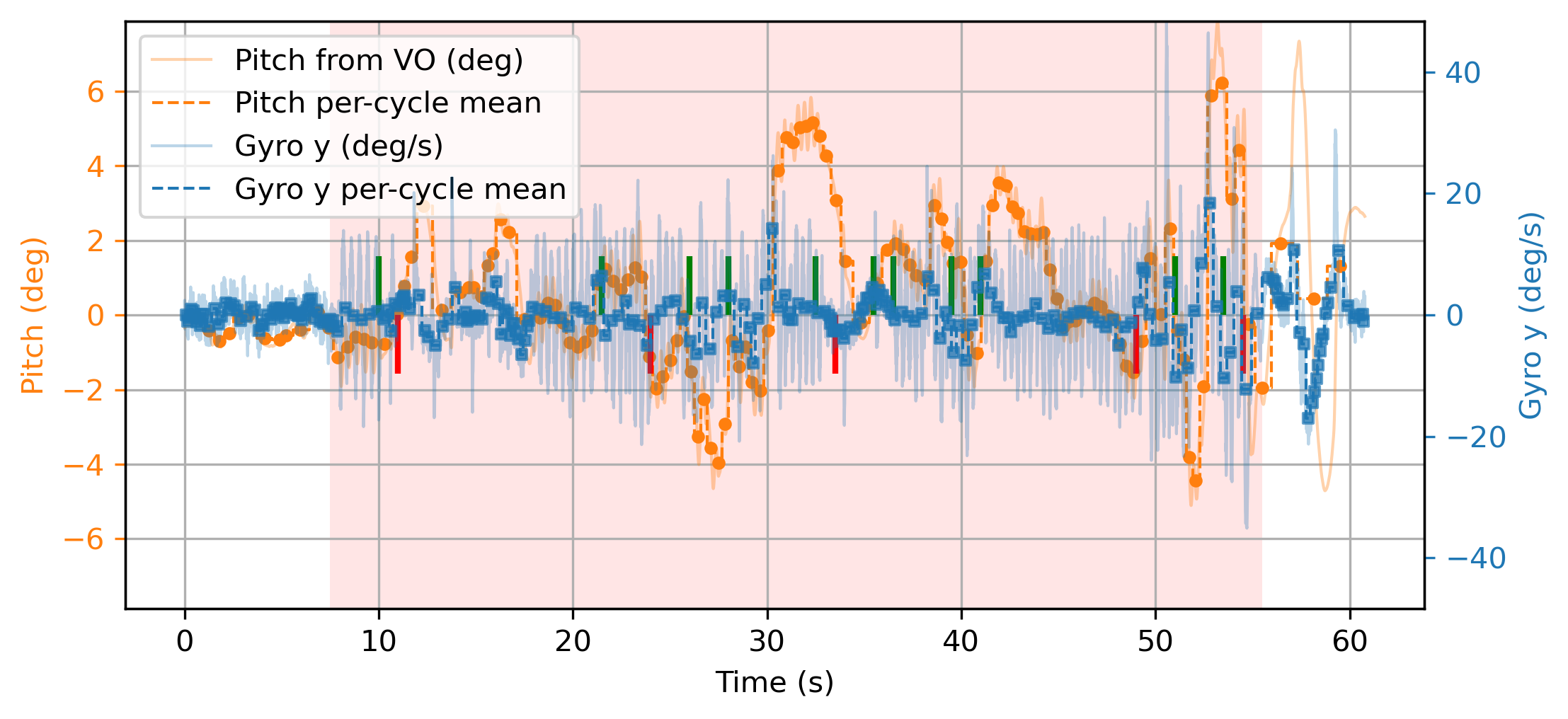}
        % \caption{}
    \end{subfigure}
    \caption{Relationship between y-axis gyroscope rate ($q$) and pitch angle ($\theta$).}

    \label{fig:outdoor_gyro_y_vo_pitch}
\end{figure}

\begin{figure}[!t]
    \centering
    \begin{subfigure}{0.49\textwidth}
    \centering
        \includegraphics[width=0.99\linewidth]{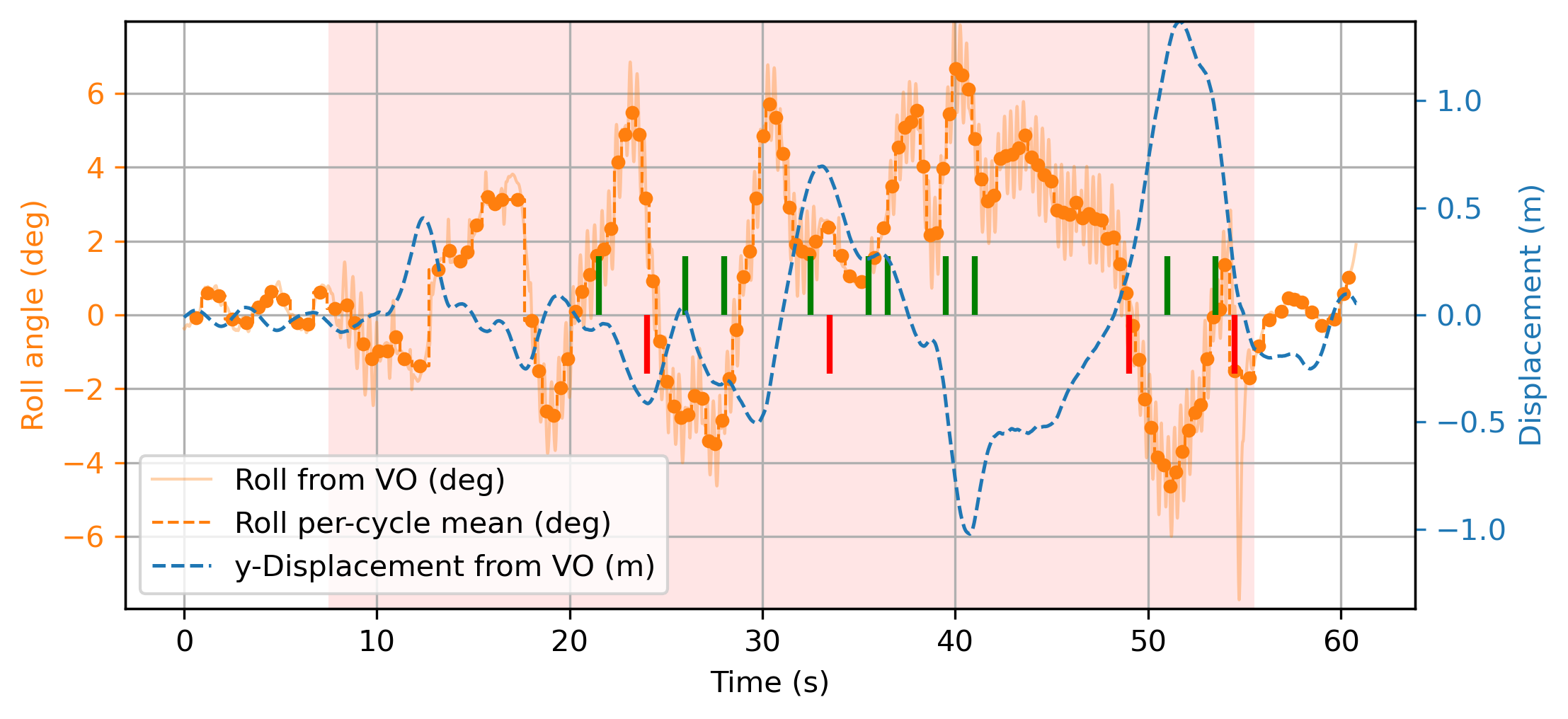}
        % \caption{}
    \end{subfigure}
    \caption{Relationship between biased roll angle and lateral $y$-displacement; red shading marks the attack window.}
    \label{fig:outdoor_roll_vo_y_vo}
\end{figure}

\begin{figure}[!t]
    \centering
    \begin{subfigure}{0.49\textwidth}
    \centering
        \includegraphics[width=0.99\linewidth]{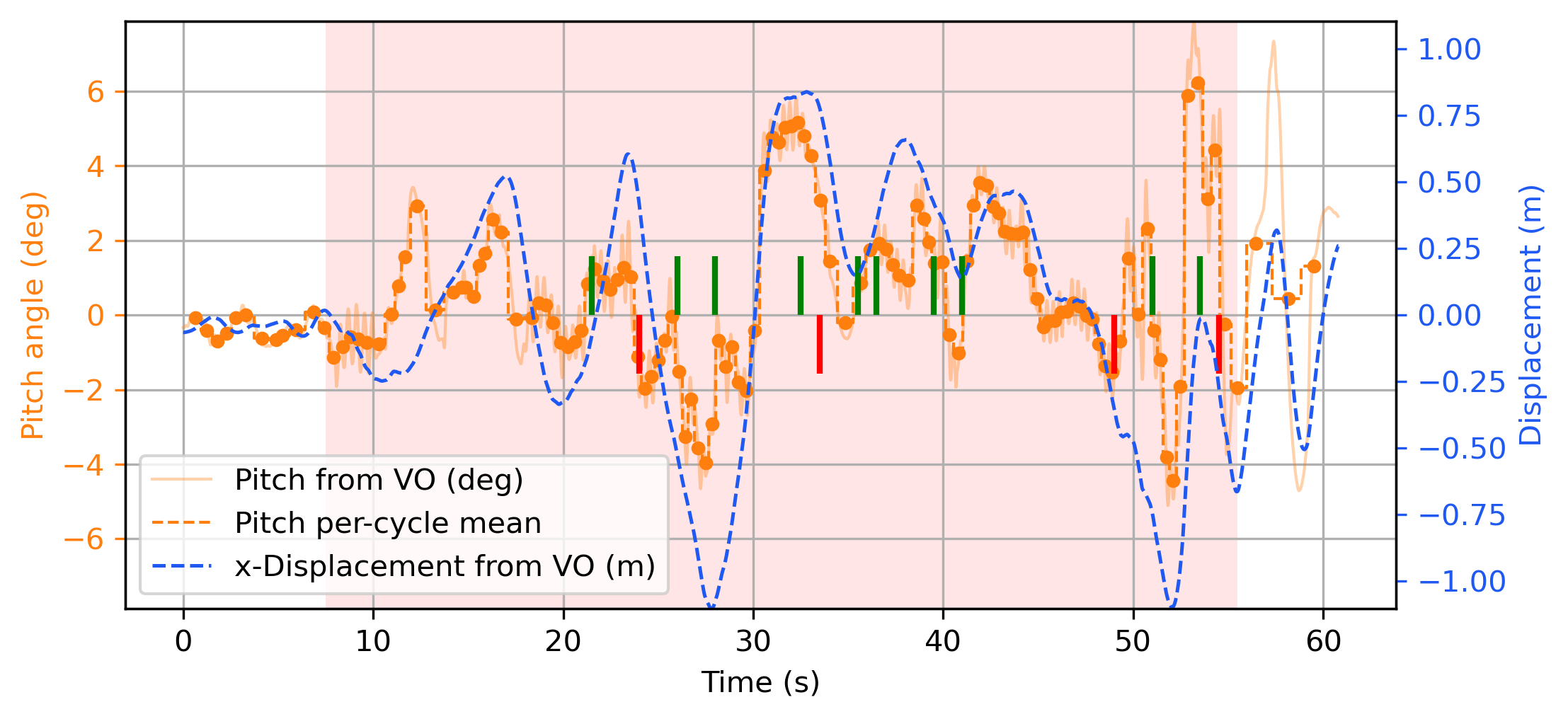}
        % \caption{}
    \end{subfigure}
   \caption{Relationship between biased pitch angle and lateral $x$-displacement; red shading marks the attack window.}

    \label{fig:outdoor_pitch_vo_x_vo}
\end{figure}

We validated \modelAbbr outdoors using the same controller gains as indoors, replacing VICON with onboard GNSS + optical flow (with magnetometer and optional barometer). The world and body frames are defined in Fig.~\ref{fig:outdoor_setting}. In a $\sim\!60$\,s run, the attack was enabled from $t_0=7$\,s to $t_1=55$\,s, during which 16 AGR triggers were issued to bias roll and induce lateral ($y$) drift. Because commodity GNSS provides only $\sim\!3$–$5$\,m absolute accuracy~\cite{DOT-VNTSC-OST-05-01}, we report relative motion from the drone's fused state, namely vehicle odometry (VO); no independent ground truth is available, and small estimator biases may remain.

As shown in Figs.~\ref{fig:outdoor_gyro_x_vo_roll}--\ref{fig:outdoor_gyro_y_vo_pitch}, injected gyro-rate biases ($p\!\to$ roll, $q\!\to$ pitch) yield the expected Euler-angle deviations, producing up to $\sim\!1.5$\,m lateral displacement (Fig.~\ref{fig:outdoor_roll_vo_y_vo}) with concurrent $x$-axis motion (Fig.~\ref{fig:outdoor_pitch_vo_x_vo}). The times and signs of the triggers fired by Alg.~\ref{alg:phase_gating} are documented in Tbl.~\ref{tab:trigger_outdoor}. Wind gusts increased vibration and reduced the SNR of the aliased oscillation, making per-cycle sine fitting less stable and occasionally forcing the AGR gate to skip cycles when quality checks failed; we discuss compute/robustness trade-offs in Sec.~\ref{sec:limitation}.

\end{document}